\documentclass[superscriptaddress, prb, noeprint, amsmath,amssymb,aps,twocolumn]{revtex4-2}

\newcommand{\figref}[2]{Fig.~\ref{#1}{\bf{}#2}}
\newcommand{\ket}[1]{\ensuremath{\left\vert\!#1\right>}}%
\newcommand{\dbz}{\ensuremath{\Delta{}B_{z}}}%
\newcommand{\w} {\ensuremath{\omega}}
\newcommand{\eps}{\ensuremath{\epsilon}}
\newcommand{\sx}{\ensuremath{\sigma_x}}
\newcommand{\sy}{\ensuremath{\sigma_y}}
\newcommand{\sz}{\ensuremath{\sigma_z}}
\newcommand{\up}{\ensuremath{\uparrow}}
\newcommand{\down}{\ensuremath{\downarrow}}
\newcommand{\tnot}{\ensuremath{\frac{1}{\sqrt{2}}\left(\ket{\up \down} + \ket {\down \up}\right)}}%
\newcommand{\sing}{\ensuremath{\frac{1}{\sqrt{2}}\left(\ket{\up \down} - \ket {\down \up}\right)}}%

\usepackage{xpatch}

\makeatletter
\xpatchcmd{\@ssect@ltx}{\@xsect}{\protected@edef\@currentlabelname{#8}\@xsect}{}{}
\xpatchcmd{\@sect@ltx}{\@xsect}{\protected@edef\@currentlabelname{#8}\@xsect}{}{}
\makeatother

\usepackage{hyperref}
\usepackage{dcolumn}
\usepackage{xcolor}
\usepackage{upgreek}
\usepackage{graphicx}
\usepackage{fixme}

\hypersetup{bookmarksnumbered=true,	unicode=false,	pdfstartview={FitH}, pdfnewwindow=true, colorlinks=true,linkcolor=blue, citecolor=blue, filecolor=blue, urlcolor=blue}

\usepackage{verbatim}

\begin{document}


\title{Quantum Dots / Spin Qubits}
\author{Shannon Harvey}

\affiliation{Stanford Institute of Materials and Energy Sciences, SLAC National Accelerator Laboratory, Menlo Park, 94025, California, USA}
\affiliation{Department of Applied Physics, Stanford University,  Stanford, 94305, California, USA}

\begin{abstract}Summary: Spin qubits in semiconductor quantum dots represent a prominent family of solid-state qubits in the effort to build a quantum computer. They are formed when electrons or holes are confined in a static potential well in a semiconductor, giving them a quantized energy spectrum. The simplest spin qubit is a single electron spin located in a quantum dot, but many additional varieties have been developed, some containing multiple spins in multiple quantum dots, each of which has different benefits and drawbacks. While these spins act as simple quantum systems in many ways, they also experience complex effects due to their semiconductor environment. They can be controlled by both magnetic and electric fields depending on their configuration and are therefore dephased by magnetic and electric field noise, with different types of spin qubits having different control mechanisms and noise susceptibilities. While initial experiments were primarily performed in gallium arsenide (GaAs) based materials, silicon qubits have developed substantially and research on qubits in metal-oxide-semiconductor (Si-MOS), silicon/silicon germanium (Si/SiGe) heterostructures, and donors in silicon is also being pursued.  An increasing number of spin qubit varieties have attained error rates that are low enough to be compatible with quantum error correction for single-qubit gates and two-qubit gates have been performed in several with success rates, or fidelities, of 90-95\%.
\end{abstract}

\maketitle
\section{Introduction}

In the current research towards a quantum computer, one type of quantum bit is the spin qubit formed in quantum dots (QDs), where the logical subspace is a function of the charge carrier's spin values. There is a zoo of spin qubit varieties spanning different numbers of dots, materials, and geometries, and electrons and holes. This article will attempt to organize this large variety of qubits, explaining the common framework they share and how each qubit varies the basic components. While electron spins in quantum dots can act as nearly prototypical quantum systems, their semiconductor environment impacts the spin state in several ways, including spin-orbit interactions, electronic band structure effects, and interactions with nuclei. There are also benefits to working in a solid-state system, such as the ease of control and clear path to scaling. The intention of this article is to provide the reader with an understanding of the major types of quantum-dot qubits in semiconductors being pursued today, and therefore does not include important work that has been done on nanowires \citep{petersson2012} and nanotubes \citep{churchill2009a}, for instance, as well as much of the initial work that focused on creating stable quantum dots that can be operated as qubits. Moreover, spin qubits and quantum dots have many different applications, including quantum sensing in defects \citep{degen2017} and single photon emission in self-assembled quantum dots \citep{warburton2013}, but this article will focus on spin qubits in quantum dots as used for quantum computing. While quantum dots can be formed with holes, for simplicity electrons will be assumed for the remainder of the article.

Quantum dots were developed after the phenomena of Coulomb blockade was discovered \citep{larkin1984, averin1986, fulton1987} in aluminum tunnel junctions at cryogenic temperatures. In Coulomb blockade, the conductance across a small, isolated metallic system oscillates with the applied gate electrode voltage (to avoid confusion with quantum gate operations, in this article gate electrode will be used to refer to the metallic gates used to define and control quantum dots). This occurs because the system is small enough that the electrostatic energy required to add an additional electron is larger than the thermal energy, and the number of electrons within the system, which is known as a single electron transistor, becomes quantized. To form a quantum dot, it is also necessary that the Fermi wavelength of the electrons or holes is of the same order size as the area where the electrons are confined, which leads to an atom-like quantized energy level spacing \citep{beenakker1991a}. These are typically formed in semiconductors due to the ability to control the charge density with gate electrodes \citep{ashoori1996, kouwenhoven2001}.
There are several approaches to creating quantum dots. One uses structures with two or more layers of semiconductors, known as semiconductor heterostructures, where the electrons are confined to a two-dimensional electron plane. There are two distinct varieties of quantum dots formed with this technique; first, vertical quantum dots \citep{tarucha1996}, where the quantum dots are confined using etching, then controlled via gate electrodes on the side of the etched region. The second design is referred to as lateral gating, where there are a series of metal gates on the surface on the semiconductor that control the energy of the quantum dot, and has become the dominant technique for quantum dots used in quantum computing. In these quantum dots, the gate electrodes create an approximately parabolic potential in which the electrons are confined. These electrons have a quantized energy spectrum comparable to that of atoms, but because of their larger length scale, from tens to hundreds of nanometers, they have proportionally smaller energy spacing. Another type of quantum dot used for quantum computing is the donor, a single dopant atom with a bound charge in a semiconductor, which is an intrinsically zero-dimensional system, and therefore the gates are needed to control the energies of the various quantum dot states. Laterally-gated quantum dots and donor qubits will be focused on this article due to their prevalence in current quantum computing research.  

While there is a wide range of quantum-dot-based spin qubits, they share the same basic set of properties. Further complexity is provided by adding additional quantum dots, which allow for more control parameters, and in particular allow the qubit state to be encoded partially in the position of the electrons within the quantum dots, endowing them with a dipole moment. The different architectures and materials each have trade-offs to their controllability, speed, noise, and other properties, which will be addressed in the section on implementations. 

The first section on \nameref{section:qd} builds a simple model describing their essential properties and explains how they are controlled. The next section, \nameref{section:spinqubits}, explains how quantum dots become qubits, describing the Hamiltonians and control techniques for the major varieties of spin qubits, and how they vary across different materials. After looking at \nameref{section:single} and after describing how they work and their early results in GaAs/AlGaAs, the subsections discuss the different types of implementations in silicon, \nameref{section:simos}, \nameref{section:sige}, and \nameref{section:donors}. The remaining subsections cover some of the multiple quantum dot spin qubits, including \nameref{section:charge}, \nameref{section:st0} and \nameref{section:triple}. In the final section, \nameref{section:design}, some topics of marked importance for the future development of spin qubits as a scalable quantum computing platform are discussed. 


\section{Quantum dots}
\label{section:qd}
\begin{figure*}
	\includegraphics[scale=0.4]{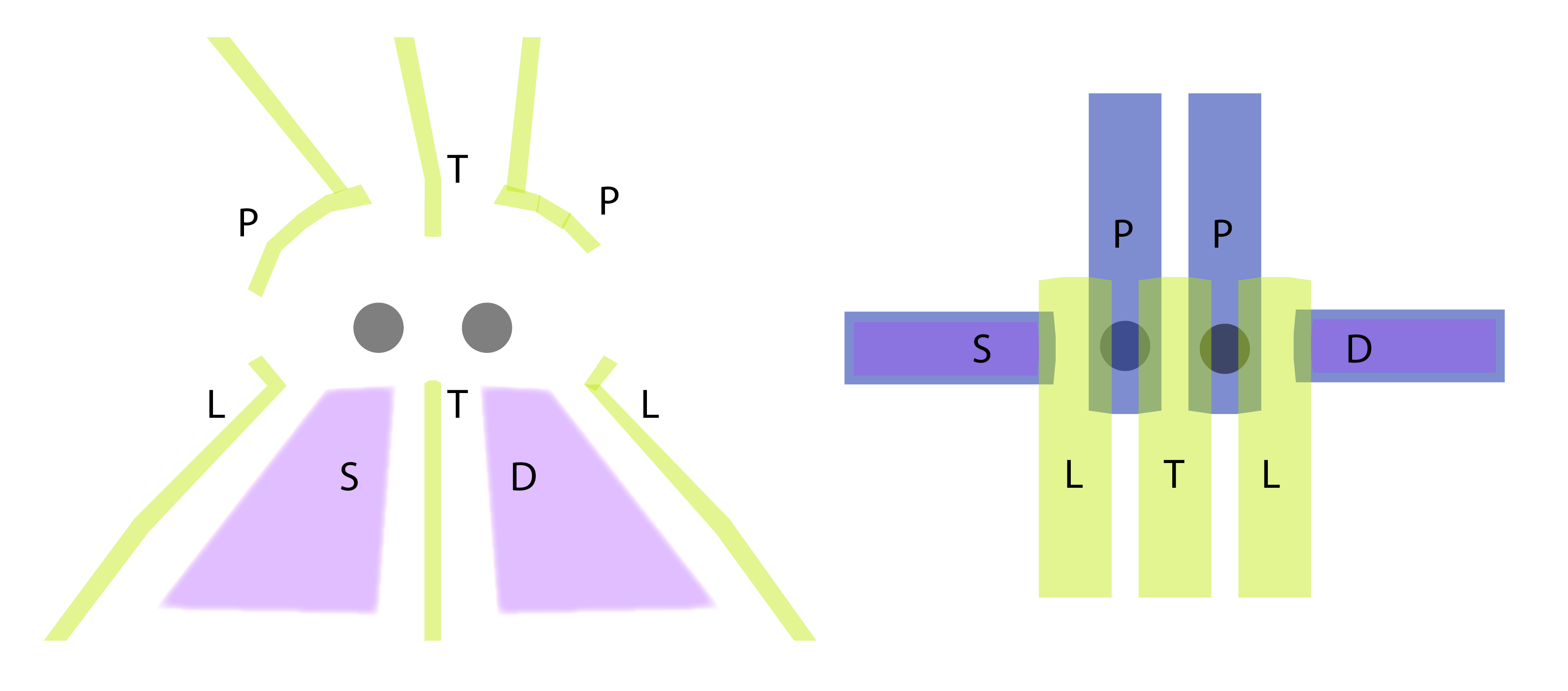}
	\caption{Drawings showing an example of the gate electrodes used to define a double quantum dot `stadium style' (left) and with overlapping gates (right). The different types of gate electrodes are labeled with P for plunger, L for lead, and T for tunnel-coupling. The P gates can be thought of as primarily affecting the chemical potential of the dot, and thereby the number of electrons, the L gates as affecting the tunnel-coupling between the dot and the leads, and the T gates as affecting the tunnel coupling between dots. The source and drain, labeled S and D, are colored purple and are regions of 2DEG that represent the leads for the two quantum dots. In the stadium style, which is named such because the gate electrodes surround the region where the quantum dot sits, gate electrodes (green) are typically biased to negative voltages to remove 2DEG locally. In the overlapping style, gate electrodes are deposited in multiple overlapping layers separated by a thin insulator, with the plunger gates electrodes (blue) biased relatively more positively to induce the quantum dots to sit directly under them, and the lead and tunneling gate electrodes (green) biased more negatively. In undoped systems, it is also necessary to put gates over the source and drain to accumulate a 2DEG to enable tunneling between the QDs and leads, as shown in this sketch in blue.}
	\label{device}
\end{figure*}

This section will examine how to isolate a two-level quantum system using quantum dots. The derivation will be for layered semiconductors, with donors addressed in the section \nameref{section:donors}. Electrons are confined to two dimensions due to the layer structure of the semiconductor, either at an interface between two different semiconductors as in GaAs/AlGaAs, in a narrow quantum well as in Si/SiGe, or between an insulator and semiconductor (MOS), with the two-dimensional conducting layer known as a two-dimensional electron gas (2DEG) \citep{davies1998b}. These layers are typically tens of nanometers below the surface. The material may be doped to accumulate the 2DEG or accumulation (positively-biased) gates may be used to induce it. Gate electrodes on the surface of the material are then used to deplete the electron gas locally around the quantum dot and create a potential well in which the remaining electrons are confined. Drawings showing two variations of a typical set of gate electrodes used to define double quantum dots (discussed further in \nameref{section:mqd}) are shown in \figref{device}. These gates also have the effect of increasing the resistance between the quantum dot and the 2DEG adjacent to it, which allows the electrons to become localized in the dot, which will be discussed later in the section. The nearby 2DEG will be referred to from now on as the leads, a reference to electrical leads, because the 2DEG is the source of electrons tunneling into the quantum dot. 

\begin{figure}
	\includegraphics[scale=0.4]{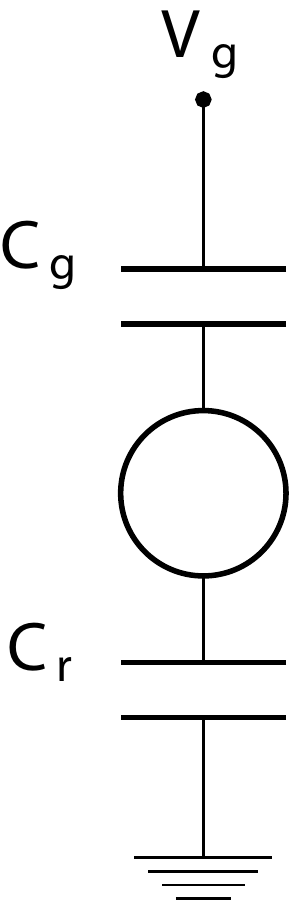}
	\caption{Schematic of a quantum dot with one gate electrode, which has voltage $V_g$, and capacitance to the dot (circle) $C_g$, and total capacitance $C_g$+$C_r$. This drawing does not represent the tunnel junctions to the leads for the quantum dot.}
	\label{quantdot}
\end{figure}

The energy of a quantum dot can be calculated by treating it like a metallic object with total capacitance $C$ and charge $Q = N e$, where $N$ is the number of electrons it contains. Ignoring tunneling with the leads for the moment, a useful simplified model has one nearby gate with voltage $V_g$ and capacitance to the quantum dot $C_g$. A schematic of this is shown in \figref{quantdot}{}, where $C_r$ is the rest of the capacitance, $C_r = C-C_g$. 
The energy of this is
\begin{align}
U(N,V_g) &= \frac{(eN - e N_0 - C_g V_g)^2}{2C} \nonumber \\ 
&= \frac{E_c}{2} \left(\frac{N - N_0 - C_g V_g}{ e}\right)^2, 
\end{align}
where $N_0$ is the number of electrons in the dot when $V_g$ is $0$, and $E_c = e^2/C$, the electrostatic cost of adding an electron to the dot or charging energy \citep{vanderwiel2002}. This can be seen by considering the chemical potential, the energy required to add the N\textsubscript{th} electron to the dot: 
\begin{align}
\mu(N,V_g) &= U(N,V_g)-U(N-1,V_g) \nonumber \\ &= E_c (N - N_0 - 1/2 - C_g V_g/e)
\end{align}

The difference between the chemical potential for adding the N-1\textsubscript{th} and N\textsubscript{th} electron is $E_c$. When $\mu=0$, states with $N$ and $N-1$ electrons have equal energy and another electron can enter the dot. Quantum dots typically have capacitance in the tens or hundreds of attofarad, leading to $E_c \approx 1-10 $ meV. When the resistance between the leads and quantum dot $R$ becomes much larger than $h/e^2$, the quantum dot can have a well-defined number of electrons. This can be understand by noting that the tunneling time onto the dot $\tau\approx R C$, and therefore through the Heisenberg uncertainty principle, $\Delta E \tau > h$, which we can rewrite as $\Delta E / E_c \ll 1 $ when $R \gg h/e^2$.

There is also a second energy scale that comes from populating different quantum dot orbitals, which can be understood to be akin to those in atoms. For the electrons to be confined in these states, the quantum dots must be designed to be near the size of the Fermi wavelength of the 2DEG, which is usually tens of nanometers. While detailed discussion of the orbital states is outside the scope of this article, the comparison to atoms is useful for establishing how electrons in quantum dots can be thought of as isolated, controllable spins. As in atoms, for principal quantum number $n>1$, the quantum dot states can have non-zero angular momentum, and quantum dots with filled s, p, d, and f shells have been measured \citep{leon2020}. This article will largely focus on electrons with $n=1$, although spin qubits with larger numbers of electrons have been studied \cite{higginbotham2014, he2019} to take advantage of increased screening. The energy of the first electron in the quantum dot can be approximated as that of an infinite well, of order $ \frac{\hbar^2}{m^* R^2}$, where $R$ is the dot's radius, and $m^*$ the effective mass. These have been measured to be about 0.05-1 meV in GaAs/AlGaAs dots and 1-10 meV in Si/SiGe \citep{reimann2002, zajac2015}. Spin qubits in quantum dots are operated at temperatures well below the point where excited states become thermally occupied, so the electrons fill the lowest energy states available. 

The electrons that are added to the quantum dot come from the leads, which are tunnel coupled to the quantum dot, allowing electrons to hop between the 2DEG and the quantum dot when the energy of an additional electron is at the Fermi level of the 2DEG. It is challenging to form quantum dots that can be fully emptied, because removing all of the electrons from a quantum dot requires large negative voltages on the gate electrodes that define it, but these will also have the effect of reducing the coupling to the 2DEG leads to zero, preventing electrons from entering and exiting the dot. To avoid this, the quantum dots must be have careful design of gate electrodes, typically by separating the gates into `plunger' gates that mostly control the electron population and `lead' gates which mostly control coupling to the nearby 2DEG \citep{ciorga2000}. This effect is particularly acute in silicon due to its larger effective mass \citep{simmons2007}. 

By applying a magnetic field in the plane of the quantum dot, $B_z$\ and including spin, the energy of the quantum dot shifts by $\pm \mu_B g B_z S_z$, where $g$ is the g-factor of the material ($g\approx-0.44$ in GaAs and $g\approx2$ in silicon) and $\mu_B$ is the Bohr magneton . The primary controllable parameters in this type of quantum dot are the number of electrons in the dot, the tunnel coupling to the leads, which is controlled by the locations of and voltages on nearby gates, and the magnetic field applied to the dot. In reality, of course, the situation is more complicated, with each gate controlling multiple parameters, disorder impacting the voltage the dot experiences, and the location and size of the dot within the gated region impacted by gate voltages (which can be taken advantage of for quantum operations, as discussed in \nameref{section:single}). The process of configuring the QD to have the desired properties by shifting the voltages of the gate electrodes that define it is referred to as `tuning,' and is often complicated. A great deal of effort has been dedicated to improving the process, which will be discussed in the section \nameref{section:tuning}. The gate electrodes that control the quantum dot may be controllable from DC to microwave frequencies depending on the wiring.

Typically, the number of electrons in the quantum dot is measured using an adjacent charge sensor, either a quantum point contact (QPC) or quantum dot that is capacitively coupled to the quantum dot and whose conductance is highly sensitive to the charge of the quantum dot \citep{field1993}. The conductance, typically by measuring the current directly or through RF reflectometry, as discussed further in \nameref{section:readout}, shows a jump at each electron transition. The absolute number of electrons can be measured by fully depleting the dot, so that there are no more transitions. These charge sensors are used both in the initial `tuning' process where the number of electrons in the quantum dot and other properties are set, and also for measuring the qubit. The techniques used to measure the qubit states will be discussed in the different sections of \nameref{section:spinqubits} and \nameref{section:readout}. 

\subsection{Multiple Quantum Dots}
\label{section:mqd}
\begin{figure}
	\includegraphics[scale=0.4]{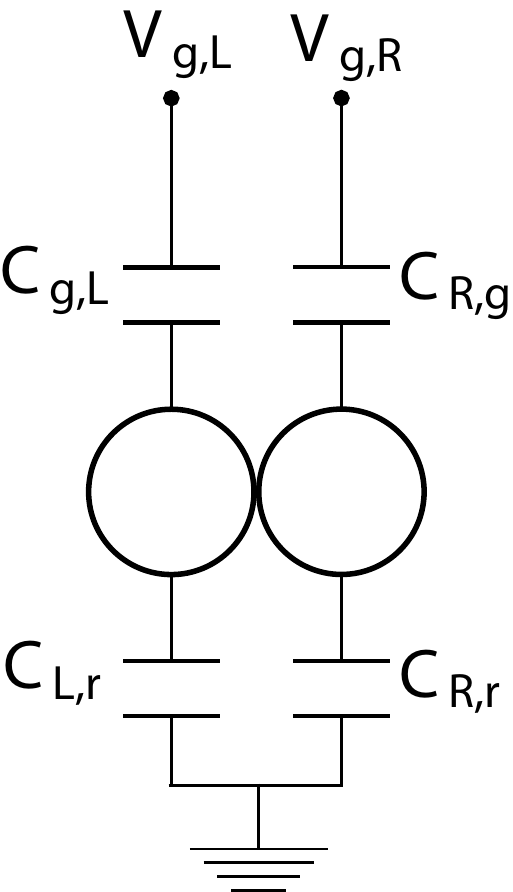}
	\caption{Schematic of a double quantum dot, with each dot having one gate electrode, which has voltage $V_{g,i}$, and capacitance to the dot (circle) $C_{g,i}$, and total capacitance $C_{g,i}$+$C_{r,i}$, with $i = \{ L, R \}$. This drawing does not represent the tunnel junctions to the leads for the quantum dots or between the quantum dots.}
	\label{quantdot2}
\end{figure}

A double quantum dot (DQD), which describes two adjacent, tunnel-coupled quantum dots, has a considerably more complicated equation for the potential energy than single quantum dots, because in addition to there being an additional dot and gate electrodes, the dots also have cross capacitance. A full derivation is presented in \citep{vanderwiel2002}, and here a minimal version is described to give some intuition. The effects of cross-capacitance are ignored because they do not affect the physics but add substantially to the complexity of the equations. Two possible charge states are considered: $(0,1)$ and $(1,0)$, where $(m,n)$ refers to the configuration with $m$ electrons in the left dot and $n$ electrons in the right dot; therefore this refers to either one electron in the right or in the left dot. Using the model from \nameref{section:qd}, the gates controlling the chemical potential of each dot are $V_{g,i}$, they have capacitance to the dot they control $C_{g,i}$, and the dots have total capacitance $C_i$, $N_{0,i}$ electrons at zero voltage and charging energy $E_{c,i}$, $i=\{L,R\}$. A cartoon of such a system is shown in \figref{quantdot2}{}. Their difference in energy is
\begin{align}
\Delta E &= U(0,1) - U(1,0) = e \frac{C_{g,L}}{C_L} V_{g,L} - e\frac{C_{g,R}}{C_R} V_{g,R} 
 \nonumber \\ & + E_{c,L}/2(1+N_{0,L}) +  E_{c,R}/2 (1-N_{0,R}) 
 \label{energydiff}
\end{align}
Here, the important takeaway is that the energy difference between the two states is a linear function of the voltage difference between the nearby gate electrodes. In the case of the multiple dot systems, it is often useful to find a single parameter that describes the important energy scale of the system. For a DQD, that is the `detuning,' or difference in potential energy between the QDs, $\eps$, 
\begin{align}
    \Delta E &= \eps = \alpha_R V_{g,R} - \alpha_L V_{g,L} + E, \qquad \alpha_i = e \frac{C_{g,i}}{C_i}
\end{align}
where $\alpha_i$ is the so-called ``lever arm" for the gate, representing the amount the QD energy shifts when the gate electrode voltage is changed, and which is dependent on the proximity and size of the gate electrode and QD; it can be measured to make quantum dot tuning easier, as discussed in \nameref{section:tuning}. $E$ is a constant voltage representing the four right terms in equation \ref{energydiff}. $\eps$ is the main control knob in performing experiments \citep{petta2005}. There is also tunnel coupling between the two dots, which takes the form $H_T = t_c | (1,0)\rangle \langle (0,1)| + h.c$. In practice, the rate $t_c$ can be controlled by adjusting the gates in between the quantum dots to change the potential barrier between the two dots. The energies for this system, centered around the point where the $(0,1)$ and $(1,0)$ states are degenerate, can be written as a Hamiltonian with the basis $(0,1)$, $(1,0)$: 

\begin{align}
H &= 
\begin{pmatrix}
	\eps/2    & t_c \\
	t_c       & -\eps/2 
	\label{eq:ham1}
\end{pmatrix}
\nonumber \\
 E_i &= \mp \sqrt{(\eps/2)^2 + t_c^2}, i = \{g,e\},
\end{align}

\begin{figure}
	\includegraphics[scale=0.3]{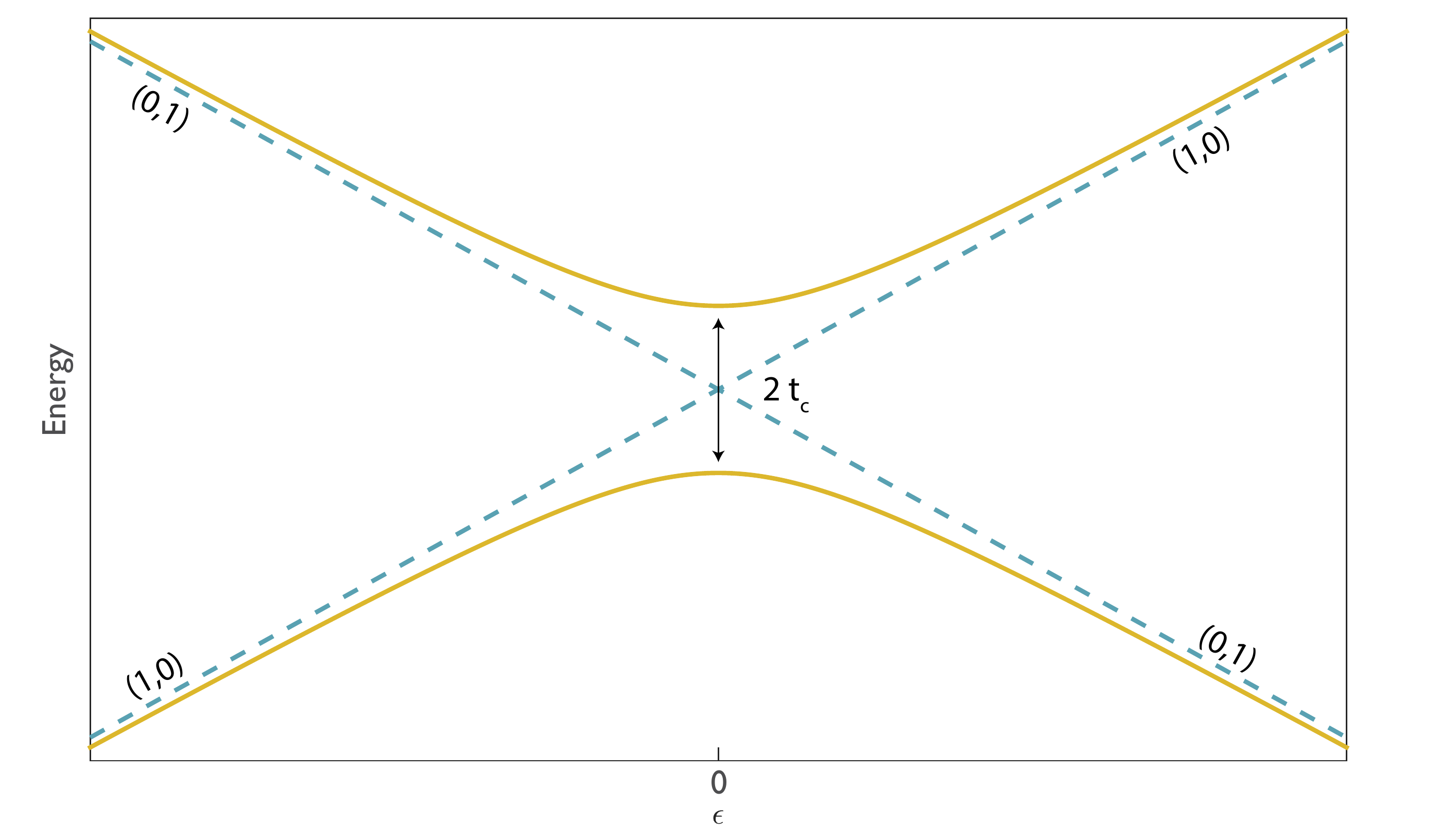}
	\caption{Energy diagram for the (1,0) and (0,1) states of a tunnel-coupled double quantum dot. The relative energy of an electron in the left or right quantum dot changes linearly with the potential energy difference between the dots, $\eps$, as as shown by the dashed blue lines. When the electron is allowed to tunnel between the dots, the left and right quantum dot occupation states hybridize, creating a ground and excited state shown in yellow, which have minimum splitting $2 t_c$. For an electron in the ground state, as $\eps$ is increased, the electron moves from the left to right QD.}
	\label{endiagdqd}
\end{figure}

Where $E_i$ are the eigenstates of the matrix and are plotted as a function of the potential energy difference between the dots $\eps$ in \figref{endiagdqd}.

The ground state charge distribution can be derived as $Q_R - Q_L = \frac{\eps}{2 E_g},$ where $Q_i$ represents the wavefunction weight in each quantum dot. This can physically be thought of as a charge dipole moment that can be controlled with gate electrodes. This approximation, of course, assumes that changing \eps{} only changes the chemical potential of the dot, not the confining potential of the dot itself. 
Next, spin is included in the model, so that the DQD's state will be represented by the product of its charge and spin state. In addition to the spatially uniform magnetic fields that impact single QDs, in a DQD a magnetic field gradient between the dots also impacts the energy diagram. Therefore the energy of electrons with the same spin state in the $(0,1)$ and $(1,0)$ will have additional offset of energy equal to $g \mu_B/2 \Delta B_z$, with $\Delta B_z = B_L-B_R$ representing the difference in the effective magnetic field experienced by the electrons in each dot \citep{petta2005}. 

The remainder of this section will focus on a DQD with two electrons, which will be the basis for the singlet-triplet qubit \citep{hu2000, levy2002}. Only the voltage region where the electrons are in $(1,1)$ and (0,2) are considered, although the analysis applies equally to the (2,0) and $(1,1)$ region. The energy diagram \figref{endiagdqd}{} also applies for the charge states $(1,1)$ and (0,2), with $(1,1)$ replacing $(1,0)$ and (0,2) replacing (0,1). Considering only electrons in the ground orbital state, there are four possible combinations of spin states for two-electrons: the singlet state and three triplet states, $S =\sing, \, T_0 = \tnot,\,  T_+ = \ket{\up \up}, \, T_- = \ket{\down \down}$

\begin{figure*}
	\includegraphics[scale=.2]{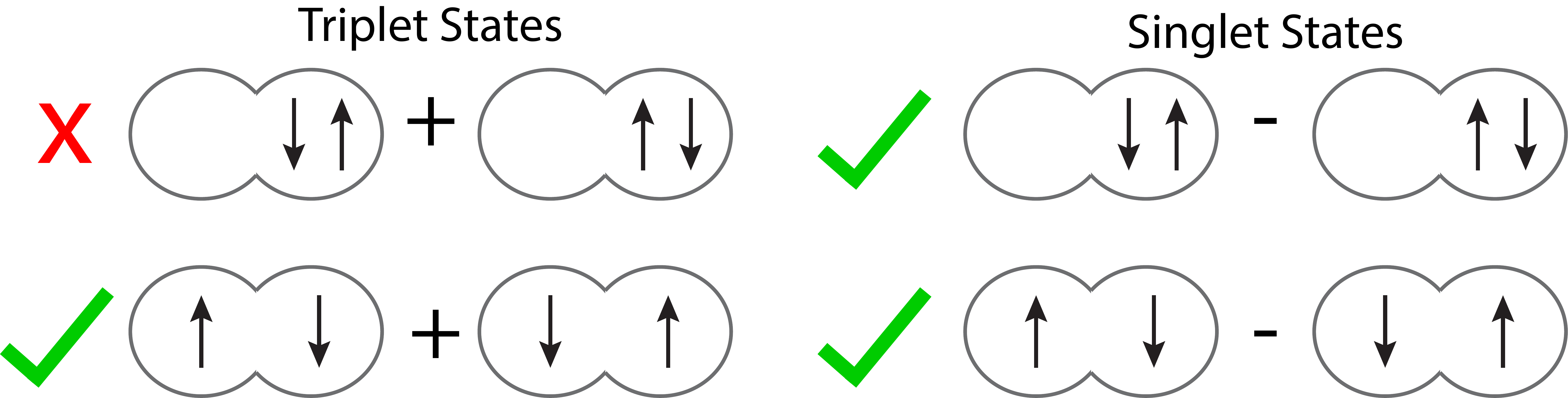}
	\caption{The possible spin states for \ket{S} and \ket{T_0} in the subspace of $(1,1)$ and (0,2) charge configurations are diagrammed here. In this case the charge configuration is represented by the number of spins in a given dot (the top row shows (0,2) states and the bottom row shows $(1,1)$ states. The states that are allowed have checkmarks and the one that isn't has a red X. This indicates that the singlet state can be in either $(1,1)$ or (0,2), but the triplet must be in (1,1).}
	\label{spin}
\end{figure*}

 The Pauli exclusion principle becomes important here because it limits the possible combinations of charge and spin states. The total wavefunction must be antisymmetric, and because the singlet spin state is antisymmetric, the charge states must be symmetric; for example in $(1,1)$ $\frac{1}{\sqrt{2}}\left(\ket{LR}+\ket{RL}\right)$ and in (0,2), \ket{RR}, where L and R represent the wavefunction for an electron sitting in the left and right quantum dot, respectively. Because the singlet has allowed $(1,1)$ and $(0,2)$ states, its ground state is the same as the ground state in \figref{endiagdqd}, a hybridized $(1,1)-(0,2)$ state. The triplet spin states, however, are symmetric, so they must have an antisymmetric charge state, which does not exist for $(0,2)$ if only the ground orbital state is used. This means that that while the singlet state can inhabit the ground state of the DQD from the Hamiltonian \eqref{eq:ham1}, the triplet states can only be in $(1,1)$, a phenomenon known as Pauli exclusion. The allowed and disallowed combination of spin and charge states are shown for the $T_0$ and S states in \figref{spin}. 

\begin{figure}
	\includegraphics[scale=.3]{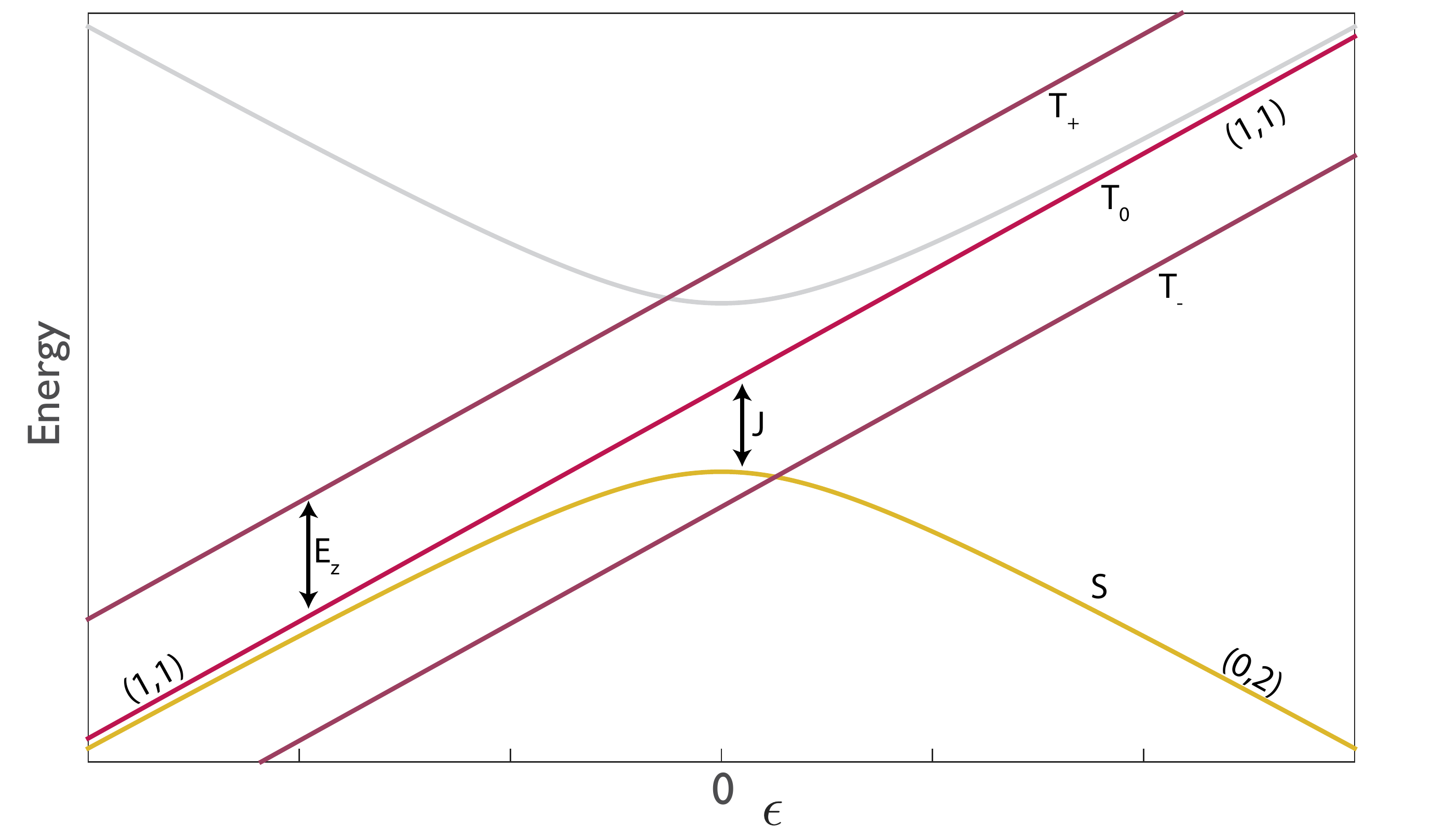}
	\caption[Energy Diagram for singlet and triplet states]{Energy diagram for singlet (\ket{S}) and triplet  (\ket{T_-}, \ket{T_0}, \ket{T_+}) states in a DQD with two electrons as a function of the detuning $\eps$. An in plane magnetic field is applied to separate the triplet states by energy $E_{\mathrm{Z}}$. The singlet is in the ground state of the DQD while the triplet states are Pauli-blockaded in (1,1).} 
	\label{endiagst0}
\end{figure}

\begin{figure}
	\includegraphics[scale=.33]{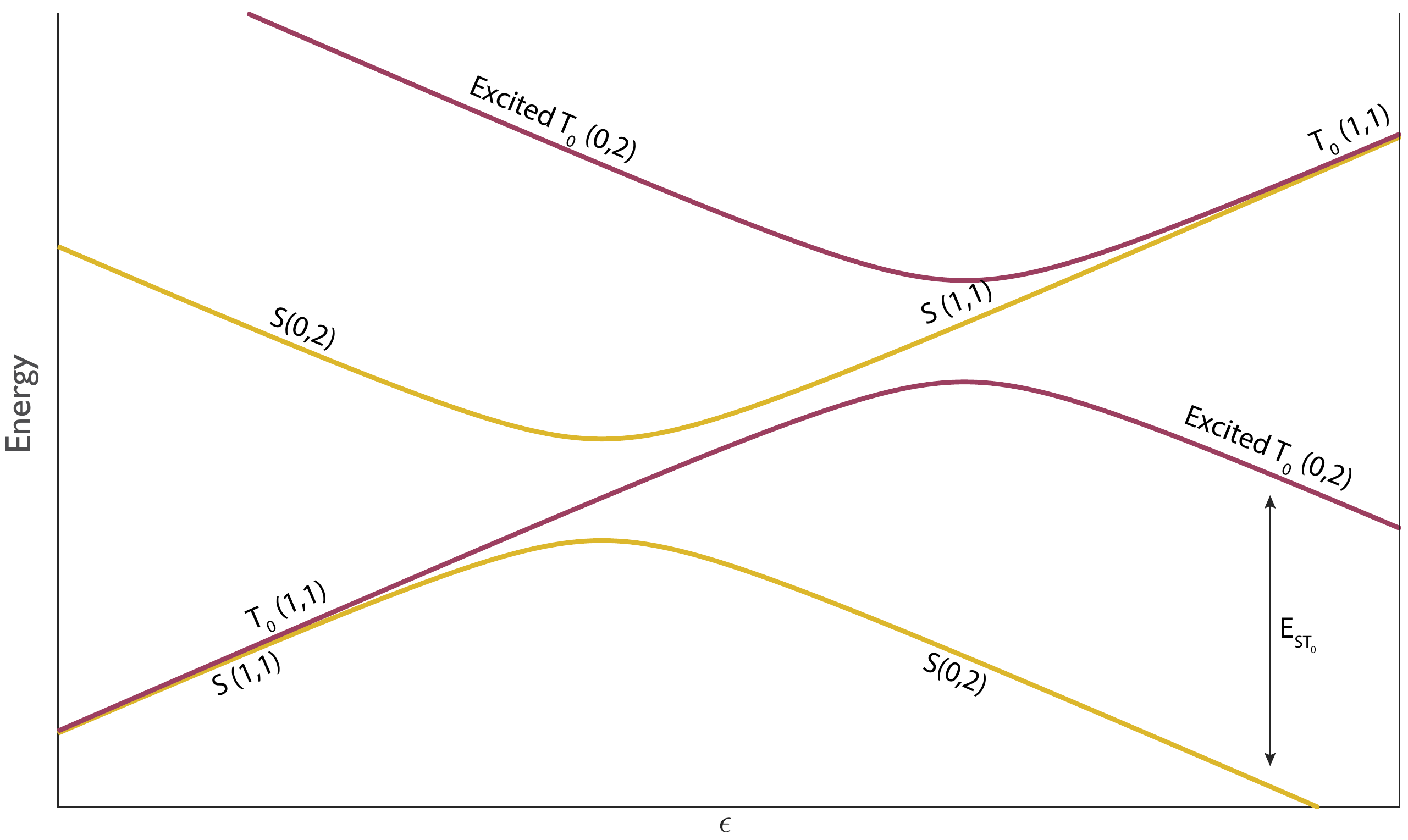}
	\caption{Energy diagram for \ket{S}) and \ket{T_0} states showing the region of $\eps$ (at the right) where it becomes energetically favorable for one of the electrons in the triplet state to enter an excited orbital state, lifting Pauli exclusion so that it can enter the (0,2) charge configuration. This state is labeled Excited $T_0$. The charge configuration is also labeled in the regions where the states are localized in a given charge configuration (i.e. away from the avoided crossings). $E_{ST_0}$ is the maximum \ket{S}-\ket{T_0} splitting, constant in the region where both \ket{S}) and \ket{T_0} are in (0,2).} 
	\label{fullst0}
\end{figure}

The energies of the four states as a function of $\eps$ are shown in \figref{endiagst0}. The plot shows that the energy splitting between the \ket{S} and \ket{T_0} states can be reduced to 0 by setting $\eps$ such that both the \ket{S} and \ket{T_0} are in the $(1,1)$ state, and then can be turned on continuously by increasing $\eps$, so that it is energetically favorable for two electrons to be in the right dot. This \ket{S}-\ket{T_0} splitting is known known as the exchange energy or $J(\eps)$. Eventually $J(\eps)$ becomes large enough that it becomes energetically favorable for one of the electrons in the triplet state to enter an excited orbital state, lifting Pauli exclusion so that it can enter the $(0,2)$ charge configuration \citep{barthel2009}. An energy diagram showing this for the \ket{T_0} state (the \ket{T_+}and \ket{T_-} are excluded to improve clarity, but they show the same crossover) is displayed in \figref{fullst0}. Because the \ket{T_0} state has the same charge configuration as the singlet state in this region, $J(\eps)$ is constant at $E_{ST_0}$ \citep{dial2013}. The \ket{S} and \ket{T_0} states can be distinguished by setting $\eps$ so that the triplet states are in $(1,1)$ and the singlet state is localized in (0,2), at which point the different charge distributions can be measured with a nearby charge sensor. Finally, the magnetic field gradient creates an energy splitting between \ket{\up \down} and \ket{\down \up} of $E_{\up \down} = g \mu_B \dbz$. Together, this can be written as a Hamiltonian $H = J/2(\eps) \sz + \dbz/2 \sx$. 

\section{Implementations}
\label{section:spinqubits}
The idea of using quantum dots for quantum computation was first described in \cite{loss1998} and \cite{kane1998}. In the years since, a variety of qubits in quantum dots have been studied, each with their own strengths and weaknesses. Several of the most common ones will be discussed in the following sections, which will explain how they are defined and controlled, their experimental results, and major challenges they face, in an attempt to give some coherence to a broad field of study. 

This article separates spin qubits into two categories, those that use a single quantum dot, described in \nameref{section:single} and those that use multiple quantum dots, described in \nameref{section:charge}, \nameref{section:st0}, \nameref{section:triple}. Those in the first category can typically be thought of as being primarily defined by their spin state, while those in the second category are usually also defined by the location of the spins in the quantum dots, or their charge state. This distinction is important because it determines how they can be controlled and measured, the noise they are susceptible to, and many of their other key features. In general, the first category of `spin-like' qubits are often controlled by magnetic fields, which which poses more stringent requirements on the control lines, but decoupling from magnetic field noise has been more achievable than decoupling from charge noise. The `charge-like' qubits in the second category are more often controlled by using voltages to shift the electron between dots applying a voltage to gates. This has the benefit of being a simple means of control, but the downside of coupling the qubit to charge noise, which is challenging to mitigate, as will be discussed in \nameref{section:decohere}.  However, most qubits are some mixture of the two, and some qubits have been developed to take advantage of the best traits of both spin and charge-coupling. 

\subsection{Single-Spin Qubits}
\label{section:single}
 A single electron in a single quantum dot with the logical subspace of \ket{\up} and \ket{\down} may be thought of as the simplest possible spin qubit. These qubits can be controlled by using magnetic resonance techniques to perform electron spin resonance. Applying a static magnetic field to split the energies of the spin states and an oscillating magnetic field perpendicular to the static one at a frequency close to  qubit splitting, drives oscillations around the Bloch sphere \citep{slichter1990}. If the static field is applied in the z direction and the AC field is applied in the x direction, the resulting Hamiltonian is 
 $$H = g \mu_B  \vec{B} \cdot \hat{S}  =g \mu_B \left( B_z \sz + B_x \sx \cos (\w t + \phi) \right )$$ 
 where $\w$ is the AC frequency and $\phi$ is the phase of the AC signal. 
 Making the rotating frame transformation, for the resonant case $\w=g \mu_B B_z/\hbar$,this becomes
 $$ H_{\mathrm{rot}} = g \mu_B \left( B_x \cos\phi \, \sx + B_x \sin\phi \, \sy \right)$$ 
 The static magnetic field is typically on the order of 1 T and is applied globally to all qubits. Several techniques have been used to apply the oscillating magnetic field, which can be far weaker in magnitude than the static one and has frequency ranging from about 5 to 50 GHz, depending on the experiment. The first technique is to fabricate a coplanar waveguide (CPW) near the qubit, which was first demonstrated in a GaAs/AlGaAs heterostructure \citep{koppens2006}. However, CPWs can cause substantial heating, due to the need to send down large voltages to create fast Rabi rates, and have large spatial extent, making it hard to address single qubits. 
 
 It is also possible to generate magnetic fields by applying a voltage with frequency $\w$ to nearby gates to drive the electron spatially within the dot. Experiments using this method in GaAs/AlGaAs relied on the spin-orbit interaction, in which the oscillating electron experiences an effective magnetic field \citep{nowack2007}. Another way to transmute electron motion into a magnetic field is to fabricate a micromagnet next to the quantum dot \citep{tokura2006, pioro-ladriere2008}. The micromagnet is designed to have a large magnetic gradient $\frac{dB_x}{dl}$, where $l$ is the direction along which the electron is driven, creating the effective field $\frac{dB_x}{dl} \delta l \cos (\w t )$, where $\delta l$ is the amplitude of the electron's motion, which is a function of the voltage applied and the quantum dot's confining potential. Micromagnets also apply a small but non-zero $B_z$ field, so if there are multiple dots close to one another, this separates the Zeeman splittings, allowing them to be addressed independently. 
 
 In experiments in GaAs/AlGaAs, the qubits decohered quickly due to magnetic field noise from uncontrolled nearby nuclear spins. The reduction of magnetic field noise in silicon, in particular with the use of isotopically-purified silicon, which has from 50-800 ppm residual $^{29}$Si \citep{tyryshkin2012}, has allowed great improvements in such devices, as will be discussed in the three following sections on \nameref{section:simos}, \nameref{section:sige}, and \nameref{section:donors}. Magnetic field noise is discussed at greater length in \nameref{section:nuclear}. Entangling gates in single-spin qubits typically rely on the exchange interaction, as described in \nameref{section:mqd}. Exchange-based gates can be extremely fast, but because they are charge-dependent they couple the qubits to charge noise, which can limit gate fidelity.
 
\begin{figure}
	\includegraphics[scale=.28]{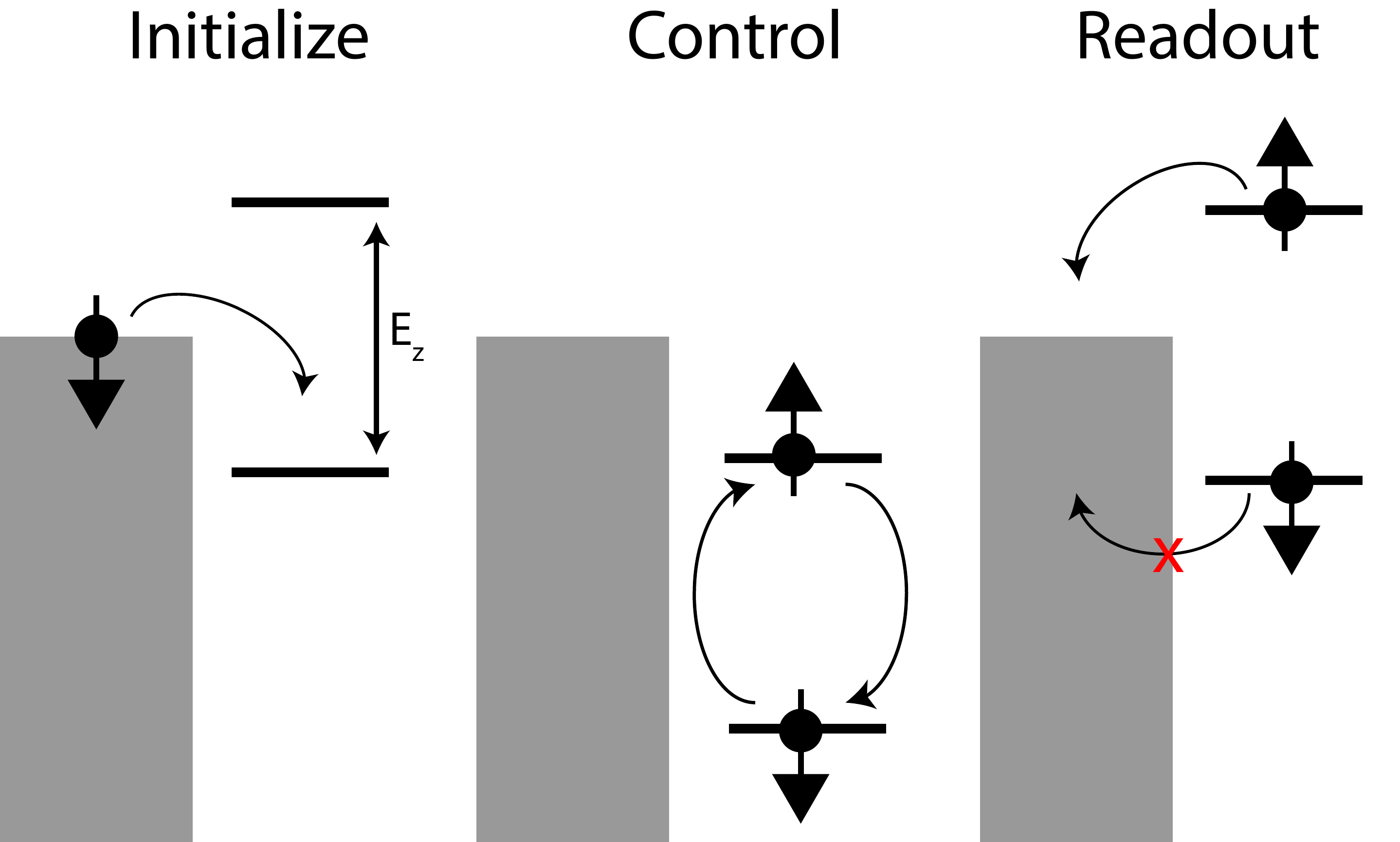}
	\caption{Cartoon of the pulse sequence for a single-spin qubit. A magnetic field is applied to separate the spin states by energy $E_{\mathrm{Z}}$ (in this figure, $g$ is treated as a positive number). Left: First, the qubit is initialized in the \ket{\down} state through tunneling with the lead (gray) . Middle: Then, a resonant magnetic field is applied to drive oscillation between spin states. Right: Finally, the qubit is measured through tunneling with the leads.} 
	\label{single}
\end{figure}

 It is useful to consider a pulse sequence for a full sequence of qubit manipulation, consisting of loading the initial state, performing oscillations around the Bloch sphere, and reading out the final state. For a single-spin qubit, this can be performed using one high-frequency gate electrode per qubit. To load the qubit in a known state, the gate electrode shifts the qubit energy so that the \ket{\down} state has a lower energy than the lead and the \ket{\up} state a higher energy, so that tunneling between the dot and leads will reset the qubit state to \ket{\down}. Next, the gate voltage is shifted so that the electron is decoupled from the lead, and the quantum gate operation is performed. Finally, readout is performed by first performing the necessary rotation to measure the correct qubit orientation, then the gate voltage is shifted so that the \ket{\up} state is above the lead energy and the \ket{\down} state is below it, so that electron can tunnel between them only if it is in the \ket{\up} state. The tunneling signal is measured by a nearby sensor, typically a QPC or QD \citep{elzerman2004}. This process is diagrammed in \figref{single}. Currently, there are three major platforms in which these qubits are being developed. The details of their implementations will be discussed in the following sections. 
 
\subsubsection{Silicon Metal-oxide-semiconductor (Si-MOS)}
\label{section:simos}

In this design, based on the dominant one for modern transistors, an insulator is deposited on silicon, and depletion and accumulation gate electrodes are deposited on top of it \citep{veldhorst2014}. The silicon is undoped, so the accumulation and depletion gate voltages are adjusted to accumulate a 2DEG at the insulator-semiconductor interface to form quantum dots. This system has increased disorder compared to semiconductor heterostructures due to increased defects in oxides compared to semiconductors and the non-epitaxial interface between the amorphous oxide and silicon, making it more challenging to form quantum dots in the desired location and to have full control over their properties. The use of overlapping gates, as shown in \figref{device}, in which gates are deposited in multiple overlapping layers with a thin insulating layer between them to prevent shorting, and with some directly over the quantum dot, better defines the potential the quantum dot experience compared to the original `stadium style,' where there is a single layer of gates at the perimeter of the quantum dot. It has been instrumental in enabling experiments with these qubits as well as some using Si/SiGe \citep{angus2007}. The increased effective mass in silicon compared to gallium arsenide decreases the Fermi wavelength, forcing dots in silicon systems to be smaller than those in GaAs; they are typically about 50 nm wide, still within the tolerance of electron beam lithography. 

One challenge in silicon qubits is valley degeneracy. This arises because silicon has six degenerate energy minima, known as valleys, in its conduction band, and electrons in quantum dots could potentially occupy any of them. To be useful qubits, the degeneracies must be lifted in the quantum dots, because small valley splittings (the energy difference between the different valleys) can cause leakage out of the logical subspace and cause $T_1$, the energy relaxation time, to be reduced. Confinement in the z direction shifts the energies of four of those states, and the final degeneracy can be lifted by applying electrical fields, for instance by gating (applying voltage to nearby gate electrodes), but it is also affected by roughness and disorder \citep{yang2013a}. In Si-MOS, electric fields applied through gating have been effective to create a robust valley splitting of up to 0.8 meV, larger than relevant qubit energy scales. $T_1$ is approximately 1 s, and therefore does not limit gate fidelities \citep{yang2019a}.

The Si-MOS implementations so far have used coplanar waveguides to perform single qubit gates. Changing the voltage of a nearby gate electrode applies a Stark shift that alters the g-factor at each quantum dot, changing the resonant frequency for each qubit by up to about 10 MHz. This enables one to apply a global static magnetic field and then use a single coplanar waveguide to control several qubits independently using multiplexing \citep{veldhorst2014}. In samples made on isotopically-purified silicon, using pulse optimization techniques \citep{yang2019a} has decreased the average single qubit gate error to fractions of a percent, with coherence times of up to 10 ms. Two-qubit gates using exchange have been measured with an average Clifford gate fidelity over 90\%, Rabi rates in the hundreds of kHz, and entangling rates close to 1 MHz \citep{huang2019}. Readout fidelities ranging from about 80-95\% and an integration time of several milliseconds are typical. 

\subsubsection{Si/SiGe}
\label{section:sige}

This system uses a quantum well of silicon, approximately 10 nm thick, between layers of Si$_{1-x}$Ge$_x$, where $x\approx0.3$ \citep{zwanenburg2013}. There are many similarities between Si/SiGe and Si-MOS systems, because both are gate-defined quantum dots whose wavefunction is primarily in silicon. Both stadium style gates and overlapping gates have been used for devices in Si/SiGe. While some devices use doping to populate the 2DEG, it has become more common to use undoped silicon. Finally, while CPWs have been used previously for qubit control, micromagnets are used more often for single qubit gates currently \citep{kawakami2013, yoneda2017}. The valley degeneracy in Si/SiGe is lifted by strain between the Si and SiGe as well as vertical confinement, but small valley splitting can pose a challenge \citep{kawakami2014}. Adjusting local gate voltages can help to lift it \citep{goswami2007, watson2018}. 

Using natural silicon and micromagnets, single qubit gate fidelities reach 99\% and Rabi rates are approximately 1 MHz \citep{kawakami2016}. In isotopically-purified silicon with improved micromagnet design, 99.9\% average gate fidelity and Rabi rates of tens of MHz have been measured \citep{yoneda2017}. Two-qubit gates using the exchange interaction have been achieved with gate rates of several MHz, and fidelities of 80-90\% in natural silicon \citep{zajac2018, watson2018, xue2019}. Improvement in these fidelities could be achieved through reducing the susceptibility to charge noise, as in \citep{zajac2018}, and by working in isotopically-purified silicon. Long distance coupling through a superconducting resonator has also been achieved in Si/SiGe devices and will be discussed in \nameref{section:cavity}. Readout fidelities of 70-85\% with integration times of several milliseconds are typical, and are limited by $T_1$ and temperature. 

\subsubsection{Donors}
\label{section:donors}
 
Donors are single atoms situated in a semiconductor and possessing bound hydrogen-like charge states. A phosphorus atom in silicon has a bound electron and nuclear spin, both with spin 1/2 \citep{tyryshkin2003, pla2012}, and which can both therefore be used as qubits. Bulk donor ensembles have shown extremely long coherence times, with $T_2=10$ s in isotopically-purified silicon, indicating that single donors might show similar coherence times \citep{tyryshkin2012}. The donors are formed either through ion implantation or by positioning them using lithography with a scanning tunneling microscope. The latter technique offers the possibility of precise placement for multiple qubit systems \citep{weber2014}. In donor-based systems, the gate electrodes are not needed to define the quantum dot potential, as in the other implementations discussed here, and instead are needed only for performing quantum gate operations. The electron state has a Bohr radius of about 2 nm, which becomes particularly important in exchange-based operations, because the two quantum dots need to be extremely close to another to have non-zero tunneling between them \citep{wang2016}.

In order to perform initialization and readout, a single-electron transistor is fabricated adjacent to the donor qubit, and it acts as both a lead as well as sensor \citep{morello2010}. Readout has shown fidelities over 98\% \citep{dehollain2016}. Single qubit gates are performed using a CPW. Measurements of the electron qubit in ion-implanted single donors in isotopically-purified silicon have shown the anticipated long coherence times, with $T_2$ of over 500 ms with dynamical decoupling \citep{muhonen2014}. Using randomized benchmarking an average single-qubit gate error of under 0.05\% was measured \citep{dehollain2016}. 

Most proposals for two-qubit gates between donor electrons rely on the exchange interaction, but this can be challenging due to the requirement for donors to be extremely close together due to their narrow wavefunction, especially for implanted donors, which are placed stochastically. Despite this, large exchange couplings have been measured between two ion-implanted electron donors \citep{dehollain2014, gonzalez-zalba2014}. For STM-placed donors, it is easier to design specific exchange couplings, but there are still challenges, such as the exchange coupling's oscillation in magnitude with donor separation due to valley interference \citep{wang2016}. An exchange gate has been measured with STM donors with gate speeds approaching 1 GHz \citep{he2019}. A number of ideas for coupling donors that don't require exchange have also been proposed, such as entangling donors with other varieties of qubits, creating hybrid donor-laterally-gated-quantum-dot qubits, and using longer-range capacitive coupling or cavity coupling. \citep{harvey-collard2017, morello2020}.

 The nuclear spin can also be used as an auxiliary qubit, and because it is decoupled from charge, it has shown coherence times of over 30 s with dynamical decoupling and average gate fidelity of 99.99\% \citep{muhonen2014, muhonen2015}. Entanglement has been achieved between the electron and nuclear spin on a single vacancy, with a Bell state fidelity of 96\% \citep{dehollain2016}. The extremely long lived states in nuclear spin qubits, combined with their high-fidelity coupling to electron spins, may enable them to act as a quantum memory. 
 
\subsection{Charge qubits}
\label{section:charge}

\begin{figure*}
	\includegraphics[scale=0.5]{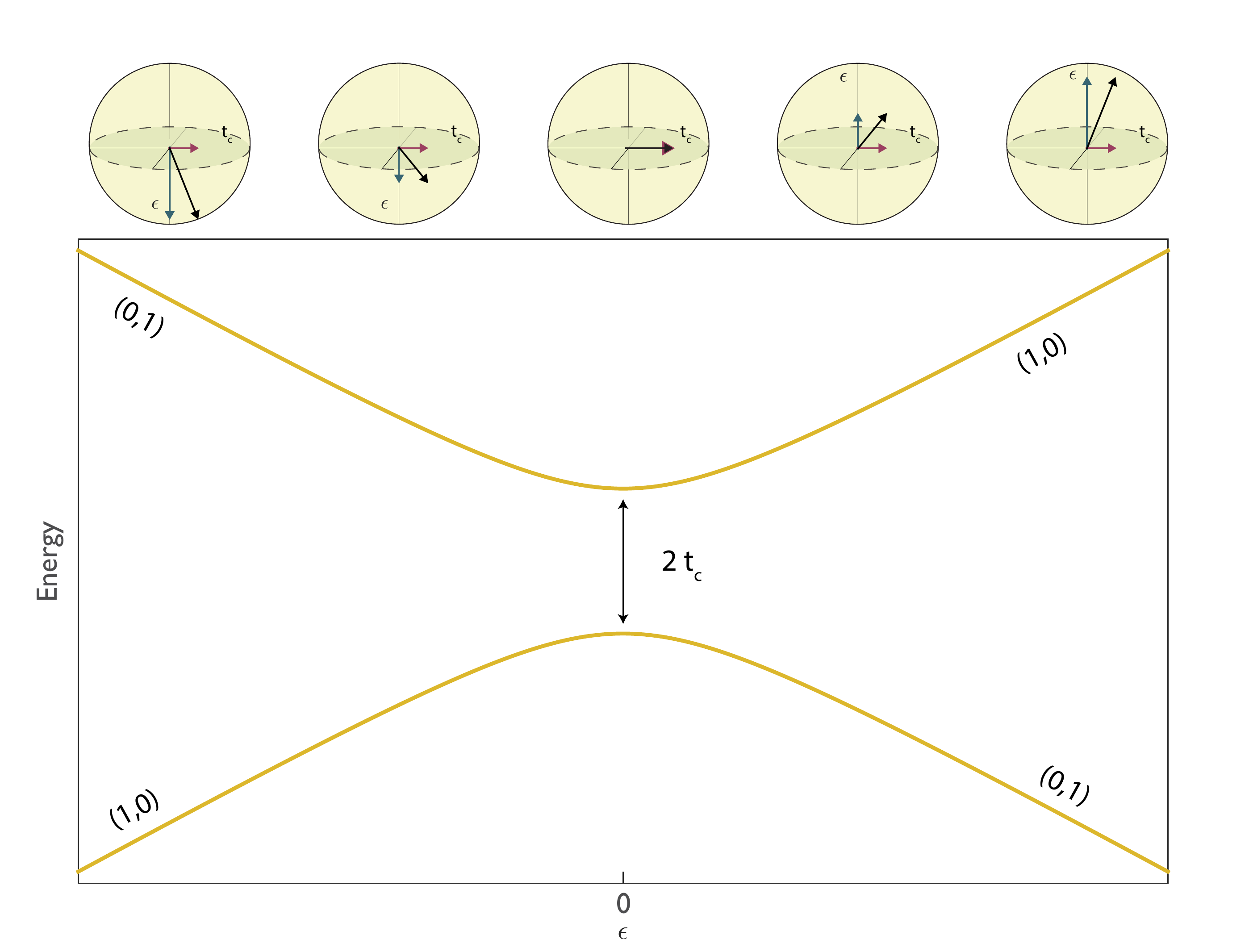}
	\caption{The energy diagram for a charge qubit (lower figure), with the Bloch sphere for each $\eps$ value superimposed on top of it, showing that the vector rotates from being largely in the -z direction, to the x direction, to the +z direction.}
	\label{charge}
\end{figure*}

Charge qubits, which have one electron in a DQD, were introduced in \nameref{section:mqd}. Their Hamiltonian is $H = \eps/2\sz + t_c \sx$, with a logical subspace of \ket{R} and \ket{L}, which represent the electron being in the right or left dot, and the energies of the ground and excited states are shown in \figref{charge}. The charge distribution of the qubit states can be changed by changing the voltage detuning between the dots $\eps$. When $\eps$ is large and positive, they approximately align with the electron being in right and left quantum dots, and $H \approx \eps \sz$ and vice versa for $\eps$ large and negative. At $\eps=0$, the left and right occupation states are degenerate and  $H\approx t_c \sx$. Therefore, a central difference between charge and single-spin qubits is that charge qubits can be controlled by gate voltages alone, without applying resonant pulses. As shown in \figref{charge}, the qubit's splitting changes magnitude and the relative amounts of $\sx$ and $\sz$ change as the local voltages are changed; this can be thought of as the Bloch sphere vector representing the Hamiltonian rotating and changing length. The minimum splitting for a charge qubit is $2 t_c$, which is typically several GHz \citep{gorman2005, petersson2010a, dovzhenko2011, shi2013}, and the splitting increases as $\eps$ is increased or decreased from zero. Because $t_c$ cannot be turned off, there is always a $\sx$ component, but by increasing $\eps$ so that it is much larger than $t_c$, nearly orthogonal control can be achieved. A second key difference between these and single-spin qubits is that charge qubits are highly coupled to charge noise, stemming from charge fluctuations in the environment and leading to decoherence. Their use in quantum computing applications so far has therefore been limited, because of the high speed of electronics required to control them and short coherence times. Charge qubits are important as a prototype for several other types of spin qubits that use multiple quantum dots, such as \nameref{section:st0} and \nameref{section:triple}, which share many properties with charge qubits, but have found ways to mitigate some of the downsides of the charge qubit. 

\subsection{Singlet-Triplet and Hybrid Qubits}
\label{section:st0}

The singlet-triplet qubit alters the charge qubit to have longer coherence times and the ability to reduce the splitting practically to zero. This reduces the coupling to charge noise enough that higher gate fidelities are achievable, but the gates can still be fast. It has two electrons in a DQD, with the logical subspace being the singlet and triplet zero-spin states. The Hamiltonian is $H = J/2(\eps) \sz + g \mu_B \dbz \sx$, and was introduced in \nameref{section:mqd}. 

As in charge qubits, gate operations are performed by changing $\eps,$ and the energy diagram is shown in \figref{endiagst0}. The singlet-triplet qubit has the somewhat unique property of being both a spin and charge qubit, depending on how it is biased. In the region of values of $\eps$ where $J\gg g \mu_B \dbz$, the singlet-triplet qubit is `charge-like,' with the two qubit states having different charge distributions than one another. Similarly to charge qubits, because \dbz{} is constant as a function of $\eps$, there is always a $\sx$ component, but $J$ can be set to be much larger than $ g \mu_B \dbz$, so nearly orthogonal control can be achieved.  By increasing $\eps$, the exchange energy $J$ can be varied from zero up to tens of GHz typically, limited by the point at which it becomes energetically favorable for the triplet state to enter an excited orbital state in the quantum dot it had been Pauli blockaded from, as shown in \figref{fullst0}{}. When $J$ is non-zero, the qubit is coupled to charge noise and the susceptibility to noise increases with $J$.

At $\eps\ll0$, $H\approx g \mu_B \dbz \sx$, and the two qubit states have the same charge state, only differing in their spin state, so the singlet-triplet qubit is effectively a spin qubit and is largely charge-insensitive. The magnetic gradient value is either set by an effective magnetic field from the nuclei, known as the Overhauser field (more typical in GaAs/AlGaAs), or by a nearby micromagnet (more typical in silicon qubits). In the case where the gradient arises from the nuclei, it fluctuates in time, but can be fixed to values from MHz to GHZ using dynamical nuclear polarization and a feedback method with $T_2^*\approx 100$ ns \citep{bluhm2010, nichol2017}. Micromagnets have lower noise but are less changeable \textit{in situ}. Due to the low-frequency character of nuclear noise, dynamical decoupling is extremely effective, with $T_2$ increasing to close to a millisecond with notch filtering of the Larmor frequencies in GaAs/AlGaAs \citep{malinowski2017a}, as discussed in \nameref{section:nuclear}.

Initializing the qubit state uses the same method as single-spin qubits: using gate electrodes, the energy of \ket{S} is set below the energy of the lead and the energy of \ket{T_0} is set above it, so that tunneling between the quantum dot and the lead loads \ket{S}. This can be performed in nanoseconds to high fidelity due to the large singlet-triplet splitting in this region. A quantum dot fabricated adjacent to the DQD so that it is sensitive to the number of electrons in each dot is used for measurement. $\eps$ is set to a large positive value where the singlet state is localized in (0,2), while the triplet remains Pauli-blockaded in (1,1). Then, by measuring the number of electrons in each dot, the singlet and triplet states can be differentiated. This charge readout can be quite fast when using a tank circuit in the measurement circuit, with 98\% fidelity in 1 $\upmu$s  \citep{barthel2009, shulman2014}. 

Randomized benchmarking has measured average single qubit gate fidelities of about 99.5\% in both GaAs/AlGaAs and Si/SiGe \citep{cerfontaine2020, takeda2020}, limited by both charge and nuclear noise. It is possible to tune singlet-triplet qubits to have `sweet spots', regions where the first-order coupling to charge noise goes to zero, thereby increasing coherence times \citep{reed2016, martins2016}. The singlet-triplet qubit has been entangled using a capacitive gate, arising from the dipole moment of the DQD when exchange is non-zero \citep{shulman2012}. The interaction is slower than the exchange gate and limited by charge noise, but using a technique to partially decouple the qubit from charge noise improved to a 90\% gate fidelity noise \citep{nichol2017}.

Hybrid qubits are also formed in DQDs, but add a third electron, which yields a Hamiltonian where both axes of control are voltage-controlled, so there is no requirement for magnetic fields \citep{kim2014}. They possess two sweet spots, allowing control around two axes that is decoupled from charge noise, and gate speeds of several GHz. Single-qubit process fidelities of about 95\% have been measured in hybrid qubits in natural Si/SiGe \citep{kim2015}. 

\subsection{Triple-dot Qubits}
\label{section:triple}

Fully electronic control can also be achieved by adding a third electron and third dot, forming the triple-dot or exchange-only qubit \citep{russ2017}. These qubits utilize the exchange between the left and middle dot and the right and middle dot to give two independent axes of control, separated by 120$^\circ$ \citep{laird2010, medford2013}. The qubits therefore have charge states with weight in the (2,0,1), (1,1,1) and (1,0,2) configurations, and the logical subspace is a singlet-like ground state and triplet-like excited state.  The Hamiltonian can be controlled either with the detuning between the left and right quantum dots or through tunnel barrier control. These qubits have the advantage of fast gates and no requirement for a magnetic field (although magnetic noise can cause dephasing). Some of their challenges include the requirement for composite pulses due to the nonorthogonality of the control axes, increasing overhead, and leakage to other three-electron states outside the logical subspace. The exchange-only qubit uses spin-to-charge conversion readout like the singlet-triplet qubit. 

Initial experiments were performed in GaAs/AlGaAs and were limited by magnetic-field noise. More recently, they have been studied in isotopically-purified Si/SiGe \citep{eng2015, andrews2019}. In these experiments, a variant of randomized benchmarking meant to capture leakage was performed, showing an average error of 0.3\%, with half of the error stemming from leakage, for symmetric gates decoupled from charge noise. Exchange-only qubits in GaAs/AlGaAs, working in the rotating frame, were used to achieve strong coupling to a microwave resonator, as discussed further in \nameref{section:cavity}. 

\section{Key Challenges and Technologies}
\label{section:design}
\subsection{Decoherence}
\label{section:decohere}
The number of coherent operations that can be performed by a quantum computer is ultimately limited by decoherence of the quantum system. This is caused by uncontrolled interactions with the environment. In spin qubits, the primary sources of decoherence are magnetic noise and charge noise. This section will explain what is known about these noise sources, how they vary across the different spin qubit types, and techniques for reducing the impact of noise on spin qubits. 

A useful approximation for modeling the noise that affects spin qubits is to separate it into two categories, quasistatic and high-frequency. Quasistatic noise is constant within a single quantum gate operation but can vary across them. It can be modelled by including a noise term in the Hamiltonian, so that $H = (H_0 + dH) \sz$, where $dH$ follows a Gaussian distribution with standard deviation $\sigma$. A simple Ramsey experiment, where a qubit is initialized in a given state and allowed to evolve under $H$, has the decay time $T_2^* =  h/(\sqrt{2} \pi \sigma)$, making it straightforward to extract the magnitude of quasistatic noise in the system.  High-frequency noise requires more sophisticated analysis and experiments, such as dynamical decoupling pulses that remove the effects of quasistatic noise. This leaves the qubit susceptible only to noise at higher frequencies, which can in turn be analyzed to extract its power spectral density \citep{cywinski2008, alvarez2011}. 

\subsubsection{Nuclear noise}
\label{section:nuclear}
In GaAs/AlGaAs heterostructures, the nuclei of Al, Ga and As all have nuclear spin $ I = 5/2$, which generate an effective magnetic field that acts on the qubit through hyperfine coupling \citep{chekhovich2013}. The noise arising from this nuclear bath was the primary factor limiting coherence in GaAs/AlGaAs qubits for many years \citep{khaetskii2001, coish2004}. The magnitude of the noise does not depend on the magnetic field at the values used in experiments, and generally $T_2^*$=10-40 ns if no correction techniques are employed \citep{petta2005, koppens2008}. Dynamical decoupling experiments have shown that nuclear noise is extremely low frequency, following a power law of $S(\w) = S_0/\w^{\beta}$ where $\beta\approx2-3$ and $S(\w)$ is the power spectral density of the noise \citep{medford2012, malinowski2017b}. This makes it highly amenable to dynamical decoupling, which was shown to increase $T_2$ to 500 ns in a single spin qubit with a Hahn echo pulse (the simplest dynamical decoupling pulse, which removes the effects of quasi-static noise) \citep{koppens2008}, and in singlet-triplet qubits to increase $T_{2}$ to 200 $\upmu$s \citep{bluhm2011} using the more complicated CPMG dynamical decoupling sequence, and 870 $\upmu$s using pulses defined to act as notch filters on the Larmor precession of the fluctuations \citep{malinowski2017a}.

In singlet-triplet qubits, which are sensitive only to the magnetic field gradient, the gradient can be fixed using dynamical nuclear polarization and a feedback method which also increase the coherence time to $T_2^*\approx 100$ ns \citep{bluhm2010}. Moreover, because the average value of the field gradient can be set to the desired value and only fluctuates slowly, the nuclear field can be measured rapidly using Hamiltonian parameter estimation and qubit operations thereby decoupled from the fluctuations, increasing $T_2^*$ to 2 $\upmu$s \citep{shulman2014}. This direct measurement of the nuclear bath finds that it diffuses as a random walk, indicating $1/f^2$ power spectral density, similar to what was measured through dynamical decoupling. These results indicate that there are many effective approaches to reducing the impact of low-frequency noise.

In natural silicon, 95\% of the isotopes are spin zero ($^{28}$Si and $^{30}$Si), and $T_2^*$ for single-spin qubits in Si/SiGe increases to 1 $\upmu$s and $T_2$ to 40 $\upmu$s \citep{zwanenburg2013, kawakami2014}, and in singlet-triplet qubits $T_2^*$ increases to 360 ns \citep{maune2012}. In isotopically-purified silicon, which usually has 800 ppm residual $^{29}$Si isotopes, $T_2^*=$120 $\upmu$s and $T_{2,\text{CPMG}}=$28 ms in Si-MOS single spin qubits, $T_2^*=$20 $\upmu$s and $T_{2,\text{CPMG}}$=3.1 ms in single-spin Si/SiGe, and in Si/SiGe singlet-triplet qubits $T_2^*=2.3 \, \upmu$s. Analyses have indicated that nuclear noise is no longer the limiting factor in these coherence times \citep{muhonen2014, chan2018, yoneda2017, eng2015}, which may also explain why they vary across the different silicon implementations.

\subsubsection{Charge noise}
The other major source of noise comes from fluctuating charges near the qubit. Their precise source is unknown, although it has been hypothesized that it stems from charge traps near the surface or in the oxide layer for qubits that have one\citep{paladino2014}.In most charge-like qubits, the sensitivity to charge noise, which is given by the derivative of the splitting with respect to detuning voltage, is voltage dependent. This makes quoting a $T_2^*$ time less useful than with nuclear noise; for instance, in singlet-triplet qubits the noise sensitivity increases proportionally with the splitting, making the gate fidelity nearly constant with respect to the splitting, although $T_2^*$ may change dramatically. 

Quasistatic noise has been measured in charge, singlet-triplet and triple-dot qubits, and has been found to have a similar magnitude across multiple materials, as noted in \citep{harvey-collard2017}, with $\sigma\approx 10 \, \upmu$eV \citep{dial2013, eng2015, shi2012, cerfontaine2020, watson2018}. Charge noise is higher frequency than nuclear noise, with most measurements finding a power spectral density of the form $S(\w)\ = S_0/\w^{\beta}$, $\beta\approx1$  \citep{yoneda2017,dial2013, chan2018, struck2020}, making the improvements from dynamical decoupling more limited than with magnetic noise.

Studying the temperature dependence of charge noise may give insight into its origins. Experiments in GaAs/AlGaAs have shown a strong temperature dependence, with $T_2$ changing quadratically with temperature and $T_2^*$ linearly between 50 mK and 250 mK. Si-MOS devices measured from 0.45 K to 1.25 K do not show temperature dependence in $T_2^*$. Measurements of the noise in the current through Si/SiGe quantum dots show an approximately linear temperature dependence from 0.1 K to 1 K, but the magnitude and functional form varies from dot to dot \citep{connors2019a}. Considered together, these results may indicate that the smaller size of the devices in silicon compared with GaAs/AlGaAs lead to interacting with a far lower number of fluctuators, changing the statistical properties of the noise. Another interesting direction of such research is studying spatial correlations to noise, though results are currently inconclusive due to competing magnetic noise \citep{boter2020}.

Increasing gate fidelities limited by charge noise has taken two main directions. The first direction is to reduce the qubit's sensitivity to charge, but this often also has the secondary impact of slowing quantum gates, and it requires ingenuity to find ways that this does not ultimately reduce the quantum gate fidelity. One method is to work at so-called `sweet spots', where the derivative of the splitting with respect to nearby gate voltages, and therefore the first-order coupling to charge noise, goes to zero, as in \cite{reed2016, martins2016,zajac2018}, and another is to design a composite gate sequence where coupling to charge noise and the qubit splitting is reduced, but the two-qubit gate speed was unchanged in \cite{nichol2017}. The second direction is to use dynamical decoupling to reduce the effect of the low-frequency component of the noise \citep{shulman2012, wang2012, gungordu2018}.

Due to the difficulty of reducing the impact of charge noise, with the advent of isotopically-purified silicon, qubits controlled by magnetic fields and largely decoupled from charge have made quick improvements in gate fidelity compared to more charge-like qubits. However, even these qubits have some coupling to charge, and their fidelities may ultimately be limited by charge noise. Furthermore, most of the two-qubit gates demonstrated so far are charge-based and thus experience decoherence through charge noise. Studying the source of charge noise and reducing it will therefore be essential for future scaling of these systems.

\subsection{Readout}
\label{section:readout}
The two main readout techniques used for spin qubits were described in the sections introducing single-spin qubits and singlet-triplet qubits, the first measuring tunneling between the quantum dot and a charge reservoir and the second converting the spin state to a charge state. In this section, there will be a more detailed discussion of the experimental implementations of these methods, as well as alternative readout technologies. 

Both methods rely on a proximal quantum sensor, either a quantum point contact or quantum dot. Connecting coaxial cables to the sensor can increase the measurement bandwidth, which can enable reduced measurement times. To capitalize on that, tank circuits are typically used to impedance match the 50 $\Omega$ coaxial lines to the sensor quantum dot or QPC, which is maximally sensitive at resistances of $h/e^2\approx26 \, \mathrm{k\Omega}$ and larger \cite{barthel2010}. In this circuit, an inductor is connected in series with the sensor and a capacitor is placed in parallel to ground. At the resonant frequency of $\w \approx 1/\sqrt{LC}$, the impedance of the circuit appears to be 50 $\Omega$ and the reflected signal is maximally sensitive to changes in the sensor's impedance at the matching resistance $R=\sqrt{L/C}$ \citep{schoelkopf1998}. In practice, it has been necessary to carefully adapt the circuit to each material and system architecture. This technique is straightforward to implement in GaAs/AlGaAs reaching 98\% fidelity in 1 $\upmu$s readout time for a singlet-triplet qubit using surface mount inductors and the parasitic capacitance of the circuit board and 2DEG to ground \citep{reilly2007, barthel2009, shulman2014}. Establishing readout with tank circuits in silicon devices has been more difficult, largely because the accumulation gates increase the parasitic capacitance in typical designs and most silicon 2DEGs, in particular the undoped varieties, are more resistive than their GaAs counterparts, both of which prevent the matching resistance from being at an optimal point for the sensor. There have now been a several successful implementations of RF readout in Si/SiGe, adjusting the device and circuit board design to improve sensitivity, achieving over 99\% fidelity in 2 $\upmu$s for singlet-triplet readout \citep{  noiri2020, connors2020, liu2021}. Most of these have performed charge readout, distinguishing the singlet and triplet states with different charge configurations in a DQD. However, it is also possible to use tank circuits to speed up tunneling measurements, as demonstrated in \cite{keith2019}, where donor spin states are measured with 97\% fidelity in 1.5 $\upmu$s.

An alternative approach is to use the qubit gate electrodes as sensors. This becomes particularly attractive as devices with more qubits are made, because gate-based readout could substantially reduce the overhead associated with qubit readout. In this method, a tank circuit is connected to one of the gate electrodes. The reflected signal is sensitive to the quantum capacitance of the qubit $dQ/dV$ a measure of how the charge in a quantum dot moves when probed with a voltage pulse \citep{petersson2010, colless2013}, make it useful for measuring electron tunneling. For this to be useful as a qubit measurement requires that the two qubit states have different quantum capacitances. These have now been used to distinguish between the singlet and triplet states in several different varieties of silicon qubits, and single-shot readout for singlets and triplets has been achieved in both Si-MOS and donors, with fidelities of 70-85\%  \citep{ahmed2018, west2019, schaal2018, pakkiam2018, crippa2019, volk2019} A variant on this is to use cavity sensing through a superconducting resonator, using the same techniques as used with superconducting qubits in the `circuit QED' approach \citep{zheng2019a, wallraff2005}. 

\subsection{Cavity Coupling}
\label{section:cavity}

A major challenge for spin qubits is that qubit-qubit interactions are typically short range, relying on either exchange or capacitive coupling, which fall off on at most the micron scale. There are practical challenges to scaling systems where all the quantum dots are adjacent, including that placing all the gate electrodes becomes nontrivial and the crosstalk between qubits makes control more complicated. Moreover, the ability to convert between stationary and flying qubits (qubits which can transmit quantum information across large distances, such as a photons) is increasingly considered essential to a functional quantum computer \citep{hughes2004}, because using photons to transmit quantum information is important for quantum communication and for the possibility of interfacing in hybrid quantum systems. Therefore, there has been a great push to develop some of the circuit QED methods that were extremely effective in superconducting qubits for spin qubits \citep{devoret2013}.

In circuit QED, qubits are coupled to an electromagnetic resonator, typically a superconducting coplanar waveguide, which transmits information between the qubits coupled to it. It has been challenging to achieve strong coupling in spin qubits due to their reduced charge dipole moment compared to superconducting qubits, and the high levels of charge noise in spin qubits that do have a large coupling to charge. Strong-coupling to charge qubits was achieved first, with examples in both Si/SiGe  \citep{mi2017} and GaAs/AlGaAs \citep{stockklauser2017}, the latter using a high-impedance SQUID resonator. High-impedance resonators have a larger voltage from each photon compared to 50 $\Omega$ resonators, leading to increased interaction rates without directly increasing coupling to noise. Using resonators of 100-1000 $\Omega$ allowed two varieties of spin qubits to reach strong coupling. In the first, a DQD variant of the micromagnet-controlled single-spin qubit in Si/SiGe was used \citep{mi2018,samkharadze2018}, which offered the low noise of a single-spin qubit and the increased gate speeds of a charge qubit. A second technique uses a triple quantum dot in GaAs, which has a large charge dipole \citep{landig2018a}.

One of the difficulties of establishing circuit QED architectures in spin qubits is the necessity of numerous gate electrodes to define the quantum dots. These can cause leakage from the resonator, leading to a low quality factor without careful filtering \citep{mi2017a, harvey-collard2020a}. The requirements of spin qubit fabrication can also conflict with those of resonator fabrication, making the use of 3D architectures with the resonator and qubits on separate chips appealing \citep{holman2020}. Attaching multiple qubits to the same resonator is a particular challenge because the resonator changes the confinement potential on qubits. With a single qubit, the resonator's voltage can be set to the value that optimizes the qubit's properties, but with more qubits a single voltage must suffice for all of them. With clever engineering, however, these effects can be reduced, enabling two-qubit correlations to be measured, although a full two-qubit gate has not yet been demonstrated \citep{borjans2019, borjans2020}. A useful review expanding on these topics is given in \cite{burkard2020}. 

\subsection{Scaling} 
\label{section:tuning}

To perform useful algorithms, spin qubits will have to scale far beyond the current systems, increasing the number of qubits and gaining the ability to implement active error control \citep{brun2020}. While their similarity to the silicon transistor suggests that it may be possible to use industrial techniques to fabricate them in huge quantities, there remains a long way to go before that is feasible, making research into the pathway there essential. 
One of the greatest difficulties in building systems with large numbers of spin qubits is tuning the quantum dots to have the correct properties so that they can act as qubits; this includes the number of electrons and the tunnel coupling to the leads and to one another. These parameters can typically be controlled by gate electrode voltages, but the control is highly non-linear and often even non-monotonic. Moreover, each parameter is controlled by multiple gates and changing one parameter often changes another. There are also large variations from sample to sample, particularly in materials with more disorder. Together, this means that automating tuning is complicated. Due to the amount of work required to create an effective automated tuning program, for many years it was more efficient to rely on experienced researchers to tune QDs, but as the number of QDs in systems has increased, there has been a concerted effort to improve tuning algorithms.

Automated tuning has made progress across all of the key tasks for spin qubits. This includes the process of defining quantum dots \citep{baart2016,lennon2019, moon2020, zwolak2020a}, adjusting the tunnel coupling \citep{vandiepen2018,teske2019, hsiao2020}, and measurement of the energy spectrum and location of charge transitions \citep{botzem2018}. Another way to approach this problem is to design qubits where crosstalk is reduced, and each gate more directly controls a qubit property in a known way, which has been a benefit of the overlapping gate structure  \citep{angus2007, zajac2015, zajac2016, borselli2015}. Simulations have also been used to aid in designing such devices \citep{frees2019}. 

Once there are larger number of qubits in each device, new design questions become important. There have been a number of papers that describe architectures for arrays of qubits incorporating readout with minimal overhead and without compromising on control \citep{jones2012, jones2018}, including for donor qubits \citep{tosi2017} and Si-MOS\citep{veldhorst2017, li2018}. As the number of qubits increases and they are used to perform more complex algorithms, issues such as heating, noise from control electronics, and classical computer overhead can pose obstacles that will require careful consideration \citep{reilly2015, vandijk2019, bluhm2019, geck2019, gonzalez-zalba2020}. One particular focus has been building electronics that can function at cryogenic temperatures \citep{hornibrook2015, xue2021, pauka2021}. One other achievement that may make these requirements less strenuous is the development of `hot' qubits in Si-MOS that work over 1 K \citep{yang2020,petit2020a}. These architecture and wiring questions were reviewed together in \cite{vandersypen2017}. 

One of the ultimate goals, of course, is to manufacture spin qubits with similar processes as computer chips. Over the last decade, there have been demonstrations of spin qubits made using CMOS processes \citep{maurand2016, ciriano-tejel2020} as well as foundry-made quantum dots \citep{ansaloni2020a}. A Si-MOS spin qubit has also been fabricated using fully industry processing and showed similar noise levels to those made with standard techniques \cite{zwerver2021}.

\section{Conclusion}

This article has tried to give order to the wide variety of spin qubits. Because spin qubits share the same basic ingredients, many of the concepts are addressed in multiple sections; for instance, while exchange is used for single qubit gates in singlet-triplet and triple-dot qubits, it is the basis for many two-qubit gates in single-quantum-dot qubits. Moreover, developments in these qubits are interdependent, with progress in one often applicable to others. The evolution of spin qubits so far has suggested that future progress will come from many sources, including improvements in materials and fabrication, new qubit types and coupling schemes, and greater theoretical understanding of the system. Therefore, as the focus turns to building effective quantum information processors that can perform more powerful and complex algorithms, it is essential to have a broad understanding of the field.

\section{Further reading} 
\begin{itemize}
\item \onlinecite{kouwenhoven2001} offers an introduction to quantum dots. 
\item \cite{hanson2007} is a thorough review of progress in spin qubits in quantum dots up to 2007. 
\item \cite{chatterjee2020} is a recent examination of spin qubits, including those in defects and applications to quantum sensing. 
\item \cite{zwanenburg2013} reviews progress in silicon electronics, with extensive focus on Si/SiGe, Si-MOS, and donors, but published in 2013 before many of the key achievements in quantum computing in silicon. 
\item \cite{ladd2018} is a recent discussion of quantum computing in silicon. 

\end{itemize}

\bibliography{bibby}

\begin{thebibliography}{164}%
\makeatletter
\providecommand \@ifxundefined [1]{%
 \@ifx{#1\undefined}
}%
\providecommand \@ifnum [1]{%
 \ifnum #1\expandafter \@firstoftwo
 \else \expandafter \@secondoftwo
 \fi
}%
\providecommand \@ifx [1]{%
 \ifx #1\expandafter \@firstoftwo
 \else \expandafter \@secondoftwo
 \fi
}%
\providecommand \natexlab [1]{#1}%
\providecommand \enquote  [1]{``#1''}%
\providecommand \bibnamefont  [1]{#1}%
\providecommand \bibfnamefont [1]{#1}%
\providecommand \citenamefont [1]{#1}%
\providecommand \href@noop [0]{\@secondoftwo}%
\providecommand \href [0]{\begingroup \@sanitize@url \@href}%
\providecommand \@href[1]{\@@startlink{#1}\@@href}%
\providecommand \@@href[1]{\endgroup#1\@@endlink}%
\providecommand \@sanitize@url [0]{\catcode `\\12\catcode `\$12\catcode
  `\&12\catcode `\#12\catcode `\^12\catcode `\_12\catcode `\%12\relax}%
\providecommand \@@startlink[1]{}%
\providecommand \@@endlink[0]{}%
\providecommand \url  [0]{\begingroup\@sanitize@url \@url }%
\providecommand \@url [1]{\endgroup\@href {#1}{\urlprefix }}%
\providecommand \urlprefix  [0]{URL }%
\providecommand \Eprint [0]{\href }%
\providecommand \doibase [0]{https://doi.org/}%
\providecommand \selectlanguage [0]{\@gobble}%
\providecommand \bibinfo  [0]{\@secondoftwo}%
\providecommand \bibfield  [0]{\@secondoftwo}%
\providecommand \translation [1]{[#1]}%
\providecommand \BibitemOpen [0]{}%
\providecommand \bibitemStop [0]{}%
\providecommand \bibitemNoStop [0]{.\EOS\space}%
\providecommand \EOS [0]{\spacefactor3000\relax}%
\providecommand \BibitemShut  [1]{\csname bibitem#1\endcsname}%
\let\auto@bib@innerbib\@empty
\bibitem [{\citenamefont {Petersson}\ \emph {et~al.}(2012)\citenamefont
  {Petersson}, \citenamefont {McFaul}, \citenamefont {Schroer}, \citenamefont
  {Jung}, \citenamefont {Taylor}, \citenamefont {Houck},\ and\ \citenamefont
  {Petta}}]{petersson2012}%
  \BibitemOpen
  \bibfield  {author} {\bibinfo {author} {\bibfnamefont {K.~D.}\ \bibnamefont
  {Petersson}}, \bibinfo {author} {\bibfnamefont {L.~W.}\ \bibnamefont
  {McFaul}}, \bibinfo {author} {\bibfnamefont {M.~D.}\ \bibnamefont {Schroer}},
  \bibinfo {author} {\bibfnamefont {M.}~\bibnamefont {Jung}}, \bibinfo {author}
  {\bibfnamefont {J.~M.}\ \bibnamefont {Taylor}}, \bibinfo {author}
  {\bibfnamefont {A.~A.}\ \bibnamefont {Houck}},\ and\ \bibinfo {author}
  {\bibfnamefont {J.~R.}\ \bibnamefont {Petta}},\ }\bibfield  {title} {\bibinfo
  {title} {Circuit quantum electrodynamics with a spin qubit},\ }\href
  {https://doi.org/10.1038/nature11559} {\bibfield  {journal} {\bibinfo
  {journal} {Nature}\ }\textbf {\bibinfo {volume} {490}},\ \bibinfo {pages}
  {380} (\bibinfo {year} {2012})}\BibitemShut {NoStop}%
\bibitem [{\citenamefont {Churchill}\ \emph {et~al.}(2009)\citenamefont
  {Churchill}, \citenamefont {Bestwick}, \citenamefont {Harlow}, \citenamefont
  {Kuemmeth}, \citenamefont {Marcos}, \citenamefont {Stwertka}, \citenamefont
  {Watson},\ and\ \citenamefont {Marcus}}]{churchill2009a}%
  \BibitemOpen
  \bibfield  {author} {\bibinfo {author} {\bibfnamefont {H.~O.~H.}\
  \bibnamefont {Churchill}}, \bibinfo {author} {\bibfnamefont {A.~J.}\
  \bibnamefont {Bestwick}}, \bibinfo {author} {\bibfnamefont {J.~W.}\
  \bibnamefont {Harlow}}, \bibinfo {author} {\bibfnamefont {F.}~\bibnamefont
  {Kuemmeth}}, \bibinfo {author} {\bibfnamefont {D.}~\bibnamefont {Marcos}},
  \bibinfo {author} {\bibfnamefont {C.~H.}\ \bibnamefont {Stwertka}}, \bibinfo
  {author} {\bibfnamefont {S.~K.}\ \bibnamefont {Watson}},\ and\ \bibinfo
  {author} {\bibfnamefont {C.~M.}\ \bibnamefont {Marcus}},\ }\bibfield  {title}
  {\bibinfo {title} {Electron\textendash nuclear interaction in 13 {{C}}
  nanotube double quantum dots},\ }\href {https://doi.org/10.1038/nphys1247}
  {\bibfield  {journal} {\bibinfo  {journal} {Nature Physics}\ }\textbf
  {\bibinfo {volume} {5}},\ \bibinfo {pages} {321} (\bibinfo {year}
  {2009})}\BibitemShut {NoStop}%
\bibitem [{\citenamefont {Degen}\ \emph {et~al.}(2017)\citenamefont {Degen},
  \citenamefont {Reinhard},\ and\ \citenamefont {Cappellaro}}]{degen2017}%
  \BibitemOpen
  \bibfield  {author} {\bibinfo {author} {\bibfnamefont {C.~L.}\ \bibnamefont
  {Degen}}, \bibinfo {author} {\bibfnamefont {F.}~\bibnamefont {Reinhard}},\
  and\ \bibinfo {author} {\bibfnamefont {P.}~\bibnamefont {Cappellaro}},\
  }\bibfield  {title} {\bibinfo {title} {Quantum sensing},\ }\href
  {https://doi.org/10.1103/RevModPhys.89.035002} {\bibfield  {journal}
  {\bibinfo  {journal} {Rev. Mod. Phys.}\ }\textbf {\bibinfo {volume} {89}},\
  \bibinfo {pages} {035002} (\bibinfo {year} {2017})}\BibitemShut {NoStop}%
\bibitem [{\citenamefont {Warburton}(2013)}]{warburton2013}%
  \BibitemOpen
  \bibfield  {author} {\bibinfo {author} {\bibfnamefont {R.~J.}\ \bibnamefont
  {Warburton}},\ }\bibfield  {title} {\bibinfo {title} {Single spins in
  self-assembled quantum dots},\ }\href {https://doi.org/10.1038/nmat3585}
  {\bibfield  {journal} {\bibinfo  {journal} {Nature Materials}\ }\textbf
  {\bibinfo {volume} {12}},\ \bibinfo {pages} {483} (\bibinfo {year}
  {2013})}\BibitemShut {NoStop}%
\bibitem [{\citenamefont {Larkin}\ \emph {et~al.}(1984)\citenamefont {Larkin},
  \citenamefont {Likharev},\ and\ \citenamefont {Ovchinnikov}}]{larkin1984}%
  \BibitemOpen
  \bibfield  {author} {\bibinfo {author} {\bibfnamefont {A.~I.}\ \bibnamefont
  {Larkin}}, \bibinfo {author} {\bibfnamefont {K.~K.}\ \bibnamefont
  {Likharev}},\ and\ \bibinfo {author} {\bibfnamefont {Y.~N.}\ \bibnamefont
  {Ovchinnikov}},\ }\bibfield  {title} {\bibinfo {title} {Secondary quantum
  macrpscopic effects in weak superconductivity},\ }\href
  {https://doi.org/10.1016/0378-4363(84)90196-7} {\bibfield  {journal}
  {\bibinfo  {journal} {Physica B+C}\ }\textbf {\bibinfo {volume} {126}},\
  \bibinfo {pages} {414} (\bibinfo {year} {1984})}\BibitemShut {NoStop}%
\bibitem [{\citenamefont {Averin}\ and\ \citenamefont
  {Likharev}(1986)}]{averin1986}%
  \BibitemOpen
  \bibfield  {author} {\bibinfo {author} {\bibfnamefont {D.~V.}\ \bibnamefont
  {Averin}}\ and\ \bibinfo {author} {\bibfnamefont {K.~K.}\ \bibnamefont
  {Likharev}},\ }\bibfield  {title} {\bibinfo {title} {Coulomb blockade of
  single-electron tunneling, and coherent oscillations in small tunnel
  junctions},\ }\href {https://doi.org/10.1007/BF00683469} {\bibfield
  {journal} {\bibinfo  {journal} {J Low Temp Phys}\ }\textbf {\bibinfo {volume}
  {62}},\ \bibinfo {pages} {345} (\bibinfo {year} {1986})}\BibitemShut
  {NoStop}%
\bibitem [{\citenamefont {Fulton}\ and\ \citenamefont
  {Dolan}(1987)}]{fulton1987}%
  \BibitemOpen
  \bibfield  {author} {\bibinfo {author} {\bibfnamefont {T.~A.}\ \bibnamefont
  {Fulton}}\ and\ \bibinfo {author} {\bibfnamefont {G.~J.}\ \bibnamefont
  {Dolan}},\ }\bibfield  {title} {\bibinfo {title} {Observation of
  single-electron charging effects in small tunnel junctions},\ }\href
  {http://journals.aps.org/prl/abstract/10.1103/PhysRevLett.59.109} {\bibfield
  {journal} {\bibinfo  {journal} {Physical review letters}\ }\textbf {\bibinfo
  {volume} {59}},\ \bibinfo {pages} {109} (\bibinfo {year} {1987})}\BibitemShut
  {NoStop}%
\bibitem [{\citenamefont {Beenakker}(1991)}]{beenakker1991a}%
  \BibitemOpen
  \bibfield  {author} {\bibinfo {author} {\bibfnamefont {C.~W.~J.}\
  \bibnamefont {Beenakker}},\ }\bibfield  {title} {\bibinfo {title} {Theory of
  {{Coulomb-blockade}} oscillations in the conductance of a quantum dot},\
  }\href {http://journals.aps.org/prb/abstract/10.1103/PhysRevB.44.1646}
  {\bibfield  {journal} {\bibinfo  {journal} {Physical Review B}\ }\textbf
  {\bibinfo {volume} {44}},\ \bibinfo {pages} {1646} (\bibinfo {year}
  {1991})}\BibitemShut {NoStop}%
\bibitem [{\citenamefont {Ashoori}(1996)}]{ashoori1996}%
  \BibitemOpen
  \bibfield  {author} {\bibinfo {author} {\bibfnamefont {R.~C.}\ \bibnamefont
  {Ashoori}},\ }\bibfield  {title} {\bibinfo {title} {Electrons in artificial
  atoms},\ }\href {https://doi.org/10.1038/379413a0} {\bibfield  {journal}
  {\bibinfo  {journal} {Nature}\ }\textbf {\bibinfo {volume} {379}},\ \bibinfo
  {pages} {413} (\bibinfo {year} {1996})}\BibitemShut {NoStop}%
\bibitem [{\citenamefont {Kouwenhoven}\ \emph {et~al.}(2001)\citenamefont
  {Kouwenhoven}, \citenamefont {Austing},\ and\ \citenamefont
  {Tarucha}}]{kouwenhoven2001}%
  \BibitemOpen
  \bibfield  {author} {\bibinfo {author} {\bibfnamefont {L.~P.}\ \bibnamefont
  {Kouwenhoven}}, \bibinfo {author} {\bibfnamefont {D.~G.}\ \bibnamefont
  {Austing}},\ and\ \bibinfo {author} {\bibfnamefont {S.}~\bibnamefont
  {Tarucha}},\ }\bibfield  {title} {\bibinfo {title} {Few-electron quantum
  dots},\ }\href {http://iopscience.iop.org/0034-4885/64/6/201} {\bibfield
  {journal} {\bibinfo  {journal} {Reports on Progress in Physics}\ }\textbf
  {\bibinfo {volume} {64}},\ \bibinfo {pages} {701} (\bibinfo {year}
  {2001})}\BibitemShut {NoStop}%
\bibitem [{\citenamefont {Tarucha}\ \emph {et~al.}(1996)\citenamefont
  {Tarucha}, \citenamefont {Austing}, \citenamefont {Honda}, \citenamefont
  {{Van der Hage}},\ and\ \citenamefont {Kouwenhoven}}]{tarucha1996}%
  \BibitemOpen
  \bibfield  {author} {\bibinfo {author} {\bibfnamefont {S.}~\bibnamefont
  {Tarucha}}, \bibinfo {author} {\bibfnamefont {D.~G.}\ \bibnamefont
  {Austing}}, \bibinfo {author} {\bibfnamefont {T.}~\bibnamefont {Honda}},
  \bibinfo {author} {\bibfnamefont {R.~J.}\ \bibnamefont {{Van der Hage}}},\
  and\ \bibinfo {author} {\bibfnamefont {L.~P.}\ \bibnamefont {Kouwenhoven}},\
  }\bibfield  {title} {\bibinfo {title} {Shell filling and spin effects in a
  few electron quantum dot},\ }\href
  {http://journals.aps.org/prl/abstract/10.1103/PhysRevLett.77.3613} {\bibfield
   {journal} {\bibinfo  {journal} {Physical Review Letters}\ }\textbf {\bibinfo
  {volume} {77}},\ \bibinfo {pages} {3613} (\bibinfo {year}
  {1996})}\BibitemShut {NoStop}%
\bibitem [{\citenamefont {Davies}(1998)}]{davies1998b}%
  \BibitemOpen
  \bibfield  {author} {\bibinfo {author} {\bibfnamefont {J.~H.}\ \bibnamefont
  {Davies}},\ }\href@noop {} {\emph {\bibinfo {title} {The {{Physics}} of
  {{Low-dimensional Semiconductors}}: {{An Introduction}}}}}\ (\bibinfo
  {publisher} {{Cambridge University Press}},\ \bibinfo {year}
  {1998})\BibitemShut {NoStop}%
\bibitem [{\citenamefont {{van der Wiel}}\ \emph {et~al.}(2002)\citenamefont
  {{van der Wiel}}, \citenamefont {De~Franceschi}, \citenamefont {Elzerman},
  \citenamefont {Fujisawa}, \citenamefont {Tarucha},\ and\ \citenamefont
  {Kouwenhoven}}]{vanderwiel2002}%
  \BibitemOpen
  \bibfield  {author} {\bibinfo {author} {\bibfnamefont {W.~G.}\ \bibnamefont
  {{van der Wiel}}}, \bibinfo {author} {\bibfnamefont {S.}~\bibnamefont
  {De~Franceschi}}, \bibinfo {author} {\bibfnamefont {J.~M.}\ \bibnamefont
  {Elzerman}}, \bibinfo {author} {\bibfnamefont {T.}~\bibnamefont {Fujisawa}},
  \bibinfo {author} {\bibfnamefont {S.}~\bibnamefont {Tarucha}},\ and\ \bibinfo
  {author} {\bibfnamefont {L.~P.}\ \bibnamefont {Kouwenhoven}},\ }\bibfield
  {title} {\bibinfo {title} {Electron transport through double quantum dots},\
  }\href {https://doi.org/10.1103/RevModPhys.75.1} {\bibfield  {journal}
  {\bibinfo  {journal} {Reviews of Modern Physics}\ }\textbf {\bibinfo {volume}
  {75}},\ \bibinfo {pages} {1} (\bibinfo {year} {2002})}\BibitemShut {NoStop}%
\bibitem [{\citenamefont {Leon}\ \emph {et~al.}(2020)\citenamefont {Leon},
  \citenamefont {Yang}, \citenamefont {Hwang}, \citenamefont {Lemyre},
  \citenamefont {Tanttu}, \citenamefont {Huang}, \citenamefont {Chan},
  \citenamefont {Tan}, \citenamefont {Hudson}, \citenamefont {Itoh},
  \citenamefont {Morello}, \citenamefont {Laucht}, \citenamefont
  {{Pioro-Ladri{\`e}re}}, \citenamefont {Saraiva},\ and\ \citenamefont
  {Dzurak}}]{leon2020}%
  \BibitemOpen
  \bibfield  {author} {\bibinfo {author} {\bibfnamefont {R.~C.~C.}\
  \bibnamefont {Leon}}, \bibinfo {author} {\bibfnamefont {C.~H.}\ \bibnamefont
  {Yang}}, \bibinfo {author} {\bibfnamefont {J.~C.~C.}\ \bibnamefont {Hwang}},
  \bibinfo {author} {\bibfnamefont {J.~C.}\ \bibnamefont {Lemyre}}, \bibinfo
  {author} {\bibfnamefont {T.}~\bibnamefont {Tanttu}}, \bibinfo {author}
  {\bibfnamefont {W.}~\bibnamefont {Huang}}, \bibinfo {author} {\bibfnamefont
  {K.~W.}\ \bibnamefont {Chan}}, \bibinfo {author} {\bibfnamefont {K.~Y.}\
  \bibnamefont {Tan}}, \bibinfo {author} {\bibfnamefont {F.~E.}\ \bibnamefont
  {Hudson}}, \bibinfo {author} {\bibfnamefont {K.~M.}\ \bibnamefont {Itoh}},
  \bibinfo {author} {\bibfnamefont {A.}~\bibnamefont {Morello}}, \bibinfo
  {author} {\bibfnamefont {A.}~\bibnamefont {Laucht}}, \bibinfo {author}
  {\bibfnamefont {M.}~\bibnamefont {{Pioro-Ladri{\`e}re}}}, \bibinfo {author}
  {\bibfnamefont {A.}~\bibnamefont {Saraiva}},\ and\ \bibinfo {author}
  {\bibfnamefont {A.~S.}\ \bibnamefont {Dzurak}},\ }\bibfield  {title}
  {\bibinfo {title} {Coherent spin control of s-, p-, d- and f-electrons in a
  silicon quantum dot},\ }\href {https://doi.org/10.1038/s41467-019-14053-w}
  {\bibfield  {journal} {\bibinfo  {journal} {Nature Communications}\ }\textbf
  {\bibinfo {volume} {11}},\ \bibinfo {pages} {797} (\bibinfo {year}
  {2020})}\BibitemShut {NoStop}%
\bibitem [{\citenamefont {Higginbotham}\ \emph {et~al.}(2014)\citenamefont
  {Higginbotham}, \citenamefont {Kuemmeth}, \citenamefont {Hanson},
  \citenamefont {Gossard},\ and\ \citenamefont {Marcus}}]{higginbotham2014}%
  \BibitemOpen
  \bibfield  {author} {\bibinfo {author} {\bibfnamefont {A.~P.}\ \bibnamefont
  {Higginbotham}}, \bibinfo {author} {\bibfnamefont {F.}~\bibnamefont
  {Kuemmeth}}, \bibinfo {author} {\bibfnamefont {M.~P.}\ \bibnamefont
  {Hanson}}, \bibinfo {author} {\bibfnamefont {A.~C.}\ \bibnamefont
  {Gossard}},\ and\ \bibinfo {author} {\bibfnamefont {C.~M.}\ \bibnamefont
  {Marcus}},\ }\bibfield  {title} {\bibinfo {title} {Coherent {{Operations}}
  and {{Screening}} in {{Multielectron Spin Qubits}}},\ }\bibfield  {journal}
  {\bibinfo  {journal} {Physical Review Letters}\ }\textbf {\bibinfo {volume}
  {112}},\ \href {https://doi.org/10.1103/PhysRevLett.112.026801}
  {10.1103/PhysRevLett.112.026801} (\bibinfo {year} {2014})\BibitemShut
  {NoStop}%
\bibitem [{\citenamefont {He}\ \emph {et~al.}(2019)\citenamefont {He},
  \citenamefont {Gorman}, \citenamefont {Keith}, \citenamefont {Kranz},
  \citenamefont {Keizer},\ and\ \citenamefont {Simmons}}]{he2019}%
  \BibitemOpen
  \bibfield  {author} {\bibinfo {author} {\bibfnamefont {Y.}~\bibnamefont
  {He}}, \bibinfo {author} {\bibfnamefont {S.~K.}\ \bibnamefont {Gorman}},
  \bibinfo {author} {\bibfnamefont {D.}~\bibnamefont {Keith}}, \bibinfo
  {author} {\bibfnamefont {L.}~\bibnamefont {Kranz}}, \bibinfo {author}
  {\bibfnamefont {J.~G.}\ \bibnamefont {Keizer}},\ and\ \bibinfo {author}
  {\bibfnamefont {M.~Y.}\ \bibnamefont {Simmons}},\ }\bibfield  {title}
  {\bibinfo {title} {A two-qubit gate between phosphorus donor electrons in
  silicon},\ }\href {https://doi.org/10.1038/s41586-019-1381-2} {\bibfield
  {journal} {\bibinfo  {journal} {Nature}\ }\textbf {\bibinfo {volume} {571}},\
  \bibinfo {pages} {371} (\bibinfo {year} {2019})}\BibitemShut {NoStop}%
\bibitem [{\citenamefont {Reimann}\ and\ \citenamefont
  {Manninen}(2002)}]{reimann2002}%
  \BibitemOpen
  \bibfield  {author} {\bibinfo {author} {\bibfnamefont {S.~M.}\ \bibnamefont
  {Reimann}}\ and\ \bibinfo {author} {\bibfnamefont {M.}~\bibnamefont
  {Manninen}},\ }\bibfield  {title} {\bibinfo {title} {Electronic structure of
  quantum dots},\ }\href {https://doi.org/10.1103/RevModPhys.74.1283}
  {\bibfield  {journal} {\bibinfo  {journal} {Rev. Mod. Phys.}\ }\textbf
  {\bibinfo {volume} {74}},\ \bibinfo {pages} {1283} (\bibinfo {year}
  {2002})}\BibitemShut {NoStop}%
\bibitem [{\citenamefont {Zajac}\ \emph {et~al.}(2015)\citenamefont {Zajac},
  \citenamefont {Hazard}, \citenamefont {Mi}, \citenamefont {Wang},\ and\
  \citenamefont {Petta}}]{zajac2015}%
  \BibitemOpen
  \bibfield  {author} {\bibinfo {author} {\bibfnamefont {D.~M.}\ \bibnamefont
  {Zajac}}, \bibinfo {author} {\bibfnamefont {T.~M.}\ \bibnamefont {Hazard}},
  \bibinfo {author} {\bibfnamefont {X.}~\bibnamefont {Mi}}, \bibinfo {author}
  {\bibfnamefont {K.}~\bibnamefont {Wang}},\ and\ \bibinfo {author}
  {\bibfnamefont {J.~R.}\ \bibnamefont {Petta}},\ }\bibfield  {title} {\bibinfo
  {title} {A reconfigurable gate architecture for {{Si}}/{{SiGe}} quantum
  dots},\ }\href {https://doi.org/10.1063/1.4922249} {\bibfield  {journal}
  {\bibinfo  {journal} {Appl. Phys. Lett.}\ }\textbf {\bibinfo {volume}
  {106}},\ \bibinfo {pages} {223507} (\bibinfo {year} {2015})}\BibitemShut
  {NoStop}%
\bibitem [{\citenamefont {Ciorga}\ \emph {et~al.}(2000)\citenamefont {Ciorga},
  \citenamefont {Sachrajda}, \citenamefont {Hawrylak}, \citenamefont {Gould},
  \citenamefont {Zawadzki}, \citenamefont {Jullian}, \citenamefont {Feng},\
  and\ \citenamefont {Wasilewski}}]{ciorga2000}%
  \BibitemOpen
  \bibfield  {author} {\bibinfo {author} {\bibfnamefont {M.}~\bibnamefont
  {Ciorga}}, \bibinfo {author} {\bibfnamefont {A.~S.}\ \bibnamefont
  {Sachrajda}}, \bibinfo {author} {\bibfnamefont {P.}~\bibnamefont {Hawrylak}},
  \bibinfo {author} {\bibfnamefont {C.}~\bibnamefont {Gould}}, \bibinfo
  {author} {\bibfnamefont {P.}~\bibnamefont {Zawadzki}}, \bibinfo {author}
  {\bibfnamefont {S.}~\bibnamefont {Jullian}}, \bibinfo {author} {\bibfnamefont
  {Y.}~\bibnamefont {Feng}},\ and\ \bibinfo {author} {\bibfnamefont
  {Z.}~\bibnamefont {Wasilewski}},\ }\bibfield  {title} {\bibinfo {title}
  {Addition spectrum of a lateral dot from {{Coulomb}} and spin-blockade
  spectroscopy},\ }\href
  {http://journals.aps.org/prb/abstract/10.1103/PhysRevB.61.R16315} {\bibfield
  {journal} {\bibinfo  {journal} {Physical Review B}\ }\textbf {\bibinfo
  {volume} {61}},\ \bibinfo {pages} {R16315} (\bibinfo {year}
  {2000})}\BibitemShut {NoStop}%
\bibitem [{\citenamefont {Simmons}\ \emph {et~al.}(2007)\citenamefont
  {Simmons}, \citenamefont {Thalakulam}, \citenamefont {Shaji}, \citenamefont
  {Klein}, \citenamefont {Qin}, \citenamefont {Blick}, \citenamefont {Savage},
  \citenamefont {Lagally}, \citenamefont {Coppersmith},\ and\ \citenamefont
  {Eriksson}}]{simmons2007}%
  \BibitemOpen
  \bibfield  {author} {\bibinfo {author} {\bibfnamefont {C.~B.}\ \bibnamefont
  {Simmons}}, \bibinfo {author} {\bibfnamefont {M.}~\bibnamefont {Thalakulam}},
  \bibinfo {author} {\bibfnamefont {N.}~\bibnamefont {Shaji}}, \bibinfo
  {author} {\bibfnamefont {L.~J.}\ \bibnamefont {Klein}}, \bibinfo {author}
  {\bibfnamefont {H.}~\bibnamefont {Qin}}, \bibinfo {author} {\bibfnamefont
  {R.~H.}\ \bibnamefont {Blick}}, \bibinfo {author} {\bibfnamefont {D.~E.}\
  \bibnamefont {Savage}}, \bibinfo {author} {\bibfnamefont {M.~G.}\
  \bibnamefont {Lagally}}, \bibinfo {author} {\bibfnamefont {S.~N.}\
  \bibnamefont {Coppersmith}},\ and\ \bibinfo {author} {\bibfnamefont {M.~A.}\
  \bibnamefont {Eriksson}},\ }\bibfield  {title} {\bibinfo {title}
  {Single-electron quantum dot in {{Si}} / {{SiGe}} with integrated charge
  sensing},\ }\href {https://doi.org/10.1063/1.2816331} {\bibfield  {journal}
  {\bibinfo  {journal} {Applied Physics Letters}\ }\textbf {\bibinfo {volume}
  {91}},\ \bibinfo {pages} {213103} (\bibinfo {year} {2007})}\BibitemShut
  {NoStop}%
\bibitem [{\citenamefont {Field}\ \emph {et~al.}(1993)\citenamefont {Field},
  \citenamefont {Smith}, \citenamefont {Pepper}, \citenamefont {Ritchie},
  \citenamefont {Frost}, \citenamefont {Jones},\ and\ \citenamefont
  {Hasko}}]{field1993}%
  \BibitemOpen
  \bibfield  {author} {\bibinfo {author} {\bibfnamefont {M.}~\bibnamefont
  {Field}}, \bibinfo {author} {\bibfnamefont {C.~G.}\ \bibnamefont {Smith}},
  \bibinfo {author} {\bibfnamefont {M.}~\bibnamefont {Pepper}}, \bibinfo
  {author} {\bibfnamefont {D.~A.}\ \bibnamefont {Ritchie}}, \bibinfo {author}
  {\bibfnamefont {J.~E.~F.}\ \bibnamefont {Frost}}, \bibinfo {author}
  {\bibfnamefont {G.~A.~C.}\ \bibnamefont {Jones}},\ and\ \bibinfo {author}
  {\bibfnamefont {D.~G.}\ \bibnamefont {Hasko}},\ }\bibfield  {title} {\bibinfo
  {title} {Measurements of {{Coulomb}} blockade with a noninvasive voltage
  probe},\ }\href
  {http://journals.aps.org/prl/abstract/10.1103/PhysRevLett.70.1311} {\bibfield
   {journal} {\bibinfo  {journal} {Physical review letters}\ }\textbf {\bibinfo
  {volume} {70}},\ \bibinfo {pages} {1311} (\bibinfo {year}
  {1993})}\BibitemShut {NoStop}%
\bibitem [{\citenamefont {Petta}\ \emph {et~al.}(2005)\citenamefont {Petta},
  \citenamefont {Johnson}, \citenamefont {Taylor}, \citenamefont {Laird},
  \citenamefont {Yacoby}, \citenamefont {Lukin}, \citenamefont {Marcus},
  \citenamefont {Hanson},\ and\ \citenamefont {Gossard}}]{petta2005}%
  \BibitemOpen
  \bibfield  {author} {\bibinfo {author} {\bibfnamefont {J.~R.}\ \bibnamefont
  {Petta}}, \bibinfo {author} {\bibfnamefont {A.~C.}\ \bibnamefont {Johnson}},
  \bibinfo {author} {\bibfnamefont {J.~M.}\ \bibnamefont {Taylor}}, \bibinfo
  {author} {\bibfnamefont {E.~A.}\ \bibnamefont {Laird}}, \bibinfo {author}
  {\bibfnamefont {A.}~\bibnamefont {Yacoby}}, \bibinfo {author} {\bibfnamefont
  {M.~D.}\ \bibnamefont {Lukin}}, \bibinfo {author} {\bibfnamefont {C.~M.}\
  \bibnamefont {Marcus}}, \bibinfo {author} {\bibfnamefont {M.~P.}\
  \bibnamefont {Hanson}},\ and\ \bibinfo {author} {\bibfnamefont {A.~C.}\
  \bibnamefont {Gossard}},\ }\bibfield  {title} {\bibinfo {title} {Coherent
  manipulation of coupled electron spins in semiconductor quantum dots},\
  }\href {https://doi.org/10.1126/science.1116955} {\bibfield  {journal}
  {\bibinfo  {journal} {Science}\ }\textbf {\bibinfo {volume} {309}},\ \bibinfo
  {pages} {2180} (\bibinfo {year} {2005})}\BibitemShut {NoStop}%
\bibitem [{\citenamefont {Hu}\ and\ \citenamefont {Sarma}(2000)}]{hu2000}%
  \BibitemOpen
  \bibfield  {author} {\bibinfo {author} {\bibfnamefont {X.}~\bibnamefont
  {Hu}}\ and\ \bibinfo {author} {\bibfnamefont {S.~D.}\ \bibnamefont {Sarma}},\
  }\bibfield  {title} {\bibinfo {title} {Hilbert-space structure of a
  solid-state quantum computer: {{Two-electron}} states of a double-quantum-dot
  artificial molecule},\ }\href
  {http://journals.aps.org/pra/abstract/10.1103/PhysRevA.61.062301} {\bibfield
  {journal} {\bibinfo  {journal} {Physical Review A}\ }\textbf {\bibinfo
  {volume} {61}},\ \bibinfo {pages} {062301} (\bibinfo {year}
  {2000})}\BibitemShut {NoStop}%
\bibitem [{\citenamefont {Levy}(2002)}]{levy2002}%
  \BibitemOpen
  \bibfield  {author} {\bibinfo {author} {\bibfnamefont {J.}~\bibnamefont
  {Levy}},\ }\bibfield  {title} {\bibinfo {title} {Universal {{Quantum
  Computation}} with {{Spin-1}}/2 {{Pairs}} and {{Heisenberg Exchange}}},\
  }\bibfield  {journal} {\bibinfo  {journal} {Physical Review Letters}\
  }\textbf {\bibinfo {volume} {89}},\ \href
  {https://doi.org/10.1103/PhysRevLett.89.147902}
  {10.1103/PhysRevLett.89.147902} (\bibinfo {year} {2002})\BibitemShut
  {NoStop}%
\bibitem [{\citenamefont {Barthel}\ \emph {et~al.}(2009)\citenamefont
  {Barthel}, \citenamefont {Reilly}, \citenamefont {Marcus}, \citenamefont
  {Hanson},\ and\ \citenamefont {Gossard}}]{barthel2009}%
  \BibitemOpen
  \bibfield  {author} {\bibinfo {author} {\bibfnamefont {C.}~\bibnamefont
  {Barthel}}, \bibinfo {author} {\bibfnamefont {D.~J.}\ \bibnamefont {Reilly}},
  \bibinfo {author} {\bibfnamefont {C.~M.}\ \bibnamefont {Marcus}}, \bibinfo
  {author} {\bibfnamefont {M.~P.}\ \bibnamefont {Hanson}},\ and\ \bibinfo
  {author} {\bibfnamefont {A.~C.}\ \bibnamefont {Gossard}},\ }\bibfield
  {title} {\bibinfo {title} {Rapid {{Single-Shot Measurement}} of a
  {{Singlet-Triplet Qubit}}},\ }\href
  {https://doi.org/10.1103/PhysRevLett.103.160503} {\bibfield  {journal}
  {\bibinfo  {journal} {Phys. Rev. Lett.}\ }\textbf {\bibinfo {volume} {103}},\
  \bibinfo {pages} {160503} (\bibinfo {year} {2009})}\BibitemShut {NoStop}%
\bibitem [{\citenamefont {Dial}\ \emph {et~al.}(2013)\citenamefont {Dial},
  \citenamefont {Shulman}, \citenamefont {Harvey}, \citenamefont {Bluhm},
  \citenamefont {Umansky},\ and\ \citenamefont {Yacoby}}]{dial2013}%
  \BibitemOpen
  \bibfield  {author} {\bibinfo {author} {\bibfnamefont {O.}~\bibnamefont
  {Dial}}, \bibinfo {author} {\bibfnamefont {M.}~\bibnamefont {Shulman}},
  \bibinfo {author} {\bibfnamefont {S.}~\bibnamefont {Harvey}}, \bibinfo
  {author} {\bibfnamefont {H.}~\bibnamefont {Bluhm}}, \bibinfo {author}
  {\bibfnamefont {V.}~\bibnamefont {Umansky}},\ and\ \bibinfo {author}
  {\bibfnamefont {A.}~\bibnamefont {Yacoby}},\ }\bibfield  {title} {\bibinfo
  {title} {Charge {{Noise Spectroscopy Using Coherent Exchange Oscillations}}
  in a {{Singlet-Triplet Qubit}}},\ }\bibfield  {journal} {\bibinfo  {journal}
  {Physical Review Letters}\ }\textbf {\bibinfo {volume} {110}},\ \href
  {https://doi.org/10.1103/PhysRevLett.110.146804}
  {10.1103/PhysRevLett.110.146804} (\bibinfo {year} {2013})\BibitemShut
  {NoStop}%
\bibitem [{\citenamefont {Loss}\ and\ \citenamefont
  {DiVincenzo}(1998)}]{loss1998}%
  \BibitemOpen
  \bibfield  {author} {\bibinfo {author} {\bibfnamefont {D.}~\bibnamefont
  {Loss}}\ and\ \bibinfo {author} {\bibfnamefont {D.~P.}\ \bibnamefont
  {DiVincenzo}},\ }\bibfield  {title} {\bibinfo {title} {Quantum computation
  with quantum dots},\ }\href {https://doi.org/10.1103/PhysRevA.57.120}
  {\bibfield  {journal} {\bibinfo  {journal} {Physical Review A}\ }\textbf
  {\bibinfo {volume} {57}},\ \bibinfo {pages} {120} (\bibinfo {year}
  {1998})}\BibitemShut {NoStop}%
\bibitem [{\citenamefont {Kane}(1998)}]{kane1998}%
  \BibitemOpen
  \bibfield  {author} {\bibinfo {author} {\bibfnamefont {B.~E.}\ \bibnamefont
  {Kane}},\ }\bibfield  {title} {\bibinfo {title} {A silicon-based nuclear spin
  quantum computer},\ }\href
  {http://www.nature.com/nature/journal/v393/n6681/abs/393133a0.html}
  {\bibfield  {journal} {\bibinfo  {journal} {nature}\ }\textbf {\bibinfo
  {volume} {393}},\ \bibinfo {pages} {133} (\bibinfo {year}
  {1998})}\BibitemShut {NoStop}%
\bibitem [{\citenamefont {Slichter}(1990)}]{slichter1990}%
  \BibitemOpen
  \bibfield  {author} {\bibinfo {author} {\bibfnamefont {C.~P.}\ \bibnamefont
  {Slichter}},\ }\href {https://doi.org/10.1007/978-3-662-09441-9} {\emph
  {\bibinfo {title} {Principles of {{Magnetic Resonance}}}}},\ \bibinfo
  {edition} {3rd}\ ed.,\ Springer {{Series}} in {{Solid-State Sciences}}\
  (\bibinfo  {publisher} {{Springer-Verlag}},\ \bibinfo {address} {{Berlin
  Heidelberg}},\ \bibinfo {year} {1990})\BibitemShut {NoStop}%
\bibitem [{\citenamefont {Koppens}\ \emph {et~al.}(2006)\citenamefont
  {Koppens}, \citenamefont {Buizert}, \citenamefont {Tielrooij}, \citenamefont
  {Vink}, \citenamefont {Nowack}, \citenamefont {Meunier}, \citenamefont
  {Kouwenhoven},\ and\ \citenamefont {Vandersypen}}]{koppens2006}%
  \BibitemOpen
  \bibfield  {author} {\bibinfo {author} {\bibfnamefont {F.~H.~L.}\
  \bibnamefont {Koppens}}, \bibinfo {author} {\bibfnamefont {C.}~\bibnamefont
  {Buizert}}, \bibinfo {author} {\bibfnamefont {K.~J.}\ \bibnamefont
  {Tielrooij}}, \bibinfo {author} {\bibfnamefont {I.~T.}\ \bibnamefont {Vink}},
  \bibinfo {author} {\bibfnamefont {K.~C.}\ \bibnamefont {Nowack}}, \bibinfo
  {author} {\bibfnamefont {T.}~\bibnamefont {Meunier}}, \bibinfo {author}
  {\bibfnamefont {L.~P.}\ \bibnamefont {Kouwenhoven}},\ and\ \bibinfo {author}
  {\bibfnamefont {L.~M.~K.}\ \bibnamefont {Vandersypen}},\ }\bibfield  {title}
  {\bibinfo {title} {Driven coherent oscillations of a single electron spin in
  a quantum dot},\ }\href {https://doi.org/10.1038/nature05065} {\bibfield
  {journal} {\bibinfo  {journal} {Nature}\ }\textbf {\bibinfo {volume} {442}},\
  \bibinfo {pages} {766} (\bibinfo {year} {2006})}\BibitemShut {NoStop}%
\bibitem [{\citenamefont {Nowack}\ \emph {et~al.}(2007)\citenamefont {Nowack},
  \citenamefont {Koppens}, \citenamefont {Nazarov},\ and\ \citenamefont
  {Vandersypen}}]{nowack2007}%
  \BibitemOpen
  \bibfield  {author} {\bibinfo {author} {\bibfnamefont {K.~C.}\ \bibnamefont
  {Nowack}}, \bibinfo {author} {\bibfnamefont {F.~H.~L.}\ \bibnamefont
  {Koppens}}, \bibinfo {author} {\bibfnamefont {Y.~V.}\ \bibnamefont
  {Nazarov}},\ and\ \bibinfo {author} {\bibfnamefont {L.~M.~K.}\ \bibnamefont
  {Vandersypen}},\ }\bibfield  {title} {\bibinfo {title} {Coherent {{Control}}
  of a {{Single Electron Spin}} with {{Electric Fields}}},\ }\href
  {https://doi.org/10.1126/science.1148092} {\bibfield  {journal} {\bibinfo
  {journal} {Science}\ }\textbf {\bibinfo {volume} {318}},\ \bibinfo {pages}
  {1430} (\bibinfo {year} {2007})}\BibitemShut {NoStop}%
\bibitem [{\citenamefont {Tokura}\ \emph {et~al.}(2006)\citenamefont {Tokura},
  \citenamefont {{van der Wiel}}, \citenamefont {Obata},\ and\ \citenamefont
  {Tarucha}}]{tokura2006}%
  \BibitemOpen
  \bibfield  {author} {\bibinfo {author} {\bibfnamefont {Y.}~\bibnamefont
  {Tokura}}, \bibinfo {author} {\bibfnamefont {W.~G.}\ \bibnamefont {{van der
  Wiel}}}, \bibinfo {author} {\bibfnamefont {T.}~\bibnamefont {Obata}},\ and\
  \bibinfo {author} {\bibfnamefont {S.}~\bibnamefont {Tarucha}},\ }\bibfield
  {title} {\bibinfo {title} {Coherent {{Single Electron Spin Control}} in a
  {{Slanting Zeeman Field}}},\ }\href
  {https://doi.org/10.1103/PhysRevLett.96.047202} {\bibfield  {journal}
  {\bibinfo  {journal} {Phys. Rev. Lett.}\ }\textbf {\bibinfo {volume} {96}},\
  \bibinfo {pages} {047202} (\bibinfo {year} {2006})}\BibitemShut {NoStop}%
\bibitem [{\citenamefont {{Pioro-Ladri{\`e}re}}\ \emph
  {et~al.}(2008)\citenamefont {{Pioro-Ladri{\`e}re}}, \citenamefont {Obata},
  \citenamefont {Tokura}, \citenamefont {Shin}, \citenamefont {Kubo},
  \citenamefont {Yoshida}, \citenamefont {Taniyama},\ and\ \citenamefont
  {Tarucha}}]{pioro-ladriere2008}%
  \BibitemOpen
  \bibfield  {author} {\bibinfo {author} {\bibfnamefont {M.}~\bibnamefont
  {{Pioro-Ladri{\`e}re}}}, \bibinfo {author} {\bibfnamefont {T.}~\bibnamefont
  {Obata}}, \bibinfo {author} {\bibfnamefont {Y.}~\bibnamefont {Tokura}},
  \bibinfo {author} {\bibfnamefont {Y.-S.}\ \bibnamefont {Shin}}, \bibinfo
  {author} {\bibfnamefont {T.}~\bibnamefont {Kubo}}, \bibinfo {author}
  {\bibfnamefont {K.}~\bibnamefont {Yoshida}}, \bibinfo {author} {\bibfnamefont
  {T.}~\bibnamefont {Taniyama}},\ and\ \bibinfo {author} {\bibfnamefont
  {S.}~\bibnamefont {Tarucha}},\ }\bibfield  {title} {\bibinfo {title}
  {Electrically driven single-electron spin resonance in a slanting {{Zeeman}}
  field},\ }\href {https://doi.org/10.1038/nphys1053} {\bibfield  {journal}
  {\bibinfo  {journal} {Nature Physics}\ }\textbf {\bibinfo {volume} {4}},\
  \bibinfo {pages} {776} (\bibinfo {year} {2008})}\BibitemShut {NoStop}%
\bibitem [{\citenamefont {Tyryshkin}\ \emph {et~al.}(2012)\citenamefont
  {Tyryshkin}, \citenamefont {Tojo}, \citenamefont {Morton}, \citenamefont
  {Riemann}, \citenamefont {Abrosimov}, \citenamefont {Becker}, \citenamefont
  {Pohl}, \citenamefont {Schenkel}, \citenamefont {Thewalt}, \citenamefont
  {Itoh},\ and\ \citenamefont {Lyon}}]{tyryshkin2012}%
  \BibitemOpen
  \bibfield  {author} {\bibinfo {author} {\bibfnamefont {A.~M.}\ \bibnamefont
  {Tyryshkin}}, \bibinfo {author} {\bibfnamefont {S.}~\bibnamefont {Tojo}},
  \bibinfo {author} {\bibfnamefont {J.~J.~L.}\ \bibnamefont {Morton}}, \bibinfo
  {author} {\bibfnamefont {H.}~\bibnamefont {Riemann}}, \bibinfo {author}
  {\bibfnamefont {N.~V.}\ \bibnamefont {Abrosimov}}, \bibinfo {author}
  {\bibfnamefont {P.}~\bibnamefont {Becker}}, \bibinfo {author} {\bibfnamefont
  {H.-J.}\ \bibnamefont {Pohl}}, \bibinfo {author} {\bibfnamefont
  {T.}~\bibnamefont {Schenkel}}, \bibinfo {author} {\bibfnamefont {M.~L.~W.}\
  \bibnamefont {Thewalt}}, \bibinfo {author} {\bibfnamefont {K.~M.}\
  \bibnamefont {Itoh}},\ and\ \bibinfo {author} {\bibfnamefont {S.~A.}\
  \bibnamefont {Lyon}},\ }\bibfield  {title} {\bibinfo {title} {Electron spin
  coherence exceeding seconds in high-purity silicon},\ }\href
  {https://doi.org/10.1038/nmat3182} {\bibfield  {journal} {\bibinfo  {journal}
  {Nature Materials}\ }\textbf {\bibinfo {volume} {11}},\ \bibinfo {pages}
  {143} (\bibinfo {year} {2012})}\BibitemShut {NoStop}%
\bibitem [{\citenamefont {Elzerman}\ \emph {et~al.}(2004)\citenamefont
  {Elzerman}, \citenamefont {Hanson}, \citenamefont {{Willems van Beveren}},
  \citenamefont {Witkamp}, \citenamefont {Vandersypen},\ and\ \citenamefont
  {Kouwenhoven}}]{elzerman2004}%
  \BibitemOpen
  \bibfield  {author} {\bibinfo {author} {\bibfnamefont {J.~M.}\ \bibnamefont
  {Elzerman}}, \bibinfo {author} {\bibfnamefont {R.}~\bibnamefont {Hanson}},
  \bibinfo {author} {\bibfnamefont {L.~H.}\ \bibnamefont {{Willems van
  Beveren}}}, \bibinfo {author} {\bibfnamefont {B.}~\bibnamefont {Witkamp}},
  \bibinfo {author} {\bibfnamefont {L.~M.~K.}\ \bibnamefont {Vandersypen}},\
  and\ \bibinfo {author} {\bibfnamefont {L.~P.}\ \bibnamefont {Kouwenhoven}},\
  }\bibfield  {title} {\bibinfo {title} {Single-shot read-out of an individual
  electron spin in a quantum dot},\ }\href
  {https://doi.org/10.1038/nature02693} {\bibfield  {journal} {\bibinfo
  {journal} {Nature}\ }\textbf {\bibinfo {volume} {430}},\ \bibinfo {pages}
  {431} (\bibinfo {year} {2004})}\BibitemShut {NoStop}%
\bibitem [{\citenamefont {Veldhorst}\ \emph {et~al.}(2014)\citenamefont
  {Veldhorst}, \citenamefont {Hwang}, \citenamefont {Yang}, \citenamefont
  {Leenstra}, \citenamefont {{de Ronde}}, \citenamefont {Dehollain},
  \citenamefont {Muhonen}, \citenamefont {Hudson}, \citenamefont {Itoh},
  \citenamefont {Morello},\ and\ \citenamefont {Dzurak}}]{veldhorst2014}%
  \BibitemOpen
  \bibfield  {author} {\bibinfo {author} {\bibfnamefont {M.}~\bibnamefont
  {Veldhorst}}, \bibinfo {author} {\bibfnamefont {J.~C.~C.}\ \bibnamefont
  {Hwang}}, \bibinfo {author} {\bibfnamefont {C.~H.}\ \bibnamefont {Yang}},
  \bibinfo {author} {\bibfnamefont {A.~W.}\ \bibnamefont {Leenstra}}, \bibinfo
  {author} {\bibfnamefont {B.}~\bibnamefont {{de Ronde}}}, \bibinfo {author}
  {\bibfnamefont {J.~P.}\ \bibnamefont {Dehollain}}, \bibinfo {author}
  {\bibfnamefont {J.~T.}\ \bibnamefont {Muhonen}}, \bibinfo {author}
  {\bibfnamefont {F.~E.}\ \bibnamefont {Hudson}}, \bibinfo {author}
  {\bibfnamefont {K.~M.}\ \bibnamefont {Itoh}}, \bibinfo {author}
  {\bibfnamefont {A.}~\bibnamefont {Morello}},\ and\ \bibinfo {author}
  {\bibfnamefont {A.~S.}\ \bibnamefont {Dzurak}},\ }\bibfield  {title}
  {\bibinfo {title} {An addressable quantum dot qubit with fault-tolerant
  control-fidelity},\ }\href {https://doi.org/10.1038/nnano.2014.216}
  {\bibfield  {journal} {\bibinfo  {journal} {Nature Nanotechnology}\ }\textbf
  {\bibinfo {volume} {9}},\ \bibinfo {pages} {981} (\bibinfo {year}
  {2014})}\BibitemShut {NoStop}%
\bibitem [{\citenamefont {Angus}\ \emph {et~al.}(2007)\citenamefont {Angus},
  \citenamefont {Ferguson}, \citenamefont {Dzurak},\ and\ \citenamefont
  {Clark}}]{angus2007}%
  \BibitemOpen
  \bibfield  {author} {\bibinfo {author} {\bibfnamefont {S.~J.}\ \bibnamefont
  {Angus}}, \bibinfo {author} {\bibfnamefont {A.~J.}\ \bibnamefont {Ferguson}},
  \bibinfo {author} {\bibfnamefont {A.~S.}\ \bibnamefont {Dzurak}},\ and\
  \bibinfo {author} {\bibfnamefont {R.~G.}\ \bibnamefont {Clark}},\ }\bibfield
  {title} {\bibinfo {title} {Gate-{{Defined Quantum Dots}} in {{Intrinsic
  Silicon}}},\ }\href {https://doi.org/10.1021/nl070949k} {\bibfield  {journal}
  {\bibinfo  {journal} {Nano Letters}\ }\textbf {\bibinfo {volume} {7}},\
  \bibinfo {pages} {2051} (\bibinfo {year} {2007})}\BibitemShut {NoStop}%
\bibitem [{\citenamefont {Yang}\ \emph {et~al.}(2013)\citenamefont {Yang},
  \citenamefont {Rossi}, \citenamefont {Ruskov}, \citenamefont {Lai},
  \citenamefont {Mohiyaddin}, \citenamefont {Lee}, \citenamefont {Tahan},
  \citenamefont {Klimeck}, \citenamefont {Morello},\ and\ \citenamefont
  {Dzurak}}]{yang2013a}%
  \BibitemOpen
  \bibfield  {author} {\bibinfo {author} {\bibfnamefont {C.~H.}\ \bibnamefont
  {Yang}}, \bibinfo {author} {\bibfnamefont {A.}~\bibnamefont {Rossi}},
  \bibinfo {author} {\bibfnamefont {R.}~\bibnamefont {Ruskov}}, \bibinfo
  {author} {\bibfnamefont {N.~S.}\ \bibnamefont {Lai}}, \bibinfo {author}
  {\bibfnamefont {F.~A.}\ \bibnamefont {Mohiyaddin}}, \bibinfo {author}
  {\bibfnamefont {S.}~\bibnamefont {Lee}}, \bibinfo {author} {\bibfnamefont
  {C.}~\bibnamefont {Tahan}}, \bibinfo {author} {\bibfnamefont
  {G.}~\bibnamefont {Klimeck}}, \bibinfo {author} {\bibfnamefont
  {A.}~\bibnamefont {Morello}},\ and\ \bibinfo {author} {\bibfnamefont {A.~S.}\
  \bibnamefont {Dzurak}},\ }\bibfield  {title} {\bibinfo {title} {Spin-valley
  lifetimes in a silicon quantum dot with tunable valley splitting},\
  }\bibfield  {journal} {\bibinfo  {journal} {Nature Communications}\ }\textbf
  {\bibinfo {volume} {4}},\ \href {https://doi.org/10.1038/ncomms3069}
  {10.1038/ncomms3069} (\bibinfo {year} {2013})\BibitemShut {NoStop}%
\bibitem [{\citenamefont {Yang}\ \emph {et~al.}(2019)\citenamefont {Yang},
  \citenamefont {Chan}, \citenamefont {Harper}, \citenamefont {Huang},
  \citenamefont {Evans}, \citenamefont {Hwang}, \citenamefont {Hensen},
  \citenamefont {Laucht}, \citenamefont {Tanttu}, \citenamefont {Hudson},
  \citenamefont {Flammia}, \citenamefont {Itoh}, \citenamefont {Morello},
  \citenamefont {Bartlett},\ and\ \citenamefont {Dzurak}}]{yang2019a}%
  \BibitemOpen
  \bibfield  {author} {\bibinfo {author} {\bibfnamefont {C.~H.}\ \bibnamefont
  {Yang}}, \bibinfo {author} {\bibfnamefont {K.~W.}\ \bibnamefont {Chan}},
  \bibinfo {author} {\bibfnamefont {R.}~\bibnamefont {Harper}}, \bibinfo
  {author} {\bibfnamefont {W.}~\bibnamefont {Huang}}, \bibinfo {author}
  {\bibfnamefont {T.}~\bibnamefont {Evans}}, \bibinfo {author} {\bibfnamefont
  {J.~C.~C.}\ \bibnamefont {Hwang}}, \bibinfo {author} {\bibfnamefont
  {B.}~\bibnamefont {Hensen}}, \bibinfo {author} {\bibfnamefont
  {A.}~\bibnamefont {Laucht}}, \bibinfo {author} {\bibfnamefont
  {T.}~\bibnamefont {Tanttu}}, \bibinfo {author} {\bibfnamefont {F.~E.}\
  \bibnamefont {Hudson}}, \bibinfo {author} {\bibfnamefont {S.~T.}\
  \bibnamefont {Flammia}}, \bibinfo {author} {\bibfnamefont {K.~M.}\
  \bibnamefont {Itoh}}, \bibinfo {author} {\bibfnamefont {A.}~\bibnamefont
  {Morello}}, \bibinfo {author} {\bibfnamefont {S.~D.}\ \bibnamefont
  {Bartlett}},\ and\ \bibinfo {author} {\bibfnamefont {A.~S.}\ \bibnamefont
  {Dzurak}},\ }\bibfield  {title} {\bibinfo {title} {Silicon qubit fidelities
  approaching incoherent noise limits via pulse engineering},\ }\href
  {https://doi.org/10.1038/s41928-019-0234-1} {\bibfield  {journal} {\bibinfo
  {journal} {Nature Electronics}\ }\textbf {\bibinfo {volume} {2}},\ \bibinfo
  {pages} {151} (\bibinfo {year} {2019})}\BibitemShut {NoStop}%
\bibitem [{\citenamefont {Huang}\ \emph {et~al.}(2019)\citenamefont {Huang},
  \citenamefont {Yang}, \citenamefont {Chan}, \citenamefont {Tanttu},
  \citenamefont {Hensen}, \citenamefont {Leon}, \citenamefont {Fogarty},
  \citenamefont {Hwang}, \citenamefont {Hudson}, \citenamefont {Itoh},
  \citenamefont {Morello}, \citenamefont {Laucht},\ and\ \citenamefont
  {Dzurak}}]{huang2019}%
  \BibitemOpen
  \bibfield  {author} {\bibinfo {author} {\bibfnamefont {W.}~\bibnamefont
  {Huang}}, \bibinfo {author} {\bibfnamefont {C.~H.}\ \bibnamefont {Yang}},
  \bibinfo {author} {\bibfnamefont {K.~W.}\ \bibnamefont {Chan}}, \bibinfo
  {author} {\bibfnamefont {T.}~\bibnamefont {Tanttu}}, \bibinfo {author}
  {\bibfnamefont {B.}~\bibnamefont {Hensen}}, \bibinfo {author} {\bibfnamefont
  {R.~C.~C.}\ \bibnamefont {Leon}}, \bibinfo {author} {\bibfnamefont {M.~A.}\
  \bibnamefont {Fogarty}}, \bibinfo {author} {\bibfnamefont {J.~C.~C.}\
  \bibnamefont {Hwang}}, \bibinfo {author} {\bibfnamefont {F.~E.}\ \bibnamefont
  {Hudson}}, \bibinfo {author} {\bibfnamefont {K.~M.}\ \bibnamefont {Itoh}},
  \bibinfo {author} {\bibfnamefont {A.}~\bibnamefont {Morello}}, \bibinfo
  {author} {\bibfnamefont {A.}~\bibnamefont {Laucht}},\ and\ \bibinfo {author}
  {\bibfnamefont {A.~S.}\ \bibnamefont {Dzurak}},\ }\bibfield  {title}
  {\bibinfo {title} {Fidelity benchmarks for two-qubit gates in silicon},\
  }\href {https://doi.org/10.1038/s41586-019-1197-0} {\bibfield  {journal}
  {\bibinfo  {journal} {Nature}\ }\textbf {\bibinfo {volume} {569}},\ \bibinfo
  {pages} {532} (\bibinfo {year} {2019})}\BibitemShut {NoStop}%
\bibitem [{\citenamefont {Zwanenburg}\ \emph {et~al.}(2013)\citenamefont
  {Zwanenburg}, \citenamefont {Dzurak}, \citenamefont {Morello}, \citenamefont
  {Simmons}, \citenamefont {Hollenberg}, \citenamefont {Klimeck}, \citenamefont
  {Rogge}, \citenamefont {Coppersmith},\ and\ \citenamefont
  {Eriksson}}]{zwanenburg2013}%
  \BibitemOpen
  \bibfield  {author} {\bibinfo {author} {\bibfnamefont {F.~A.}\ \bibnamefont
  {Zwanenburg}}, \bibinfo {author} {\bibfnamefont {A.~S.}\ \bibnamefont
  {Dzurak}}, \bibinfo {author} {\bibfnamefont {A.}~\bibnamefont {Morello}},
  \bibinfo {author} {\bibfnamefont {M.~Y.}\ \bibnamefont {Simmons}}, \bibinfo
  {author} {\bibfnamefont {L.~C.~L.}\ \bibnamefont {Hollenberg}}, \bibinfo
  {author} {\bibfnamefont {G.}~\bibnamefont {Klimeck}}, \bibinfo {author}
  {\bibfnamefont {S.}~\bibnamefont {Rogge}}, \bibinfo {author} {\bibfnamefont
  {S.~N.}\ \bibnamefont {Coppersmith}},\ and\ \bibinfo {author} {\bibfnamefont
  {M.~A.}\ \bibnamefont {Eriksson}},\ }\bibfield  {title} {\bibinfo {title}
  {Silicon quantum electronics},\ }\href
  {https://doi.org/10.1103/RevModPhys.85.961} {\bibfield  {journal} {\bibinfo
  {journal} {Reviews of Modern Physics}\ }\textbf {\bibinfo {volume} {85}},\
  \bibinfo {pages} {961} (\bibinfo {year} {2013})}\BibitemShut {NoStop}%
\bibitem [{\citenamefont {Kawakami}\ \emph {et~al.}(2013)\citenamefont
  {Kawakami}, \citenamefont {Scarlino}, \citenamefont {Schreiber},
  \citenamefont {Prance}, \citenamefont {Savage}, \citenamefont {Lagally},
  \citenamefont {Eriksson},\ and\ \citenamefont {Vandersypen}}]{kawakami2013}%
  \BibitemOpen
  \bibfield  {author} {\bibinfo {author} {\bibfnamefont {E.}~\bibnamefont
  {Kawakami}}, \bibinfo {author} {\bibfnamefont {P.}~\bibnamefont {Scarlino}},
  \bibinfo {author} {\bibfnamefont {L.~R.}\ \bibnamefont {Schreiber}}, \bibinfo
  {author} {\bibfnamefont {J.~R.}\ \bibnamefont {Prance}}, \bibinfo {author}
  {\bibfnamefont {D.~E.}\ \bibnamefont {Savage}}, \bibinfo {author}
  {\bibfnamefont {M.~G.}\ \bibnamefont {Lagally}}, \bibinfo {author}
  {\bibfnamefont {M.~A.}\ \bibnamefont {Eriksson}},\ and\ \bibinfo {author}
  {\bibfnamefont {L.~M.~K.}\ \bibnamefont {Vandersypen}},\ }\bibfield  {title}
  {\bibinfo {title} {Excitation of a {{Si}}/{{SiGe}} quantum dot using an
  on-chip microwave antenna},\ }\href {https://doi.org/10.1063/1.4821995}
  {\bibfield  {journal} {\bibinfo  {journal} {Applied Physics Letters}\
  }\textbf {\bibinfo {volume} {103}},\ \bibinfo {pages} {132410} (\bibinfo
  {year} {2013})}\BibitemShut {NoStop}%
\bibitem [{\citenamefont {Yoneda}\ \emph {et~al.}(2017)\citenamefont {Yoneda},
  \citenamefont {Takeda}, \citenamefont {Otsuka}, \citenamefont {Nakajima},
  \citenamefont {Delbecq}, \citenamefont {Allison}, \citenamefont {Honda},
  \citenamefont {Kodera}, \citenamefont {Oda}, \citenamefont {Hoshi},
  \citenamefont {Usami}, \citenamefont {Itoh},\ and\ \citenamefont
  {Tarucha}}]{yoneda2017}%
  \BibitemOpen
  \bibfield  {author} {\bibinfo {author} {\bibfnamefont {J.}~\bibnamefont
  {Yoneda}}, \bibinfo {author} {\bibfnamefont {K.}~\bibnamefont {Takeda}},
  \bibinfo {author} {\bibfnamefont {T.}~\bibnamefont {Otsuka}}, \bibinfo
  {author} {\bibfnamefont {T.}~\bibnamefont {Nakajima}}, \bibinfo {author}
  {\bibfnamefont {M.~R.}\ \bibnamefont {Delbecq}}, \bibinfo {author}
  {\bibfnamefont {G.}~\bibnamefont {Allison}}, \bibinfo {author} {\bibfnamefont
  {T.}~\bibnamefont {Honda}}, \bibinfo {author} {\bibfnamefont
  {T.}~\bibnamefont {Kodera}}, \bibinfo {author} {\bibfnamefont
  {S.}~\bibnamefont {Oda}}, \bibinfo {author} {\bibfnamefont {Y.}~\bibnamefont
  {Hoshi}}, \bibinfo {author} {\bibfnamefont {N.}~\bibnamefont {Usami}},
  \bibinfo {author} {\bibfnamefont {K.~M.}\ \bibnamefont {Itoh}},\ and\
  \bibinfo {author} {\bibfnamefont {S.}~\bibnamefont {Tarucha}},\ }\bibfield
  {title} {\bibinfo {title} {A quantum-dot spin qubit with coherence limited by
  charge noise and fidelity higher than 99.9{$\%$}},\ }\href
  {https://doi.org/10.1038/s41565-017-0014-x} {\bibfield  {journal} {\bibinfo
  {journal} {Nature Nanotechnology}\ ,\ \bibinfo {pages} {1}} (\bibinfo {year}
  {2017})}\BibitemShut {NoStop}%
\bibitem [{\citenamefont {Kawakami}\ \emph {et~al.}(2014)\citenamefont
  {Kawakami}, \citenamefont {Scarlino}, \citenamefont {Ward}, \citenamefont
  {Braakman}, \citenamefont {Savage}, \citenamefont {Lagally}, \citenamefont
  {Friesen}, \citenamefont {Coppersmith}, \citenamefont {Eriksson},\ and\
  \citenamefont {Vandersypen}}]{kawakami2014}%
  \BibitemOpen
  \bibfield  {author} {\bibinfo {author} {\bibfnamefont {E.}~\bibnamefont
  {Kawakami}}, \bibinfo {author} {\bibfnamefont {P.}~\bibnamefont {Scarlino}},
  \bibinfo {author} {\bibfnamefont {D.~R.}\ \bibnamefont {Ward}}, \bibinfo
  {author} {\bibfnamefont {F.~R.}\ \bibnamefont {Braakman}}, \bibinfo {author}
  {\bibfnamefont {D.~E.}\ \bibnamefont {Savage}}, \bibinfo {author}
  {\bibfnamefont {M.~G.}\ \bibnamefont {Lagally}}, \bibinfo {author}
  {\bibfnamefont {M.}~\bibnamefont {Friesen}}, \bibinfo {author} {\bibfnamefont
  {S.~N.}\ \bibnamefont {Coppersmith}}, \bibinfo {author} {\bibfnamefont
  {M.~A.}\ \bibnamefont {Eriksson}},\ and\ \bibinfo {author} {\bibfnamefont
  {L.~M.~K.}\ \bibnamefont {Vandersypen}},\ }\bibfield  {title} {\bibinfo
  {title} {Electrical control of a long-lived spin qubit in a {{Si}}/{{SiGe}}
  quantum dot},\ }\href {https://doi.org/10.1038/nnano.2014.153} {\bibfield
  {journal} {\bibinfo  {journal} {Nature Nanotechnology}\ }\textbf {\bibinfo
  {volume} {9}},\ \bibinfo {pages} {666} (\bibinfo {year} {2014})}\BibitemShut
  {NoStop}%
\bibitem [{\citenamefont {Goswami}\ \emph {et~al.}(2007)\citenamefont
  {Goswami}, \citenamefont {Slinker}, \citenamefont {Friesen}, \citenamefont
  {McGuire}, \citenamefont {Truitt}, \citenamefont {Tahan}, \citenamefont
  {Klein}, \citenamefont {Chu}, \citenamefont {Mooney}, \citenamefont {{van der
  Weide}}, \citenamefont {Joynt}, \citenamefont {Coppersmith},\ and\
  \citenamefont {Eriksson}}]{goswami2007}%
  \BibitemOpen
  \bibfield  {author} {\bibinfo {author} {\bibfnamefont {S.}~\bibnamefont
  {Goswami}}, \bibinfo {author} {\bibfnamefont {K.~A.}\ \bibnamefont
  {Slinker}}, \bibinfo {author} {\bibfnamefont {M.}~\bibnamefont {Friesen}},
  \bibinfo {author} {\bibfnamefont {L.~M.}\ \bibnamefont {McGuire}}, \bibinfo
  {author} {\bibfnamefont {J.~L.}\ \bibnamefont {Truitt}}, \bibinfo {author}
  {\bibfnamefont {C.}~\bibnamefont {Tahan}}, \bibinfo {author} {\bibfnamefont
  {L.~J.}\ \bibnamefont {Klein}}, \bibinfo {author} {\bibfnamefont {J.~O.}\
  \bibnamefont {Chu}}, \bibinfo {author} {\bibfnamefont {P.~M.}\ \bibnamefont
  {Mooney}}, \bibinfo {author} {\bibfnamefont {D.~W.}\ \bibnamefont {{van der
  Weide}}}, \bibinfo {author} {\bibfnamefont {R.}~\bibnamefont {Joynt}},
  \bibinfo {author} {\bibfnamefont {S.~N.}\ \bibnamefont {Coppersmith}},\ and\
  \bibinfo {author} {\bibfnamefont {M.~A.}\ \bibnamefont {Eriksson}},\
  }\bibfield  {title} {\bibinfo {title} {Controllable valley splitting in
  silicon quantum devices},\ }\href {https://doi.org/10.1038/nphys475}
  {\bibfield  {journal} {\bibinfo  {journal} {Nature Physics}\ }\textbf
  {\bibinfo {volume} {3}},\ \bibinfo {pages} {41} (\bibinfo {year}
  {2007})}\BibitemShut {NoStop}%
\bibitem [{\citenamefont {Watson}\ \emph {et~al.}(2018)\citenamefont {Watson},
  \citenamefont {Philips}, \citenamefont {Kawakami}, \citenamefont {Ward},
  \citenamefont {Scarlino}, \citenamefont {Veldhorst}, \citenamefont {Savage},
  \citenamefont {Lagally}, \citenamefont {Friesen}, \citenamefont
  {Coppersmith}, \citenamefont {Eriksson},\ and\ \citenamefont
  {Vandersypen}}]{watson2018}%
  \BibitemOpen
  \bibfield  {author} {\bibinfo {author} {\bibfnamefont {T.~F.}\ \bibnamefont
  {Watson}}, \bibinfo {author} {\bibfnamefont {S.~G.~J.}\ \bibnamefont
  {Philips}}, \bibinfo {author} {\bibfnamefont {E.}~\bibnamefont {Kawakami}},
  \bibinfo {author} {\bibfnamefont {D.~R.}\ \bibnamefont {Ward}}, \bibinfo
  {author} {\bibfnamefont {P.}~\bibnamefont {Scarlino}}, \bibinfo {author}
  {\bibfnamefont {M.}~\bibnamefont {Veldhorst}}, \bibinfo {author}
  {\bibfnamefont {D.~E.}\ \bibnamefont {Savage}}, \bibinfo {author}
  {\bibfnamefont {M.~G.}\ \bibnamefont {Lagally}}, \bibinfo {author}
  {\bibfnamefont {M.}~\bibnamefont {Friesen}}, \bibinfo {author} {\bibfnamefont
  {S.~N.}\ \bibnamefont {Coppersmith}}, \bibinfo {author} {\bibfnamefont
  {M.~A.}\ \bibnamefont {Eriksson}},\ and\ \bibinfo {author} {\bibfnamefont
  {L.~M.~K.}\ \bibnamefont {Vandersypen}},\ }\bibfield  {title} {\bibinfo
  {title} {A programmable two-qubit quantum processor in silicon},\ }\href
  {https://doi.org/10.1038/nature25766} {\bibfield  {journal} {\bibinfo
  {journal} {Nature}\ }\textbf {\bibinfo {volume} {555}},\ \bibinfo {pages}
  {633} (\bibinfo {year} {2018})}\BibitemShut {NoStop}%
\bibitem [{\citenamefont {Kawakami}\ \emph {et~al.}(2016)\citenamefont
  {Kawakami}, \citenamefont {Jullien}, \citenamefont {Scarlino}, \citenamefont
  {Ward}, \citenamefont {Savage}, \citenamefont {Lagally}, \citenamefont
  {Dobrovitski}, \citenamefont {Friesen}, \citenamefont {Coppersmith},
  \citenamefont {Eriksson},\ and\ \citenamefont {Vandersypen}}]{kawakami2016}%
  \BibitemOpen
  \bibfield  {author} {\bibinfo {author} {\bibfnamefont {E.}~\bibnamefont
  {Kawakami}}, \bibinfo {author} {\bibfnamefont {T.}~\bibnamefont {Jullien}},
  \bibinfo {author} {\bibfnamefont {P.}~\bibnamefont {Scarlino}}, \bibinfo
  {author} {\bibfnamefont {D.~R.}\ \bibnamefont {Ward}}, \bibinfo {author}
  {\bibfnamefont {D.~E.}\ \bibnamefont {Savage}}, \bibinfo {author}
  {\bibfnamefont {M.~G.}\ \bibnamefont {Lagally}}, \bibinfo {author}
  {\bibfnamefont {V.~V.}\ \bibnamefont {Dobrovitski}}, \bibinfo {author}
  {\bibfnamefont {M.}~\bibnamefont {Friesen}}, \bibinfo {author} {\bibfnamefont
  {S.~N.}\ \bibnamefont {Coppersmith}}, \bibinfo {author} {\bibfnamefont
  {M.~A.}\ \bibnamefont {Eriksson}},\ and\ \bibinfo {author} {\bibfnamefont
  {L.~M.~K.}\ \bibnamefont {Vandersypen}},\ }\bibfield  {title} {\bibinfo
  {title} {Gate fidelity and coherence of an electron spin in an
  {{Si}}/{{SiGe}} quantum dot with micromagnet},\ }\href
  {https://doi.org/10.1073/pnas.1603251113} {\bibfield  {journal} {\bibinfo
  {journal} {PNAS}\ }\textbf {\bibinfo {volume} {113}},\ \bibinfo {pages}
  {11738} (\bibinfo {year} {2016})}\BibitemShut {NoStop}%
\bibitem [{\citenamefont {Zajac}\ \emph {et~al.}(2018)\citenamefont {Zajac},
  \citenamefont {Sigillito}, \citenamefont {Russ}, \citenamefont {Borjans},
  \citenamefont {Taylor}, \citenamefont {Burkard},\ and\ \citenamefont
  {Petta}}]{zajac2018}%
  \BibitemOpen
  \bibfield  {author} {\bibinfo {author} {\bibfnamefont {D.~M.}\ \bibnamefont
  {Zajac}}, \bibinfo {author} {\bibfnamefont {A.~J.}\ \bibnamefont
  {Sigillito}}, \bibinfo {author} {\bibfnamefont {M.}~\bibnamefont {Russ}},
  \bibinfo {author} {\bibfnamefont {F.}~\bibnamefont {Borjans}}, \bibinfo
  {author} {\bibfnamefont {J.~M.}\ \bibnamefont {Taylor}}, \bibinfo {author}
  {\bibfnamefont {G.}~\bibnamefont {Burkard}},\ and\ \bibinfo {author}
  {\bibfnamefont {J.~R.}\ \bibnamefont {Petta}},\ }\bibfield  {title} {\bibinfo
  {title} {Resonantly driven {{CNOT}} gate for electron spins},\ }\href
  {https://doi.org/10.1126/science.aao5965} {\bibfield  {journal} {\bibinfo
  {journal} {Science}\ }\textbf {\bibinfo {volume} {359}},\ \bibinfo {pages}
  {439} (\bibinfo {year} {2018})}\BibitemShut {NoStop}%
\bibitem [{\citenamefont {Xue}\ \emph {et~al.}(2019)\citenamefont {Xue},
  \citenamefont {Watson}, \citenamefont {Helsen}, \citenamefont {Ward},
  \citenamefont {Savage}, \citenamefont {Lagally}, \citenamefont {Coppersmith},
  \citenamefont {Eriksson}, \citenamefont {Wehner},\ and\ \citenamefont
  {Vandersypen}}]{xue2019}%
  \BibitemOpen
  \bibfield  {author} {\bibinfo {author} {\bibfnamefont {X.}~\bibnamefont
  {Xue}}, \bibinfo {author} {\bibfnamefont {T.~F.}\ \bibnamefont {Watson}},
  \bibinfo {author} {\bibfnamefont {J.}~\bibnamefont {Helsen}}, \bibinfo
  {author} {\bibfnamefont {D.~R.}\ \bibnamefont {Ward}}, \bibinfo {author}
  {\bibfnamefont {D.~E.}\ \bibnamefont {Savage}}, \bibinfo {author}
  {\bibfnamefont {M.~G.}\ \bibnamefont {Lagally}}, \bibinfo {author}
  {\bibfnamefont {S.~N.}\ \bibnamefont {Coppersmith}}, \bibinfo {author}
  {\bibfnamefont {M.~A.}\ \bibnamefont {Eriksson}}, \bibinfo {author}
  {\bibfnamefont {S.}~\bibnamefont {Wehner}},\ and\ \bibinfo {author}
  {\bibfnamefont {L.~M.~K.}\ \bibnamefont {Vandersypen}},\ }\bibfield  {title}
  {\bibinfo {title} {Benchmarking {{Gate Fidelities}} in a {{Si}} / {{SiGe}}
  {{Two-Qubit Device}}},\ }\href {https://doi.org/10.1103/PhysRevX.9.021011}
  {\bibfield  {journal} {\bibinfo  {journal} {Phys. Rev. X}\ }\textbf {\bibinfo
  {volume} {9}},\ \bibinfo {pages} {021011} (\bibinfo {year}
  {2019})}\BibitemShut {NoStop}%
\bibitem [{\citenamefont {Tyryshkin}\ \emph {et~al.}(2003)\citenamefont
  {Tyryshkin}, \citenamefont {Lyon}, \citenamefont {Astashkin},\ and\
  \citenamefont {Raitsimring}}]{tyryshkin2003}%
  \BibitemOpen
  \bibfield  {author} {\bibinfo {author} {\bibfnamefont {A.}~\bibnamefont
  {Tyryshkin}}, \bibinfo {author} {\bibfnamefont {S.}~\bibnamefont {Lyon}},
  \bibinfo {author} {\bibfnamefont {A.}~\bibnamefont {Astashkin}},\ and\
  \bibinfo {author} {\bibfnamefont {A.}~\bibnamefont {Raitsimring}},\
  }\bibfield  {title} {\bibinfo {title} {Electron spin relaxation times of
  phosphorus donors in silicon},\ }\bibfield  {journal} {\bibinfo  {journal}
  {Physical Review B}\ }\textbf {\bibinfo {volume} {68}},\ \href
  {https://doi.org/10.1103/PhysRevB.68.193207} {10.1103/PhysRevB.68.193207}
  (\bibinfo {year} {2003})\BibitemShut {NoStop}%
\bibitem [{\citenamefont {Pla}\ \emph {et~al.}(2012)\citenamefont {Pla},
  \citenamefont {Tan}, \citenamefont {Dehollain}, \citenamefont {Lim},
  \citenamefont {Morton}, \citenamefont {Jamieson}, \citenamefont {Dzurak},\
  and\ \citenamefont {Morello}}]{pla2012}%
  \BibitemOpen
  \bibfield  {author} {\bibinfo {author} {\bibfnamefont {J.~J.}\ \bibnamefont
  {Pla}}, \bibinfo {author} {\bibfnamefont {K.~Y.}\ \bibnamefont {Tan}},
  \bibinfo {author} {\bibfnamefont {J.~P.}\ \bibnamefont {Dehollain}}, \bibinfo
  {author} {\bibfnamefont {W.~H.}\ \bibnamefont {Lim}}, \bibinfo {author}
  {\bibfnamefont {J.~J.~L.}\ \bibnamefont {Morton}}, \bibinfo {author}
  {\bibfnamefont {D.~N.}\ \bibnamefont {Jamieson}}, \bibinfo {author}
  {\bibfnamefont {A.~S.}\ \bibnamefont {Dzurak}},\ and\ \bibinfo {author}
  {\bibfnamefont {A.}~\bibnamefont {Morello}},\ }\bibfield  {title} {\bibinfo
  {title} {A single-atom electron spin qubit in silicon},\ }\href
  {https://doi.org/10.1038/nature11449} {\bibfield  {journal} {\bibinfo
  {journal} {Nature}\ }\textbf {\bibinfo {volume} {489}},\ \bibinfo {pages}
  {541} (\bibinfo {year} {2012})}\BibitemShut {NoStop}%
\bibitem [{\citenamefont {Weber}\ \emph {et~al.}(2014)\citenamefont {Weber},
  \citenamefont {Tan}, \citenamefont {Mahapatra}, \citenamefont {Watson},
  \citenamefont {Ryu}, \citenamefont {Rahman}, \citenamefont {Hollenberg},
  \citenamefont {Klimeck},\ and\ \citenamefont {Simmons}}]{weber2014}%
  \BibitemOpen
  \bibfield  {author} {\bibinfo {author} {\bibfnamefont {B.}~\bibnamefont
  {Weber}}, \bibinfo {author} {\bibfnamefont {Y.~H.~M.}\ \bibnamefont {Tan}},
  \bibinfo {author} {\bibfnamefont {S.}~\bibnamefont {Mahapatra}}, \bibinfo
  {author} {\bibfnamefont {T.~F.}\ \bibnamefont {Watson}}, \bibinfo {author}
  {\bibfnamefont {H.}~\bibnamefont {Ryu}}, \bibinfo {author} {\bibfnamefont
  {R.}~\bibnamefont {Rahman}}, \bibinfo {author} {\bibfnamefont {L.~C.~L.}\
  \bibnamefont {Hollenberg}}, \bibinfo {author} {\bibfnamefont
  {G.}~\bibnamefont {Klimeck}},\ and\ \bibinfo {author} {\bibfnamefont {M.~Y.}\
  \bibnamefont {Simmons}},\ }\bibfield  {title} {\bibinfo {title} {Spin
  blockade and exchange in {{Coulomb-confined}} silicon double quantum dots},\
  }\href {https://doi.org/10.1038/nnano.2014.63} {\bibfield  {journal}
  {\bibinfo  {journal} {Nature Nanotechnology}\ }\textbf {\bibinfo {volume}
  {9}},\ \bibinfo {pages} {430} (\bibinfo {year} {2014})}\BibitemShut {NoStop}%
\bibitem [{\citenamefont {Wang}\ \emph {et~al.}(2016)\citenamefont {Wang},
  \citenamefont {Tankasala}, \citenamefont {Hollenberg}, \citenamefont
  {Klimeck}, \citenamefont {Simmons},\ and\ \citenamefont {Rahman}}]{wang2016}%
  \BibitemOpen
  \bibfield  {author} {\bibinfo {author} {\bibfnamefont {Y.}~\bibnamefont
  {Wang}}, \bibinfo {author} {\bibfnamefont {A.}~\bibnamefont {Tankasala}},
  \bibinfo {author} {\bibfnamefont {L.~C.~L.}\ \bibnamefont {Hollenberg}},
  \bibinfo {author} {\bibfnamefont {G.}~\bibnamefont {Klimeck}}, \bibinfo
  {author} {\bibfnamefont {M.~Y.}\ \bibnamefont {Simmons}},\ and\ \bibinfo
  {author} {\bibfnamefont {R.}~\bibnamefont {Rahman}},\ }\bibfield  {title}
  {\bibinfo {title} {Highly tunable exchange in donor qubits in silicon},\
  }\href {https://doi.org/10.1038/npjqi.2016.8} {\bibfield  {journal} {\bibinfo
   {journal} {npj Quantum Information}\ }\textbf {\bibinfo {volume} {2}},\
  \bibinfo {pages} {npjqi20168} (\bibinfo {year} {2016})}\BibitemShut {NoStop}%
\bibitem [{\citenamefont {Morello}\ \emph {et~al.}(2010)\citenamefont
  {Morello}, \citenamefont {Pla}, \citenamefont {Zwanenburg}, \citenamefont
  {Chan}, \citenamefont {Tan}, \citenamefont {Huebl}, \citenamefont
  {M{\"o}tt{\"o}nen}, \citenamefont {Nugroho}, \citenamefont {Yang},
  \citenamefont {{van Donkelaar}}, \citenamefont {Alves}, \citenamefont
  {Jamieson}, \citenamefont {Escott}, \citenamefont {Hollenberg}, \citenamefont
  {Clark},\ and\ \citenamefont {Dzurak}}]{morello2010}%
  \BibitemOpen
  \bibfield  {author} {\bibinfo {author} {\bibfnamefont {A.}~\bibnamefont
  {Morello}}, \bibinfo {author} {\bibfnamefont {J.~J.}\ \bibnamefont {Pla}},
  \bibinfo {author} {\bibfnamefont {F.~A.}\ \bibnamefont {Zwanenburg}},
  \bibinfo {author} {\bibfnamefont {K.~W.}\ \bibnamefont {Chan}}, \bibinfo
  {author} {\bibfnamefont {K.~Y.}\ \bibnamefont {Tan}}, \bibinfo {author}
  {\bibfnamefont {H.}~\bibnamefont {Huebl}}, \bibinfo {author} {\bibfnamefont
  {M.}~\bibnamefont {M{\"o}tt{\"o}nen}}, \bibinfo {author} {\bibfnamefont
  {C.~D.}\ \bibnamefont {Nugroho}}, \bibinfo {author} {\bibfnamefont
  {C.}~\bibnamefont {Yang}}, \bibinfo {author} {\bibfnamefont {J.~A.}\
  \bibnamefont {{van Donkelaar}}}, \bibinfo {author} {\bibfnamefont {A.~D.~C.}\
  \bibnamefont {Alves}}, \bibinfo {author} {\bibfnamefont {D.~N.}\ \bibnamefont
  {Jamieson}}, \bibinfo {author} {\bibfnamefont {C.~C.}\ \bibnamefont
  {Escott}}, \bibinfo {author} {\bibfnamefont {L.~C.~L.}\ \bibnamefont
  {Hollenberg}}, \bibinfo {author} {\bibfnamefont {R.~G.}\ \bibnamefont
  {Clark}},\ and\ \bibinfo {author} {\bibfnamefont {A.~S.}\ \bibnamefont
  {Dzurak}},\ }\bibfield  {title} {\bibinfo {title} {Single-shot readout of an
  electron spin in silicon},\ }\href {https://doi.org/10.1038/nature09392}
  {\bibfield  {journal} {\bibinfo  {journal} {Nature}\ }\textbf {\bibinfo
  {volume} {467}},\ \bibinfo {pages} {687} (\bibinfo {year}
  {2010})}\BibitemShut {NoStop}%
\bibitem [{\citenamefont {Dehollain}\ \emph {et~al.}(2016)\citenamefont
  {Dehollain}, \citenamefont {Simmons}, \citenamefont {Muhonen}, \citenamefont
  {Kalra}, \citenamefont {Laucht}, \citenamefont {Hudson}, \citenamefont
  {Itoh}, \citenamefont {Jamieson}, \citenamefont {McCallum}, \citenamefont
  {Dzurak},\ and\ \citenamefont {Morello}}]{dehollain2016}%
  \BibitemOpen
  \bibfield  {author} {\bibinfo {author} {\bibfnamefont {J.~P.}\ \bibnamefont
  {Dehollain}}, \bibinfo {author} {\bibfnamefont {S.}~\bibnamefont {Simmons}},
  \bibinfo {author} {\bibfnamefont {J.~T.}\ \bibnamefont {Muhonen}}, \bibinfo
  {author} {\bibfnamefont {R.}~\bibnamefont {Kalra}}, \bibinfo {author}
  {\bibfnamefont {A.}~\bibnamefont {Laucht}}, \bibinfo {author} {\bibfnamefont
  {F.}~\bibnamefont {Hudson}}, \bibinfo {author} {\bibfnamefont {K.~M.}\
  \bibnamefont {Itoh}}, \bibinfo {author} {\bibfnamefont {D.~N.}\ \bibnamefont
  {Jamieson}}, \bibinfo {author} {\bibfnamefont {J.~C.}\ \bibnamefont
  {McCallum}}, \bibinfo {author} {\bibfnamefont {A.~S.}\ \bibnamefont
  {Dzurak}},\ and\ \bibinfo {author} {\bibfnamefont {A.}~\bibnamefont
  {Morello}},\ }\bibfield  {title} {\bibinfo {title} {Bell's inequality
  violation with spins in silicon},\ }\href
  {https://doi.org/10.1038/nnano.2015.262} {\bibfield  {journal} {\bibinfo
  {journal} {Nature Nanotechnology}\ }\textbf {\bibinfo {volume} {11}},\
  \bibinfo {pages} {242} (\bibinfo {year} {2016})}\BibitemShut {NoStop}%
\bibitem [{\citenamefont {Muhonen}\ \emph {et~al.}(2014)\citenamefont
  {Muhonen}, \citenamefont {Dehollain}, \citenamefont {Laucht}, \citenamefont
  {Hudson}, \citenamefont {Kalra}, \citenamefont {Sekiguchi}, \citenamefont
  {Itoh}, \citenamefont {Jamieson}, \citenamefont {McCallum}, \citenamefont
  {Dzurak},\ and\ \citenamefont {Morello}}]{muhonen2014}%
  \BibitemOpen
  \bibfield  {author} {\bibinfo {author} {\bibfnamefont {J.~T.}\ \bibnamefont
  {Muhonen}}, \bibinfo {author} {\bibfnamefont {J.~P.}\ \bibnamefont
  {Dehollain}}, \bibinfo {author} {\bibfnamefont {A.}~\bibnamefont {Laucht}},
  \bibinfo {author} {\bibfnamefont {F.~E.}\ \bibnamefont {Hudson}}, \bibinfo
  {author} {\bibfnamefont {R.}~\bibnamefont {Kalra}}, \bibinfo {author}
  {\bibfnamefont {T.}~\bibnamefont {Sekiguchi}}, \bibinfo {author}
  {\bibfnamefont {K.~M.}\ \bibnamefont {Itoh}}, \bibinfo {author}
  {\bibfnamefont {D.~N.}\ \bibnamefont {Jamieson}}, \bibinfo {author}
  {\bibfnamefont {J.~C.}\ \bibnamefont {McCallum}}, \bibinfo {author}
  {\bibfnamefont {A.~S.}\ \bibnamefont {Dzurak}},\ and\ \bibinfo {author}
  {\bibfnamefont {A.}~\bibnamefont {Morello}},\ }\bibfield  {title} {\bibinfo
  {title} {Storing quantum information for 30 seconds in a nanoelectronic
  device},\ }\href {https://doi.org/10.1038/nnano.2014.211} {\bibfield
  {journal} {\bibinfo  {journal} {Nature Nanotechnology}\ }\textbf {\bibinfo
  {volume} {9}},\ \bibinfo {pages} {986} (\bibinfo {year} {2014})}\BibitemShut
  {NoStop}%
\bibitem [{\citenamefont {Dehollain}\ \emph {et~al.}(2014)\citenamefont
  {Dehollain}, \citenamefont {Muhonen}, \citenamefont {Tan}, \citenamefont
  {Saraiva}, \citenamefont {Jamieson}, \citenamefont {Dzurak},\ and\
  \citenamefont {Morello}}]{dehollain2014}%
  \BibitemOpen
  \bibfield  {author} {\bibinfo {author} {\bibfnamefont {J.~P.}\ \bibnamefont
  {Dehollain}}, \bibinfo {author} {\bibfnamefont {J.~T.}\ \bibnamefont
  {Muhonen}}, \bibinfo {author} {\bibfnamefont {K.~Y.}\ \bibnamefont {Tan}},
  \bibinfo {author} {\bibfnamefont {A.}~\bibnamefont {Saraiva}}, \bibinfo
  {author} {\bibfnamefont {D.~N.}\ \bibnamefont {Jamieson}}, \bibinfo {author}
  {\bibfnamefont {A.~S.}\ \bibnamefont {Dzurak}},\ and\ \bibinfo {author}
  {\bibfnamefont {A.}~\bibnamefont {Morello}},\ }\bibfield  {title} {\bibinfo
  {title} {Single-{{Shot Readout}} and {{Relaxation}} of {{Singlet}} and
  {{Triplet States}} in {{Exchange-Coupled Electron Spins}} in {{Silicon}}},\
  }\bibfield  {journal} {\bibinfo  {journal} {Physical Review Letters}\
  }\textbf {\bibinfo {volume} {112}},\ \href
  {https://doi.org/10.1103/PhysRevLett.112.236801}
  {10.1103/PhysRevLett.112.236801} (\bibinfo {year} {2014})\BibitemShut
  {NoStop}%
\bibitem [{\citenamefont {{Gonzalez-Zalba}}\ \emph {et~al.}(2014)\citenamefont
  {{Gonzalez-Zalba}}, \citenamefont {Saraiva}, \citenamefont {Calder{\'o}n},
  \citenamefont {Heiss}, \citenamefont {Koiller},\ and\ \citenamefont
  {Ferguson}}]{gonzalez-zalba2014}%
  \BibitemOpen
  \bibfield  {author} {\bibinfo {author} {\bibfnamefont {M.~F.}\ \bibnamefont
  {{Gonzalez-Zalba}}}, \bibinfo {author} {\bibfnamefont {A.}~\bibnamefont
  {Saraiva}}, \bibinfo {author} {\bibfnamefont {M.~J.}\ \bibnamefont
  {Calder{\'o}n}}, \bibinfo {author} {\bibfnamefont {D.}~\bibnamefont {Heiss}},
  \bibinfo {author} {\bibfnamefont {B.}~\bibnamefont {Koiller}},\ and\ \bibinfo
  {author} {\bibfnamefont {A.~J.}\ \bibnamefont {Ferguson}},\ }\bibfield
  {title} {\bibinfo {title} {An {{Exchange-Coupled Donor Molecule}} in
  {{Silicon}}},\ }\href {https://doi.org/10.1021/nl5023942} {\bibfield
  {journal} {\bibinfo  {journal} {Nano Lett.}\ }\textbf {\bibinfo {volume}
  {14}},\ \bibinfo {pages} {5672} (\bibinfo {year} {2014})}\BibitemShut
  {NoStop}%
\bibitem [{\citenamefont {{Harvey-Collard}}\ \emph {et~al.}(2017)\citenamefont
  {{Harvey-Collard}}, \citenamefont {Jacobson}, \citenamefont {Rudolph},
  \citenamefont {Dominguez}, \citenamefont {Ten~Eyck}, \citenamefont {Wendt},
  \citenamefont {Pluym}, \citenamefont {Gamble}, \citenamefont {Lilly},
  \citenamefont {{Pioro-Ladri{\`e}re}},\ and\ \citenamefont
  {Carroll}}]{harvey-collard2017}%
  \BibitemOpen
  \bibfield  {author} {\bibinfo {author} {\bibfnamefont {P.}~\bibnamefont
  {{Harvey-Collard}}}, \bibinfo {author} {\bibfnamefont {N.~T.}\ \bibnamefont
  {Jacobson}}, \bibinfo {author} {\bibfnamefont {M.}~\bibnamefont {Rudolph}},
  \bibinfo {author} {\bibfnamefont {J.}~\bibnamefont {Dominguez}}, \bibinfo
  {author} {\bibfnamefont {G.~A.}\ \bibnamefont {Ten~Eyck}}, \bibinfo {author}
  {\bibfnamefont {J.~R.}\ \bibnamefont {Wendt}}, \bibinfo {author}
  {\bibfnamefont {T.}~\bibnamefont {Pluym}}, \bibinfo {author} {\bibfnamefont
  {J.~K.}\ \bibnamefont {Gamble}}, \bibinfo {author} {\bibfnamefont {M.~P.}\
  \bibnamefont {Lilly}}, \bibinfo {author} {\bibfnamefont {M.}~\bibnamefont
  {{Pioro-Ladri{\`e}re}}},\ and\ \bibinfo {author} {\bibfnamefont {M.~S.}\
  \bibnamefont {Carroll}},\ }\bibfield  {title} {\bibinfo {title} {Coherent
  coupling between a quantum dot and a donor in silicon},\ }\href
  {https://doi.org/10.1038/s41467-017-01113-2} {\bibfield  {journal} {\bibinfo
  {journal} {Nature Communications}\ }\textbf {\bibinfo {volume} {8}},\
  \bibinfo {pages} {1029} (\bibinfo {year} {2017})}\BibitemShut {NoStop}%
\bibitem [{\citenamefont {Morello}\ \emph {et~al.}(2020)\citenamefont
  {Morello}, \citenamefont {Pla}, \citenamefont {Bertet},\ and\ \citenamefont
  {Jamieson}}]{morello2020}%
  \BibitemOpen
  \bibfield  {author} {\bibinfo {author} {\bibfnamefont {A.}~\bibnamefont
  {Morello}}, \bibinfo {author} {\bibfnamefont {J.~J.}\ \bibnamefont {Pla}},
  \bibinfo {author} {\bibfnamefont {P.}~\bibnamefont {Bertet}},\ and\ \bibinfo
  {author} {\bibfnamefont {D.~N.}\ \bibnamefont {Jamieson}},\ }\bibfield
  {title} {\bibinfo {title} {Donor {{Spins}} in {{Silicon}} for {{Quantum
  Technologies}}},\ }\href {https://doi.org/10.1002/qute.202000005} {\bibfield
  {journal} {\bibinfo  {journal} {Advanced Quantum Technologies}\ }\textbf
  {\bibinfo {volume} {3}},\ \bibinfo {pages} {2000005} (\bibinfo {year}
  {2020})}\BibitemShut {NoStop}%
\bibitem [{\citenamefont {Muhonen}\ \emph {et~al.}(2015)\citenamefont
  {Muhonen}, \citenamefont {Laucht}, \citenamefont {Simmons}, \citenamefont
  {Dehollain}, \citenamefont {Kalra}, \citenamefont {Hudson}, \citenamefont
  {Freer}, \citenamefont {Itoh}, \citenamefont {Jamieson}, \citenamefont
  {McCallum}, \citenamefont {Dzurak},\ and\ \citenamefont
  {Morello}}]{muhonen2015}%
  \BibitemOpen
  \bibfield  {author} {\bibinfo {author} {\bibfnamefont {J.~T.}\ \bibnamefont
  {Muhonen}}, \bibinfo {author} {\bibfnamefont {A.}~\bibnamefont {Laucht}},
  \bibinfo {author} {\bibfnamefont {S.}~\bibnamefont {Simmons}}, \bibinfo
  {author} {\bibfnamefont {J.~P.}\ \bibnamefont {Dehollain}}, \bibinfo {author}
  {\bibfnamefont {R.}~\bibnamefont {Kalra}}, \bibinfo {author} {\bibfnamefont
  {F.~E.}\ \bibnamefont {Hudson}}, \bibinfo {author} {\bibfnamefont
  {S.}~\bibnamefont {Freer}}, \bibinfo {author} {\bibfnamefont {K.~M.}\
  \bibnamefont {Itoh}}, \bibinfo {author} {\bibfnamefont {D.~N.}\ \bibnamefont
  {Jamieson}}, \bibinfo {author} {\bibfnamefont {J.~C.}\ \bibnamefont
  {McCallum}}, \bibinfo {author} {\bibfnamefont {A.~S.}\ \bibnamefont
  {Dzurak}},\ and\ \bibinfo {author} {\bibfnamefont {A.}~\bibnamefont
  {Morello}},\ }\bibfield  {title} {\bibinfo {title} {Quantifying the quantum
  gate fidelity of single-atom spin qubits in silicon by randomized
  benchmarking},\ }\href {https://doi.org/10.1088/0953-8984/27/15/154205}
  {\bibfield  {journal} {\bibinfo  {journal} {J. Phys.: Condens. Matter}\
  }\textbf {\bibinfo {volume} {27}},\ \bibinfo {pages} {154205} (\bibinfo
  {year} {2015})}\BibitemShut {NoStop}%
\bibitem [{\citenamefont {Gorman}\ \emph {et~al.}(2005)\citenamefont {Gorman},
  \citenamefont {Hasko},\ and\ \citenamefont {Williams}}]{gorman2005}%
  \BibitemOpen
  \bibfield  {author} {\bibinfo {author} {\bibfnamefont {J.}~\bibnamefont
  {Gorman}}, \bibinfo {author} {\bibfnamefont {D.~G.}\ \bibnamefont {Hasko}},\
  and\ \bibinfo {author} {\bibfnamefont {D.~A.}\ \bibnamefont {Williams}},\
  }\bibfield  {title} {\bibinfo {title} {Charge-{{Qubit Operation}} of an
  {{Isolated Double Quantum Dot}}},\ }\href
  {https://doi.org/10.1103/PhysRevLett.95.090502} {\bibfield  {journal}
  {\bibinfo  {journal} {Phys. Rev. Lett.}\ }\textbf {\bibinfo {volume} {95}},\
  \bibinfo {pages} {090502} (\bibinfo {year} {2005})}\BibitemShut {NoStop}%
\bibitem [{\citenamefont {Petersson}\ \emph
  {et~al.}(2010{\natexlab{a}})\citenamefont {Petersson}, \citenamefont {Petta},
  \citenamefont {Lu},\ and\ \citenamefont {Gossard}}]{petersson2010a}%
  \BibitemOpen
  \bibfield  {author} {\bibinfo {author} {\bibfnamefont {K.~D.}\ \bibnamefont
  {Petersson}}, \bibinfo {author} {\bibfnamefont {J.~R.}\ \bibnamefont
  {Petta}}, \bibinfo {author} {\bibfnamefont {H.}~\bibnamefont {Lu}},\ and\
  \bibinfo {author} {\bibfnamefont {A.~C.}\ \bibnamefont {Gossard}},\
  }\bibfield  {title} {\bibinfo {title} {Quantum {{Coherence}} in a
  {{One-Electron Semiconductor Charge Qubit}}},\ }\bibfield  {journal}
  {\bibinfo  {journal} {Physical Review Letters}\ }\textbf {\bibinfo {volume}
  {105}},\ \href {https://doi.org/10.1103/PhysRevLett.105.246804}
  {10.1103/PhysRevLett.105.246804} (\bibinfo {year}
  {2010}{\natexlab{a}})\BibitemShut {NoStop}%
\bibitem [{\citenamefont {Dovzhenko}\ \emph {et~al.}(2011)\citenamefont
  {Dovzhenko}, \citenamefont {Stehlik}, \citenamefont {Petersson},
  \citenamefont {Petta}, \citenamefont {Lu},\ and\ \citenamefont
  {Gossard}}]{dovzhenko2011}%
  \BibitemOpen
  \bibfield  {author} {\bibinfo {author} {\bibfnamefont {Y.}~\bibnamefont
  {Dovzhenko}}, \bibinfo {author} {\bibfnamefont {J.}~\bibnamefont {Stehlik}},
  \bibinfo {author} {\bibfnamefont {K.~D.}\ \bibnamefont {Petersson}}, \bibinfo
  {author} {\bibfnamefont {J.~R.}\ \bibnamefont {Petta}}, \bibinfo {author}
  {\bibfnamefont {H.}~\bibnamefont {Lu}},\ and\ \bibinfo {author}
  {\bibfnamefont {A.~C.}\ \bibnamefont {Gossard}},\ }\bibfield  {title}
  {\bibinfo {title} {Nonadiabatic quantum control of a semiconductor charge
  qubit},\ }\href {https://doi.org/10.1103/PhysRevB.84.161302} {\bibfield
  {journal} {\bibinfo  {journal} {Phys. Rev. B}\ }\textbf {\bibinfo {volume}
  {84}},\ \bibinfo {pages} {161302} (\bibinfo {year} {2011})}\BibitemShut
  {NoStop}%
\bibitem [{\citenamefont {Shi}\ \emph {et~al.}(2013)\citenamefont {Shi},
  \citenamefont {Simmons}, \citenamefont {Ward}, \citenamefont {Prance},
  \citenamefont {Mohr}, \citenamefont {Koh}, \citenamefont {Gamble},
  \citenamefont {Wu}, \citenamefont {Savage}, \citenamefont {Lagally},
  \citenamefont {Friesen}, \citenamefont {Coppersmith},\ and\ \citenamefont
  {Eriksson}}]{shi2013}%
  \BibitemOpen
  \bibfield  {author} {\bibinfo {author} {\bibfnamefont {Z.}~\bibnamefont
  {Shi}}, \bibinfo {author} {\bibfnamefont {C.~B.}\ \bibnamefont {Simmons}},
  \bibinfo {author} {\bibfnamefont {D.~R.}\ \bibnamefont {Ward}}, \bibinfo
  {author} {\bibfnamefont {J.~R.}\ \bibnamefont {Prance}}, \bibinfo {author}
  {\bibfnamefont {R.~T.}\ \bibnamefont {Mohr}}, \bibinfo {author}
  {\bibfnamefont {T.~S.}\ \bibnamefont {Koh}}, \bibinfo {author} {\bibfnamefont
  {J.~K.}\ \bibnamefont {Gamble}}, \bibinfo {author} {\bibfnamefont
  {X.}~\bibnamefont {Wu}}, \bibinfo {author} {\bibfnamefont {D.~E.}\
  \bibnamefont {Savage}}, \bibinfo {author} {\bibfnamefont {M.~G.}\
  \bibnamefont {Lagally}}, \bibinfo {author} {\bibfnamefont {M.}~\bibnamefont
  {Friesen}}, \bibinfo {author} {\bibfnamefont {S.~N.}\ \bibnamefont
  {Coppersmith}},\ and\ \bibinfo {author} {\bibfnamefont {M.~A.}\ \bibnamefont
  {Eriksson}},\ }\bibfield  {title} {\bibinfo {title} {Coherent quantum
  oscillations and echo measurements of a {{Si}} charge qubit},\ }\href
  {https://doi.org/10.1103/PhysRevB.88.075416} {\bibfield  {journal} {\bibinfo
  {journal} {Phys. Rev. B}\ }\textbf {\bibinfo {volume} {88}},\ \bibinfo
  {pages} {075416} (\bibinfo {year} {2013})}\BibitemShut {NoStop}%
\bibitem [{\citenamefont {Bluhm}\ \emph {et~al.}(2010)\citenamefont {Bluhm},
  \citenamefont {Foletti}, \citenamefont {Mahalu}, \citenamefont {Umansky},\
  and\ \citenamefont {Yacoby}}]{bluhm2010}%
  \BibitemOpen
  \bibfield  {author} {\bibinfo {author} {\bibfnamefont {H.}~\bibnamefont
  {Bluhm}}, \bibinfo {author} {\bibfnamefont {S.}~\bibnamefont {Foletti}},
  \bibinfo {author} {\bibfnamefont {D.}~\bibnamefont {Mahalu}}, \bibinfo
  {author} {\bibfnamefont {V.}~\bibnamefont {Umansky}},\ and\ \bibinfo {author}
  {\bibfnamefont {A.}~\bibnamefont {Yacoby}},\ }\bibfield  {title} {\bibinfo
  {title} {Enhancing the {{Coherence}} of a {{Spin Qubit}} by {{Operating}} it
  as a {{Feedback Loop That Controls}} its {{Nuclear Spin Bath}}},\ }\href
  {https://doi.org/10.1103/PhysRevLett.105.216803} {\bibfield  {journal}
  {\bibinfo  {journal} {Phys. Rev. Lett.}\ }\textbf {\bibinfo {volume} {105}},\
  \bibinfo {pages} {216803} (\bibinfo {year} {2010})}\BibitemShut {NoStop}%
\bibitem [{\citenamefont {Nichol}\ \emph {et~al.}(2017)\citenamefont {Nichol},
  \citenamefont {Orona}, \citenamefont {Harvey}, \citenamefont {Fallahi},
  \citenamefont {Gardner}, \citenamefont {Manfra},\ and\ \citenamefont
  {Yacoby}}]{nichol2017}%
  \BibitemOpen
  \bibfield  {author} {\bibinfo {author} {\bibfnamefont {J.~M.}\ \bibnamefont
  {Nichol}}, \bibinfo {author} {\bibfnamefont {L.~A.}\ \bibnamefont {Orona}},
  \bibinfo {author} {\bibfnamefont {S.~P.}\ \bibnamefont {Harvey}}, \bibinfo
  {author} {\bibfnamefont {S.}~\bibnamefont {Fallahi}}, \bibinfo {author}
  {\bibfnamefont {G.~C.}\ \bibnamefont {Gardner}}, \bibinfo {author}
  {\bibfnamefont {M.~J.}\ \bibnamefont {Manfra}},\ and\ \bibinfo {author}
  {\bibfnamefont {A.}~\bibnamefont {Yacoby}},\ }\bibfield  {title} {\bibinfo
  {title} {High-fidelity entangling gate for double-quantum-dot spin qubits},\
  }\href {https://doi.org/10.1038/s41534-016-0003-1} {\bibfield  {journal}
  {\bibinfo  {journal} {npj Quantum Information}\ }\textbf {\bibinfo {volume}
  {3}},\ \bibinfo {pages} {3} (\bibinfo {year} {2017})}\BibitemShut {NoStop}%
\bibitem [{\citenamefont {Malinowski}\ \emph
  {et~al.}(2017{\natexlab{a}})\citenamefont {Malinowski}, \citenamefont
  {Martins}, \citenamefont {Nissen}, \citenamefont {Barnes}, \citenamefont
  {Cywi{\'n}ski}, \citenamefont {Rudner}, \citenamefont {Fallahi},
  \citenamefont {Gardner}, \citenamefont {Manfra}, \citenamefont {Marcus},\
  and\ \citenamefont {Kuemmeth}}]{malinowski2017a}%
  \BibitemOpen
  \bibfield  {author} {\bibinfo {author} {\bibfnamefont {F.~K.}\ \bibnamefont
  {Malinowski}}, \bibinfo {author} {\bibfnamefont {F.}~\bibnamefont {Martins}},
  \bibinfo {author} {\bibfnamefont {P.~D.}\ \bibnamefont {Nissen}}, \bibinfo
  {author} {\bibfnamefont {E.}~\bibnamefont {Barnes}}, \bibinfo {author}
  {\bibfnamefont {{\L}.}~\bibnamefont {Cywi{\'n}ski}}, \bibinfo {author}
  {\bibfnamefont {M.~S.}\ \bibnamefont {Rudner}}, \bibinfo {author}
  {\bibfnamefont {S.}~\bibnamefont {Fallahi}}, \bibinfo {author} {\bibfnamefont
  {G.~C.}\ \bibnamefont {Gardner}}, \bibinfo {author} {\bibfnamefont {M.~J.}\
  \bibnamefont {Manfra}}, \bibinfo {author} {\bibfnamefont {C.~M.}\
  \bibnamefont {Marcus}},\ and\ \bibinfo {author} {\bibfnamefont
  {F.}~\bibnamefont {Kuemmeth}},\ }\bibfield  {title} {\bibinfo {title} {Notch
  filtering the nuclear environment of a spin qubit},\ }\href
  {https://doi.org/10.1038/nnano.2016.170} {\bibfield  {journal} {\bibinfo
  {journal} {Nat Nano}\ }\textbf {\bibinfo {volume} {12}},\ \bibinfo {pages}
  {16} (\bibinfo {year} {2017}{\natexlab{a}})}\BibitemShut {NoStop}%
\bibitem [{\citenamefont {Shulman}\ \emph {et~al.}(2014)\citenamefont
  {Shulman}, \citenamefont {Harvey}, \citenamefont {Nichol}, \citenamefont
  {Bartlett}, \citenamefont {Doherty}, \citenamefont {Umansky},\ and\
  \citenamefont {Yacoby}}]{shulman2014}%
  \BibitemOpen
  \bibfield  {author} {\bibinfo {author} {\bibfnamefont {M.~D.}\ \bibnamefont
  {Shulman}}, \bibinfo {author} {\bibfnamefont {S.~P.}\ \bibnamefont {Harvey}},
  \bibinfo {author} {\bibfnamefont {J.~M.}\ \bibnamefont {Nichol}}, \bibinfo
  {author} {\bibfnamefont {S.~D.}\ \bibnamefont {Bartlett}}, \bibinfo {author}
  {\bibfnamefont {A.~C.}\ \bibnamefont {Doherty}}, \bibinfo {author}
  {\bibfnamefont {V.}~\bibnamefont {Umansky}},\ and\ \bibinfo {author}
  {\bibfnamefont {A.}~\bibnamefont {Yacoby}},\ }\bibfield  {title} {\bibinfo
  {title} {Suppressing qubit dephasing using real-time {{Hamiltonian}}
  estimation},\ }\href {https://doi.org/10.1038/ncomms6156} {\bibfield
  {journal} {\bibinfo  {journal} {Nature Communications}\ }\textbf {\bibinfo
  {volume} {5}},\ \bibinfo {pages} {5156} (\bibinfo {year} {2014})}\BibitemShut
  {NoStop}%
\bibitem [{\citenamefont {Cerfontaine}\ \emph {et~al.}(2020)\citenamefont
  {Cerfontaine}, \citenamefont {Botzem}, \citenamefont {Ritzmann},
  \citenamefont {Humpohl}, \citenamefont {Ludwig}, \citenamefont {Schuh},
  \citenamefont {Bougeard}, \citenamefont {Wieck},\ and\ \citenamefont
  {Bluhm}}]{cerfontaine2020}%
  \BibitemOpen
  \bibfield  {author} {\bibinfo {author} {\bibfnamefont {P.}~\bibnamefont
  {Cerfontaine}}, \bibinfo {author} {\bibfnamefont {T.}~\bibnamefont {Botzem}},
  \bibinfo {author} {\bibfnamefont {J.}~\bibnamefont {Ritzmann}}, \bibinfo
  {author} {\bibfnamefont {S.~S.}\ \bibnamefont {Humpohl}}, \bibinfo {author}
  {\bibfnamefont {A.}~\bibnamefont {Ludwig}}, \bibinfo {author} {\bibfnamefont
  {D.}~\bibnamefont {Schuh}}, \bibinfo {author} {\bibfnamefont
  {D.}~\bibnamefont {Bougeard}}, \bibinfo {author} {\bibfnamefont {A.~D.}\
  \bibnamefont {Wieck}},\ and\ \bibinfo {author} {\bibfnamefont
  {H.}~\bibnamefont {Bluhm}},\ }\bibfield  {title} {\bibinfo {title}
  {Closed-loop control of a {{GaAs-based}} singlet-triplet spin qubit with
  99.5{$\%$} gate fidelity and low leakage},\ }\href
  {https://doi.org/10.1038/s41467-020-17865-3} {\bibfield  {journal} {\bibinfo
  {journal} {Nature Communications}\ }\textbf {\bibinfo {volume} {11}},\
  \bibinfo {pages} {4144} (\bibinfo {year} {2020})}\BibitemShut {NoStop}%
\bibitem [{\citenamefont {Takeda}\ \emph {et~al.}(2020)\citenamefont {Takeda},
  \citenamefont {Noiri}, \citenamefont {Yoneda}, \citenamefont {Nakajima},\
  and\ \citenamefont {Tarucha}}]{takeda2020}%
  \BibitemOpen
  \bibfield  {author} {\bibinfo {author} {\bibfnamefont {K.}~\bibnamefont
  {Takeda}}, \bibinfo {author} {\bibfnamefont {A.}~\bibnamefont {Noiri}},
  \bibinfo {author} {\bibfnamefont {J.}~\bibnamefont {Yoneda}}, \bibinfo
  {author} {\bibfnamefont {T.}~\bibnamefont {Nakajima}},\ and\ \bibinfo
  {author} {\bibfnamefont {S.}~\bibnamefont {Tarucha}},\ }\bibfield  {title}
  {\bibinfo {title} {Resonantly {{Driven Singlet-Triplet Spin Qubit}} in
  {{Silicon}}},\ }\href {https://doi.org/10.1103/PhysRevLett.124.117701}
  {\bibfield  {journal} {\bibinfo  {journal} {Phys. Rev. Lett.}\ }\textbf
  {\bibinfo {volume} {124}},\ \bibinfo {pages} {117701} (\bibinfo {year}
  {2020})}\BibitemShut {NoStop}%
\bibitem [{\citenamefont {Reed}\ \emph {et~al.}(2016)\citenamefont {Reed},
  \citenamefont {Maune}, \citenamefont {Andrews}, \citenamefont {Borselli},
  \citenamefont {Eng}, \citenamefont {Jura}, \citenamefont {Kiselev},
  \citenamefont {Ladd}, \citenamefont {Merkel}, \citenamefont {Milosavljevic},
  \citenamefont {Pritchett}, \citenamefont {Rakher}, \citenamefont {Ross},
  \citenamefont {Schmitz}, \citenamefont {Smith}, \citenamefont {Wright},
  \citenamefont {Gyure},\ and\ \citenamefont {Hunter}}]{reed2016}%
  \BibitemOpen
  \bibfield  {author} {\bibinfo {author} {\bibfnamefont {M.~D.}\ \bibnamefont
  {Reed}}, \bibinfo {author} {\bibfnamefont {B.~M.}\ \bibnamefont {Maune}},
  \bibinfo {author} {\bibfnamefont {R.~W.}\ \bibnamefont {Andrews}}, \bibinfo
  {author} {\bibfnamefont {M.~G.}\ \bibnamefont {Borselli}}, \bibinfo {author}
  {\bibfnamefont {K.}~\bibnamefont {Eng}}, \bibinfo {author} {\bibfnamefont
  {M.~P.}\ \bibnamefont {Jura}}, \bibinfo {author} {\bibfnamefont {A.~A.}\
  \bibnamefont {Kiselev}}, \bibinfo {author} {\bibfnamefont {T.~D.}\
  \bibnamefont {Ladd}}, \bibinfo {author} {\bibfnamefont {S.~T.}\ \bibnamefont
  {Merkel}}, \bibinfo {author} {\bibfnamefont {I.}~\bibnamefont
  {Milosavljevic}}, \bibinfo {author} {\bibfnamefont {E.~J.}\ \bibnamefont
  {Pritchett}}, \bibinfo {author} {\bibfnamefont {M.~T.}\ \bibnamefont
  {Rakher}}, \bibinfo {author} {\bibfnamefont {R.~S.}\ \bibnamefont {Ross}},
  \bibinfo {author} {\bibfnamefont {A.~E.}\ \bibnamefont {Schmitz}}, \bibinfo
  {author} {\bibfnamefont {A.}~\bibnamefont {Smith}}, \bibinfo {author}
  {\bibfnamefont {J.~A.}\ \bibnamefont {Wright}}, \bibinfo {author}
  {\bibfnamefont {M.~F.}\ \bibnamefont {Gyure}},\ and\ \bibinfo {author}
  {\bibfnamefont {A.~T.}\ \bibnamefont {Hunter}},\ }\bibfield  {title}
  {\bibinfo {title} {Reduced {{Sensitivity}} to {{Charge Noise}} in
  {{Semiconductor Spin Qubits}} via {{Symmetric Operation}}},\ }\href
  {https://doi.org/10.1103/PhysRevLett.116.110402} {\bibfield  {journal}
  {\bibinfo  {journal} {Phys. Rev. Lett.}\ }\textbf {\bibinfo {volume} {116}},\
  \bibinfo {pages} {110402} (\bibinfo {year} {2016})}\BibitemShut {NoStop}%
\bibitem [{\citenamefont {Martins}\ \emph {et~al.}(2016)\citenamefont
  {Martins}, \citenamefont {Malinowski}, \citenamefont {Nissen}, \citenamefont
  {Barnes}, \citenamefont {Fallahi}, \citenamefont {Gardner}, \citenamefont
  {Manfra}, \citenamefont {Marcus},\ and\ \citenamefont
  {Kuemmeth}}]{martins2016}%
  \BibitemOpen
  \bibfield  {author} {\bibinfo {author} {\bibfnamefont {F.}~\bibnamefont
  {Martins}}, \bibinfo {author} {\bibfnamefont {F.~K.}\ \bibnamefont
  {Malinowski}}, \bibinfo {author} {\bibfnamefont {P.~D.}\ \bibnamefont
  {Nissen}}, \bibinfo {author} {\bibfnamefont {E.}~\bibnamefont {Barnes}},
  \bibinfo {author} {\bibfnamefont {S.}~\bibnamefont {Fallahi}}, \bibinfo
  {author} {\bibfnamefont {G.~C.}\ \bibnamefont {Gardner}}, \bibinfo {author}
  {\bibfnamefont {M.~J.}\ \bibnamefont {Manfra}}, \bibinfo {author}
  {\bibfnamefont {C.~M.}\ \bibnamefont {Marcus}},\ and\ \bibinfo {author}
  {\bibfnamefont {F.}~\bibnamefont {Kuemmeth}},\ }\bibfield  {title} {\bibinfo
  {title} {Noise {{Suppression Using Symmetric Exchange Gates}} in {{Spin
  Qubits}}},\ }\href {https://doi.org/10.1103/PhysRevLett.116.116801}
  {\bibfield  {journal} {\bibinfo  {journal} {Phys. Rev. Lett.}\ }\textbf
  {\bibinfo {volume} {116}},\ \bibinfo {pages} {116801} (\bibinfo {year}
  {2016})}\BibitemShut {NoStop}%
\bibitem [{\citenamefont {Shulman}\ \emph {et~al.}(2012)\citenamefont
  {Shulman}, \citenamefont {Dial}, \citenamefont {Harvey}, \citenamefont
  {Bluhm}, \citenamefont {Umansky},\ and\ \citenamefont
  {Yacoby}}]{shulman2012}%
  \BibitemOpen
  \bibfield  {author} {\bibinfo {author} {\bibfnamefont {M.~D.}\ \bibnamefont
  {Shulman}}, \bibinfo {author} {\bibfnamefont {O.~E.}\ \bibnamefont {Dial}},
  \bibinfo {author} {\bibfnamefont {S.~P.}\ \bibnamefont {Harvey}}, \bibinfo
  {author} {\bibfnamefont {H.}~\bibnamefont {Bluhm}}, \bibinfo {author}
  {\bibfnamefont {V.}~\bibnamefont {Umansky}},\ and\ \bibinfo {author}
  {\bibfnamefont {A.}~\bibnamefont {Yacoby}},\ }\bibfield  {title} {\bibinfo
  {title} {Demonstration of {{Entanglement}} of {{Electrostatically Coupled
  Singlet-Triplet Qubits}}},\ }\href {https://doi.org/10.1126/science.1217692}
  {\bibfield  {journal} {\bibinfo  {journal} {Science}\ }\textbf {\bibinfo
  {volume} {336}},\ \bibinfo {pages} {202} (\bibinfo {year}
  {2012})}\BibitemShut {NoStop}%
\bibitem [{\citenamefont {Kim}\ \emph {et~al.}(2014)\citenamefont {Kim},
  \citenamefont {Shi}, \citenamefont {Simmons}, \citenamefont {Ward},
  \citenamefont {Prance}, \citenamefont {Koh}, \citenamefont {Gamble},
  \citenamefont {Savage}, \citenamefont {Lagally}, \citenamefont {Friesen},
  \citenamefont {Coppersmith},\ and\ \citenamefont {Eriksson}}]{kim2014}%
  \BibitemOpen
  \bibfield  {author} {\bibinfo {author} {\bibfnamefont {D.}~\bibnamefont
  {Kim}}, \bibinfo {author} {\bibfnamefont {Z.}~\bibnamefont {Shi}}, \bibinfo
  {author} {\bibfnamefont {C.~B.}\ \bibnamefont {Simmons}}, \bibinfo {author}
  {\bibfnamefont {D.~R.}\ \bibnamefont {Ward}}, \bibinfo {author}
  {\bibfnamefont {J.~R.}\ \bibnamefont {Prance}}, \bibinfo {author}
  {\bibfnamefont {T.~S.}\ \bibnamefont {Koh}}, \bibinfo {author} {\bibfnamefont
  {J.~K.}\ \bibnamefont {Gamble}}, \bibinfo {author} {\bibfnamefont {D.~E.}\
  \bibnamefont {Savage}}, \bibinfo {author} {\bibfnamefont {M.~G.}\
  \bibnamefont {Lagally}}, \bibinfo {author} {\bibfnamefont {M.}~\bibnamefont
  {Friesen}}, \bibinfo {author} {\bibfnamefont {S.~N.}\ \bibnamefont
  {Coppersmith}},\ and\ \bibinfo {author} {\bibfnamefont {M.~A.}\ \bibnamefont
  {Eriksson}},\ }\bibfield  {title} {\bibinfo {title} {Quantum control and
  process tomography of a semiconductor quantum dot hybrid qubit},\ }\href
  {https://doi.org/10.1038/nature13407} {\bibfield  {journal} {\bibinfo
  {journal} {Nature}\ }\textbf {\bibinfo {volume} {511}},\ \bibinfo {pages}
  {70} (\bibinfo {year} {2014})}\BibitemShut {NoStop}%
\bibitem [{\citenamefont {Kim}\ \emph {et~al.}(2015)\citenamefont {Kim},
  \citenamefont {Ward}, \citenamefont {Simmons}, \citenamefont {Savage},
  \citenamefont {Lagally}, \citenamefont {Friesen}, \citenamefont
  {Coppersmith},\ and\ \citenamefont {Eriksson}}]{kim2015}%
  \BibitemOpen
  \bibfield  {author} {\bibinfo {author} {\bibfnamefont {D.}~\bibnamefont
  {Kim}}, \bibinfo {author} {\bibfnamefont {D.~R.}\ \bibnamefont {Ward}},
  \bibinfo {author} {\bibfnamefont {C.~B.}\ \bibnamefont {Simmons}}, \bibinfo
  {author} {\bibfnamefont {D.~E.}\ \bibnamefont {Savage}}, \bibinfo {author}
  {\bibfnamefont {M.~G.}\ \bibnamefont {Lagally}}, \bibinfo {author}
  {\bibfnamefont {M.}~\bibnamefont {Friesen}}, \bibinfo {author} {\bibfnamefont
  {S.~N.}\ \bibnamefont {Coppersmith}},\ and\ \bibinfo {author} {\bibfnamefont
  {M.~A.}\ \bibnamefont {Eriksson}},\ }\bibfield  {title} {\bibinfo {title}
  {High-fidelity resonant gating of a silicon-based quantum dot hybrid qubit},\
  }\href {https://doi.org/10.1038/npjqi.2015.4} {\bibfield  {journal} {\bibinfo
   {journal} {npj Quantum Information}\ }\textbf {\bibinfo {volume} {1}},\
  \bibinfo {pages} {15004} (\bibinfo {year} {2015})}\BibitemShut {NoStop}%
\bibitem [{\citenamefont {Russ}\ and\ \citenamefont
  {Burkard}(2017)}]{russ2017}%
  \BibitemOpen
  \bibfield  {author} {\bibinfo {author} {\bibfnamefont {M.}~\bibnamefont
  {Russ}}\ and\ \bibinfo {author} {\bibfnamefont {G.}~\bibnamefont {Burkard}},\
  }\bibfield  {title} {\bibinfo {title} {Three-electron spin qubits},\ }\href
  {https://doi.org/10.1088/1361-648X/aa761f} {\bibfield  {journal} {\bibinfo
  {journal} {J. Phys.: Condens. Matter}\ }\textbf {\bibinfo {volume} {29}},\
  \bibinfo {pages} {393001} (\bibinfo {year} {2017})}\BibitemShut {NoStop}%
\bibitem [{\citenamefont {Laird}\ \emph {et~al.}(2010)\citenamefont {Laird},
  \citenamefont {Taylor}, \citenamefont {DiVincenzo}, \citenamefont {Marcus},
  \citenamefont {Hanson},\ and\ \citenamefont {Gossard}}]{laird2010}%
  \BibitemOpen
  \bibfield  {author} {\bibinfo {author} {\bibfnamefont {E.~A.}\ \bibnamefont
  {Laird}}, \bibinfo {author} {\bibfnamefont {J.~M.}\ \bibnamefont {Taylor}},
  \bibinfo {author} {\bibfnamefont {D.~P.}\ \bibnamefont {DiVincenzo}},
  \bibinfo {author} {\bibfnamefont {C.~M.}\ \bibnamefont {Marcus}}, \bibinfo
  {author} {\bibfnamefont {M.~P.}\ \bibnamefont {Hanson}},\ and\ \bibinfo
  {author} {\bibfnamefont {A.~C.}\ \bibnamefont {Gossard}},\ }\bibfield
  {title} {\bibinfo {title} {Coherent spin manipulation in an exchange-only
  qubit},\ }\bibfield  {journal} {\bibinfo  {journal} {Physical Review B}\
  }\textbf {\bibinfo {volume} {82}},\ \href
  {https://doi.org/10.1103/PhysRevB.82.075403} {10.1103/PhysRevB.82.075403}
  (\bibinfo {year} {2010})\BibitemShut {NoStop}%
\bibitem [{\citenamefont {Medford}\ \emph {et~al.}(2013)\citenamefont
  {Medford}, \citenamefont {Beil}, \citenamefont {Taylor}, \citenamefont
  {Bartlett}, \citenamefont {Doherty}, \citenamefont {Rashba}, \citenamefont
  {DiVincenzo}, \citenamefont {Lu}, \citenamefont {Gossard},\ and\
  \citenamefont {Marcus}}]{medford2013}%
  \BibitemOpen
  \bibfield  {author} {\bibinfo {author} {\bibfnamefont {J.}~\bibnamefont
  {Medford}}, \bibinfo {author} {\bibfnamefont {J.}~\bibnamefont {Beil}},
  \bibinfo {author} {\bibfnamefont {J.~M.}\ \bibnamefont {Taylor}}, \bibinfo
  {author} {\bibfnamefont {S.~D.}\ \bibnamefont {Bartlett}}, \bibinfo {author}
  {\bibfnamefont {A.~C.}\ \bibnamefont {Doherty}}, \bibinfo {author}
  {\bibfnamefont {E.~I.}\ \bibnamefont {Rashba}}, \bibinfo {author}
  {\bibfnamefont {D.~P.}\ \bibnamefont {DiVincenzo}}, \bibinfo {author}
  {\bibfnamefont {H.}~\bibnamefont {Lu}}, \bibinfo {author} {\bibfnamefont
  {A.~C.}\ \bibnamefont {Gossard}},\ and\ \bibinfo {author} {\bibfnamefont
  {C.~M.}\ \bibnamefont {Marcus}},\ }\bibfield  {title} {\bibinfo {title}
  {Self-consistent measurement and state tomography of an exchange-only spin
  qubit},\ }\href {https://doi.org/10.1038/nnano.2013.168} {\bibfield
  {journal} {\bibinfo  {journal} {Nature Nanotechnology}\ }\textbf {\bibinfo
  {volume} {8}},\ \bibinfo {pages} {654} (\bibinfo {year} {2013})}\BibitemShut
  {NoStop}%
\bibitem [{\citenamefont {Eng}\ \emph {et~al.}(2015)\citenamefont {Eng},
  \citenamefont {Ladd}, \citenamefont {Smith}, \citenamefont {Borselli},
  \citenamefont {Kiselev}, \citenamefont {Fong}, \citenamefont {Holabird},
  \citenamefont {Hazard}, \citenamefont {Huang}, \citenamefont {Deelman},
  \citenamefont {Milosavljevic}, \citenamefont {Schmitz}, \citenamefont {Ross},
  \citenamefont {Gyure},\ and\ \citenamefont {Hunter}}]{eng2015}%
  \BibitemOpen
  \bibfield  {author} {\bibinfo {author} {\bibfnamefont {K.}~\bibnamefont
  {Eng}}, \bibinfo {author} {\bibfnamefont {T.~D.}\ \bibnamefont {Ladd}},
  \bibinfo {author} {\bibfnamefont {A.}~\bibnamefont {Smith}}, \bibinfo
  {author} {\bibfnamefont {M.~G.}\ \bibnamefont {Borselli}}, \bibinfo {author}
  {\bibfnamefont {A.~A.}\ \bibnamefont {Kiselev}}, \bibinfo {author}
  {\bibfnamefont {B.~H.}\ \bibnamefont {Fong}}, \bibinfo {author}
  {\bibfnamefont {K.~S.}\ \bibnamefont {Holabird}}, \bibinfo {author}
  {\bibfnamefont {T.~M.}\ \bibnamefont {Hazard}}, \bibinfo {author}
  {\bibfnamefont {B.}~\bibnamefont {Huang}}, \bibinfo {author} {\bibfnamefont
  {P.~W.}\ \bibnamefont {Deelman}}, \bibinfo {author} {\bibfnamefont
  {I.}~\bibnamefont {Milosavljevic}}, \bibinfo {author} {\bibfnamefont {A.~E.}\
  \bibnamefont {Schmitz}}, \bibinfo {author} {\bibfnamefont {R.~S.}\
  \bibnamefont {Ross}}, \bibinfo {author} {\bibfnamefont {M.~F.}\ \bibnamefont
  {Gyure}},\ and\ \bibinfo {author} {\bibfnamefont {A.~T.}\ \bibnamefont
  {Hunter}},\ }\bibfield  {title} {\bibinfo {title} {Isotopically enhanced
  triple-quantum-dot qubit},\ }\href {https://doi.org/10.1126/sciadv.1500214}
  {\bibfield  {journal} {\bibinfo  {journal} {Science Advances}\ }\textbf
  {\bibinfo {volume} {1}},\ \bibinfo {pages} {e1500214} (\bibinfo {year}
  {2015})}\BibitemShut {NoStop}%
\bibitem [{\citenamefont {Andrews}\ \emph {et~al.}(2019)\citenamefont
  {Andrews}, \citenamefont {Jones}, \citenamefont {Reed}, \citenamefont
  {Jones}, \citenamefont {Ha}, \citenamefont {Jura}, \citenamefont {Kerckhoff},
  \citenamefont {Levendorf}, \citenamefont {Meenehan}, \citenamefont {Merkel},
  \citenamefont {Smith}, \citenamefont {Sun}, \citenamefont {Weinstein},
  \citenamefont {Rakher}, \citenamefont {Ladd},\ and\ \citenamefont
  {Borselli}}]{andrews2019}%
  \BibitemOpen
  \bibfield  {author} {\bibinfo {author} {\bibfnamefont {R.~W.}\ \bibnamefont
  {Andrews}}, \bibinfo {author} {\bibfnamefont {C.}~\bibnamefont {Jones}},
  \bibinfo {author} {\bibfnamefont {M.~D.}\ \bibnamefont {Reed}}, \bibinfo
  {author} {\bibfnamefont {A.~M.}\ \bibnamefont {Jones}}, \bibinfo {author}
  {\bibfnamefont {S.~D.}\ \bibnamefont {Ha}}, \bibinfo {author} {\bibfnamefont
  {M.~P.}\ \bibnamefont {Jura}}, \bibinfo {author} {\bibfnamefont
  {J.}~\bibnamefont {Kerckhoff}}, \bibinfo {author} {\bibfnamefont
  {M.}~\bibnamefont {Levendorf}}, \bibinfo {author} {\bibfnamefont
  {S.}~\bibnamefont {Meenehan}}, \bibinfo {author} {\bibfnamefont {S.~T.}\
  \bibnamefont {Merkel}}, \bibinfo {author} {\bibfnamefont {A.}~\bibnamefont
  {Smith}}, \bibinfo {author} {\bibfnamefont {B.}~\bibnamefont {Sun}}, \bibinfo
  {author} {\bibfnamefont {A.~J.}\ \bibnamefont {Weinstein}}, \bibinfo {author}
  {\bibfnamefont {M.~T.}\ \bibnamefont {Rakher}}, \bibinfo {author}
  {\bibfnamefont {T.~D.}\ \bibnamefont {Ladd}},\ and\ \bibinfo {author}
  {\bibfnamefont {M.~G.}\ \bibnamefont {Borselli}},\ }\bibfield  {title}
  {\bibinfo {title} {Quantifying error and leakage in an encoded
  {{Si}}/{{SiGe}} triple-dot qubit},\ }\href
  {https://doi.org/10.1038/s41565-019-0500-4} {\bibfield  {journal} {\bibinfo
  {journal} {Nat. Nanotechnol.}\ }\textbf {\bibinfo {volume} {14}},\ \bibinfo
  {pages} {747} (\bibinfo {year} {2019})}\BibitemShut {NoStop}%
\bibitem [{\citenamefont {Cywi{\'n}ski}\ \emph {et~al.}(2008)\citenamefont
  {Cywi{\'n}ski}, \citenamefont {Lutchyn}, \citenamefont {Nave},\ and\
  \citenamefont {Das~Sarma}}]{cywinski2008}%
  \BibitemOpen
  \bibfield  {author} {\bibinfo {author} {\bibfnamefont {{\L}.}~\bibnamefont
  {Cywi{\'n}ski}}, \bibinfo {author} {\bibfnamefont {R.}~\bibnamefont
  {Lutchyn}}, \bibinfo {author} {\bibfnamefont {C.}~\bibnamefont {Nave}},\ and\
  \bibinfo {author} {\bibfnamefont {S.}~\bibnamefont {Das~Sarma}},\ }\bibfield
  {title} {\bibinfo {title} {How to enhance dephasing time in superconducting
  qubits},\ }\bibfield  {journal} {\bibinfo  {journal} {Physical Review B}\
  }\textbf {\bibinfo {volume} {77}},\ \href
  {https://doi.org/10.1103/PhysRevB.77.174509} {10.1103/PhysRevB.77.174509}
  (\bibinfo {year} {2008})\BibitemShut {NoStop}%
\bibitem [{\citenamefont {{\'A}lvarez}\ and\ \citenamefont
  {Suter}(2011)}]{alvarez2011}%
  \BibitemOpen
  \bibfield  {author} {\bibinfo {author} {\bibfnamefont {G.~A.}\ \bibnamefont
  {{\'A}lvarez}}\ and\ \bibinfo {author} {\bibfnamefont {D.}~\bibnamefont
  {Suter}},\ }\bibfield  {title} {\bibinfo {title} {Measuring the {{Spectrum}}
  of {{Colored Noise}} by {{Dynamical Decoupling}}},\ }\href
  {https://doi.org/10.1103/PhysRevLett.107.230501} {\bibfield  {journal}
  {\bibinfo  {journal} {Phys. Rev. Lett.}\ }\textbf {\bibinfo {volume} {107}},\
  \bibinfo {pages} {230501} (\bibinfo {year} {2011})}\BibitemShut {NoStop}%
\bibitem [{\citenamefont {Chekhovich}\ \emph {et~al.}(2013)\citenamefont
  {Chekhovich}, \citenamefont {Makhonin}, \citenamefont {Tartakovskii},
  \citenamefont {Yacoby}, \citenamefont {Bluhm}, \citenamefont {Nowack},\ and\
  \citenamefont {Vandersypen}}]{chekhovich2013}%
  \BibitemOpen
  \bibfield  {author} {\bibinfo {author} {\bibfnamefont {E.~A.}\ \bibnamefont
  {Chekhovich}}, \bibinfo {author} {\bibfnamefont {M.~N.}\ \bibnamefont
  {Makhonin}}, \bibinfo {author} {\bibfnamefont {A.~I.}\ \bibnamefont
  {Tartakovskii}}, \bibinfo {author} {\bibfnamefont {A.}~\bibnamefont
  {Yacoby}}, \bibinfo {author} {\bibfnamefont {H.}~\bibnamefont {Bluhm}},
  \bibinfo {author} {\bibfnamefont {K.~C.}\ \bibnamefont {Nowack}},\ and\
  \bibinfo {author} {\bibfnamefont {L.~M.~K.}\ \bibnamefont {Vandersypen}},\
  }\bibfield  {title} {\bibinfo {title} {Nuclear spin effects in semiconductor
  quantum dots},\ }\href {https://doi.org/10.1038/nmat3652} {\bibfield
  {journal} {\bibinfo  {journal} {Nature Materials}\ }\textbf {\bibinfo
  {volume} {12}},\ \bibinfo {pages} {494} (\bibinfo {year} {2013})}\BibitemShut
  {NoStop}%
\bibitem [{\citenamefont {Khaetskii}\ and\ \citenamefont
  {Nazarov}(2001)}]{khaetskii2001}%
  \BibitemOpen
  \bibfield  {author} {\bibinfo {author} {\bibfnamefont {A.}~\bibnamefont
  {Khaetskii}}\ and\ \bibinfo {author} {\bibfnamefont {Y.}~\bibnamefont
  {Nazarov}},\ }\bibfield  {title} {\bibinfo {title} {Spin-flip transitions
  between {{Zeeman}} sublevels in semiconductor quantum dots},\ }\bibfield
  {journal} {\bibinfo  {journal} {Physical Review B}\ }\textbf {\bibinfo
  {volume} {64}},\ \href {https://doi.org/10.1103/PhysRevB.64.125316}
  {10.1103/PhysRevB.64.125316} (\bibinfo {year} {2001})\BibitemShut {NoStop}%
\bibitem [{\citenamefont {Coish}\ and\ \citenamefont {Loss}(2004)}]{coish2004}%
  \BibitemOpen
  \bibfield  {author} {\bibinfo {author} {\bibfnamefont {W.}~\bibnamefont
  {Coish}}\ and\ \bibinfo {author} {\bibfnamefont {D.}~\bibnamefont {Loss}},\
  }\bibfield  {title} {\bibinfo {title} {Hyperfine interaction in a quantum
  dot: {{Non-Markovian}} electron spin dynamics},\ }\bibfield  {journal}
  {\bibinfo  {journal} {Physical Review B}\ }\textbf {\bibinfo {volume} {70}},\
  \href {https://doi.org/10.1103/PhysRevB.70.195340}
  {10.1103/PhysRevB.70.195340} (\bibinfo {year} {2004})\BibitemShut {NoStop}%
\bibitem [{\citenamefont {Koppens}\ \emph {et~al.}(2008)\citenamefont
  {Koppens}, \citenamefont {Nowack},\ and\ \citenamefont
  {Vandersypen}}]{koppens2008}%
  \BibitemOpen
  \bibfield  {author} {\bibinfo {author} {\bibfnamefont {F.}~\bibnamefont
  {Koppens}}, \bibinfo {author} {\bibfnamefont {K.}~\bibnamefont {Nowack}},\
  and\ \bibinfo {author} {\bibfnamefont {L.}~\bibnamefont {Vandersypen}},\
  }\bibfield  {title} {\bibinfo {title} {Spin {{Echo}} of a {{Single Electron
  Spin}} in a {{Quantum Dot}}},\ }\bibfield  {journal} {\bibinfo  {journal}
  {Physical Review Letters}\ }\textbf {\bibinfo {volume} {100}},\ \href
  {https://doi.org/10.1103/PhysRevLett.100.236802}
  {10.1103/PhysRevLett.100.236802} (\bibinfo {year} {2008})\BibitemShut
  {NoStop}%
\bibitem [{\citenamefont {Medford}\ \emph {et~al.}(2012)\citenamefont
  {Medford}, \citenamefont {Cywi{\'n}ski}, \citenamefont {Barthel},
  \citenamefont {Marcus}, \citenamefont {Hanson},\ and\ \citenamefont
  {Gossard}}]{medford2012}%
  \BibitemOpen
  \bibfield  {author} {\bibinfo {author} {\bibfnamefont {J.}~\bibnamefont
  {Medford}}, \bibinfo {author} {\bibfnamefont {{\l}.}~\bibnamefont
  {Cywi{\'n}ski}}, \bibinfo {author} {\bibfnamefont {C.}~\bibnamefont
  {Barthel}}, \bibinfo {author} {\bibfnamefont {C.~M.}\ \bibnamefont {Marcus}},
  \bibinfo {author} {\bibfnamefont {M.~P.}\ \bibnamefont {Hanson}},\ and\
  \bibinfo {author} {\bibfnamefont {A.~C.}\ \bibnamefont {Gossard}},\
  }\bibfield  {title} {\bibinfo {title} {Scaling of {{Dynamical Decoupling}}
  for {{Spin Qubits}}},\ }\bibfield  {journal} {\bibinfo  {journal} {Physical
  Review Letters}\ }\textbf {\bibinfo {volume} {108}},\ \href
  {https://doi.org/10.1103/PhysRevLett.108.086802}
  {10.1103/PhysRevLett.108.086802} (\bibinfo {year} {2012})\BibitemShut
  {NoStop}%
\bibitem [{\citenamefont {Malinowski}\ \emph
  {et~al.}(2017{\natexlab{b}})\citenamefont {Malinowski}, \citenamefont
  {Martins}, \citenamefont {Cywi{\'n}ski}, \citenamefont {Rudner},
  \citenamefont {Nissen}, \citenamefont {Fallahi}, \citenamefont {Gardner},
  \citenamefont {Manfra}, \citenamefont {Marcus},\ and\ \citenamefont
  {Kuemmeth}}]{malinowski2017b}%
  \BibitemOpen
  \bibfield  {author} {\bibinfo {author} {\bibfnamefont {F.~K.}\ \bibnamefont
  {Malinowski}}, \bibinfo {author} {\bibfnamefont {F.}~\bibnamefont {Martins}},
  \bibinfo {author} {\bibfnamefont {{\L}.}~\bibnamefont {Cywi{\'n}ski}},
  \bibinfo {author} {\bibfnamefont {M.~S.}\ \bibnamefont {Rudner}}, \bibinfo
  {author} {\bibfnamefont {P.~D.}\ \bibnamefont {Nissen}}, \bibinfo {author}
  {\bibfnamefont {S.}~\bibnamefont {Fallahi}}, \bibinfo {author} {\bibfnamefont
  {G.~C.}\ \bibnamefont {Gardner}}, \bibinfo {author} {\bibfnamefont {M.~J.}\
  \bibnamefont {Manfra}}, \bibinfo {author} {\bibfnamefont {C.~M.}\
  \bibnamefont {Marcus}},\ and\ \bibinfo {author} {\bibfnamefont
  {F.}~\bibnamefont {Kuemmeth}},\ }\bibfield  {title} {\bibinfo {title}
  {Spectrum of the {{Nuclear Environment}} for {{GaAs Spin Qubits}}},\ }\href
  {https://doi.org/10.1103/PhysRevLett.118.177702} {\bibfield  {journal}
  {\bibinfo  {journal} {Phys. Rev. Lett.}\ }\textbf {\bibinfo {volume} {118}},\
  \bibinfo {pages} {177702} (\bibinfo {year} {2017}{\natexlab{b}})}\BibitemShut
  {NoStop}%
\bibitem [{\citenamefont {Bluhm}\ \emph {et~al.}(2011)\citenamefont {Bluhm},
  \citenamefont {Foletti}, \citenamefont {Neder}, \citenamefont {Rudner},
  \citenamefont {Mahalu}, \citenamefont {Umansky},\ and\ \citenamefont
  {Yacoby}}]{bluhm2011}%
  \BibitemOpen
  \bibfield  {author} {\bibinfo {author} {\bibfnamefont {H.}~\bibnamefont
  {Bluhm}}, \bibinfo {author} {\bibfnamefont {S.}~\bibnamefont {Foletti}},
  \bibinfo {author} {\bibfnamefont {I.}~\bibnamefont {Neder}}, \bibinfo
  {author} {\bibfnamefont {M.}~\bibnamefont {Rudner}}, \bibinfo {author}
  {\bibfnamefont {D.}~\bibnamefont {Mahalu}}, \bibinfo {author} {\bibfnamefont
  {V.}~\bibnamefont {Umansky}},\ and\ \bibinfo {author} {\bibfnamefont
  {A.}~\bibnamefont {Yacoby}},\ }\bibfield  {title} {\bibinfo {title}
  {Dephasing time of {{GaAs}} electron-spin qubits coupled to a nuclear bath
  exceeding 200 {$M$}s},\ }\href {https://doi.org/10.1038/nphys1856} {\bibfield
   {journal} {\bibinfo  {journal} {Nature Physics}\ }\textbf {\bibinfo {volume}
  {7}},\ \bibinfo {pages} {109} (\bibinfo {year} {2011})}\BibitemShut {NoStop}%
\bibitem [{\citenamefont {Maune}\ \emph {et~al.}(2012)\citenamefont {Maune},
  \citenamefont {Borselli}, \citenamefont {Huang}, \citenamefont {Ladd},
  \citenamefont {Deelman}, \citenamefont {Holabird}, \citenamefont {Kiselev},
  \citenamefont {{Alvarado-Rodriguez}}, \citenamefont {Ross}, \citenamefont
  {Schmitz}, \citenamefont {Sokolich}, \citenamefont {Watson}, \citenamefont
  {Gyure},\ and\ \citenamefont {Hunter}}]{maune2012}%
  \BibitemOpen
  \bibfield  {author} {\bibinfo {author} {\bibfnamefont {B.~M.}\ \bibnamefont
  {Maune}}, \bibinfo {author} {\bibfnamefont {M.~G.}\ \bibnamefont {Borselli}},
  \bibinfo {author} {\bibfnamefont {B.}~\bibnamefont {Huang}}, \bibinfo
  {author} {\bibfnamefont {T.~D.}\ \bibnamefont {Ladd}}, \bibinfo {author}
  {\bibfnamefont {P.~W.}\ \bibnamefont {Deelman}}, \bibinfo {author}
  {\bibfnamefont {K.~S.}\ \bibnamefont {Holabird}}, \bibinfo {author}
  {\bibfnamefont {A.~A.}\ \bibnamefont {Kiselev}}, \bibinfo {author}
  {\bibfnamefont {I.}~\bibnamefont {{Alvarado-Rodriguez}}}, \bibinfo {author}
  {\bibfnamefont {R.~S.}\ \bibnamefont {Ross}}, \bibinfo {author}
  {\bibfnamefont {A.~E.}\ \bibnamefont {Schmitz}}, \bibinfo {author}
  {\bibfnamefont {M.}~\bibnamefont {Sokolich}}, \bibinfo {author}
  {\bibfnamefont {C.~A.}\ \bibnamefont {Watson}}, \bibinfo {author}
  {\bibfnamefont {M.~F.}\ \bibnamefont {Gyure}},\ and\ \bibinfo {author}
  {\bibfnamefont {A.~T.}\ \bibnamefont {Hunter}},\ }\bibfield  {title}
  {\bibinfo {title} {Coherent singlet-triplet oscillations in a silicon-based
  double quantum dot},\ }\href {https://doi.org/10.1038/nature10707} {\bibfield
   {journal} {\bibinfo  {journal} {Nature}\ }\textbf {\bibinfo {volume}
  {481}},\ \bibinfo {pages} {344} (\bibinfo {year} {2012})}\BibitemShut
  {NoStop}%
\bibitem [{\citenamefont {Chan}\ \emph {et~al.}(2018)\citenamefont {Chan},
  \citenamefont {Huang}, \citenamefont {Yang}, \citenamefont {Hwang},
  \citenamefont {Hensen}, \citenamefont {Tanttu}, \citenamefont {Hudson},
  \citenamefont {Itoh}, \citenamefont {Laucht}, \citenamefont {Morello},\ and\
  \citenamefont {Dzurak}}]{chan2018}%
  \BibitemOpen
  \bibfield  {author} {\bibinfo {author} {\bibfnamefont {K.~W.}\ \bibnamefont
  {Chan}}, \bibinfo {author} {\bibfnamefont {W.}~\bibnamefont {Huang}},
  \bibinfo {author} {\bibfnamefont {C.~H.}\ \bibnamefont {Yang}}, \bibinfo
  {author} {\bibfnamefont {J.~C.~C.}\ \bibnamefont {Hwang}}, \bibinfo {author}
  {\bibfnamefont {B.}~\bibnamefont {Hensen}}, \bibinfo {author} {\bibfnamefont
  {T.}~\bibnamefont {Tanttu}}, \bibinfo {author} {\bibfnamefont {F.~E.}\
  \bibnamefont {Hudson}}, \bibinfo {author} {\bibfnamefont {K.~M.}\
  \bibnamefont {Itoh}}, \bibinfo {author} {\bibfnamefont {A.}~\bibnamefont
  {Laucht}}, \bibinfo {author} {\bibfnamefont {A.}~\bibnamefont {Morello}},\
  and\ \bibinfo {author} {\bibfnamefont {A.~S.}\ \bibnamefont {Dzurak}},\
  }\bibfield  {title} {\bibinfo {title} {Assessment of a {{Silicon Quantum Dot
  Spin Qubit Environment}} via {{Noise Spectroscopy}}},\ }\href
  {https://doi.org/10.1103/PhysRevApplied.10.044017} {\bibfield  {journal}
  {\bibinfo  {journal} {Phys. Rev. Applied}\ }\textbf {\bibinfo {volume}
  {10}},\ \bibinfo {pages} {044017} (\bibinfo {year} {2018})}\BibitemShut
  {NoStop}%
\bibitem [{\citenamefont {Paladino}\ \emph {et~al.}(2014)\citenamefont
  {Paladino}, \citenamefont {Galperin}, \citenamefont {Falci},\ and\
  \citenamefont {Altshuler}}]{paladino2014}%
  \BibitemOpen
  \bibfield  {author} {\bibinfo {author} {\bibfnamefont {E.}~\bibnamefont
  {Paladino}}, \bibinfo {author} {\bibfnamefont {Y.~M.}\ \bibnamefont
  {Galperin}}, \bibinfo {author} {\bibfnamefont {G.}~\bibnamefont {Falci}},\
  and\ \bibinfo {author} {\bibfnamefont {B.~L.}\ \bibnamefont {Altshuler}},\
  }\bibfield  {title} {\bibinfo {title} {Noise: {{Implications}} for
  solid-state quantum information},\ }\href
  {https://doi.org/10.1103/RevModPhys.86.361} {\bibfield  {journal} {\bibinfo
  {journal} {Reviews of Modern Physics}\ }\textbf {\bibinfo {volume} {86}},\
  \bibinfo {pages} {361} (\bibinfo {year} {2014})}\BibitemShut {NoStop}%
\bibitem [{\citenamefont {Shi}\ \emph {et~al.}(2012)\citenamefont {Shi},
  \citenamefont {Simmons}, \citenamefont {Prance}, \citenamefont {Gamble},
  \citenamefont {Koh}, \citenamefont {Shim}, \citenamefont {Hu}, \citenamefont
  {Savage}, \citenamefont {Lagally}, \citenamefont {Eriksson}, \citenamefont
  {Friesen},\ and\ \citenamefont {Coppersmith}}]{shi2012}%
  \BibitemOpen
  \bibfield  {author} {\bibinfo {author} {\bibfnamefont {Z.}~\bibnamefont
  {Shi}}, \bibinfo {author} {\bibfnamefont {C.~B.}\ \bibnamefont {Simmons}},
  \bibinfo {author} {\bibfnamefont {J.~R.}\ \bibnamefont {Prance}}, \bibinfo
  {author} {\bibfnamefont {J.~K.}\ \bibnamefont {Gamble}}, \bibinfo {author}
  {\bibfnamefont {T.~S.}\ \bibnamefont {Koh}}, \bibinfo {author} {\bibfnamefont
  {Y.-P.}\ \bibnamefont {Shim}}, \bibinfo {author} {\bibfnamefont
  {X.}~\bibnamefont {Hu}}, \bibinfo {author} {\bibfnamefont {D.~E.}\
  \bibnamefont {Savage}}, \bibinfo {author} {\bibfnamefont {M.~G.}\
  \bibnamefont {Lagally}}, \bibinfo {author} {\bibfnamefont {M.~A.}\
  \bibnamefont {Eriksson}}, \bibinfo {author} {\bibfnamefont {M.}~\bibnamefont
  {Friesen}},\ and\ \bibinfo {author} {\bibfnamefont {S.~N.}\ \bibnamefont
  {Coppersmith}},\ }\bibfield  {title} {\bibinfo {title} {Fast {{Hybrid Silicon
  Double-Quantum-Dot Qubit}}},\ }\bibfield  {journal} {\bibinfo  {journal}
  {Physical Review Letters}\ }\textbf {\bibinfo {volume} {108}},\ \href
  {https://doi.org/10.1103/PhysRevLett.108.140503}
  {10.1103/PhysRevLett.108.140503} (\bibinfo {year} {2012})\BibitemShut
  {NoStop}%
\bibitem [{\citenamefont {Struck}\ \emph {et~al.}(2020)\citenamefont {Struck},
  \citenamefont {Hollmann}, \citenamefont {Schauer}, \citenamefont {Fedorets},
  \citenamefont {Schmidbauer}, \citenamefont {Sawano}, \citenamefont {Riemann},
  \citenamefont {Abrosimov}, \citenamefont {Cywi{\'n}ski}, \citenamefont
  {Bougeard},\ and\ \citenamefont {Schreiber}}]{struck2020}%
  \BibitemOpen
  \bibfield  {author} {\bibinfo {author} {\bibfnamefont {T.}~\bibnamefont
  {Struck}}, \bibinfo {author} {\bibfnamefont {A.}~\bibnamefont {Hollmann}},
  \bibinfo {author} {\bibfnamefont {F.}~\bibnamefont {Schauer}}, \bibinfo
  {author} {\bibfnamefont {O.}~\bibnamefont {Fedorets}}, \bibinfo {author}
  {\bibfnamefont {A.}~\bibnamefont {Schmidbauer}}, \bibinfo {author}
  {\bibfnamefont {K.}~\bibnamefont {Sawano}}, \bibinfo {author} {\bibfnamefont
  {H.}~\bibnamefont {Riemann}}, \bibinfo {author} {\bibfnamefont {N.~V.}\
  \bibnamefont {Abrosimov}}, \bibinfo {author} {\bibfnamefont
  {{\L}.}~\bibnamefont {Cywi{\'n}ski}}, \bibinfo {author} {\bibfnamefont
  {D.}~\bibnamefont {Bougeard}},\ and\ \bibinfo {author} {\bibfnamefont
  {L.~R.}\ \bibnamefont {Schreiber}},\ }\bibfield  {title} {\bibinfo {title}
  {Low-frequency spin qubit energy splitting noise in highly purified 28
  {{Si}}/{{SiGe}}},\ }\href {https://doi.org/10.1038/s41534-020-0276-2}
  {\bibfield  {journal} {\bibinfo  {journal} {npj Quantum Information}\
  }\textbf {\bibinfo {volume} {6}},\ \bibinfo {pages} {1} (\bibinfo {year}
  {2020})}\BibitemShut {NoStop}%
\bibitem [{\citenamefont {Connors}\ \emph {et~al.}(2019)\citenamefont
  {Connors}, \citenamefont {Nelson}, \citenamefont {Qiao}, \citenamefont
  {Edge},\ and\ \citenamefont {Nichol}}]{connors2019a}%
  \BibitemOpen
  \bibfield  {author} {\bibinfo {author} {\bibfnamefont {E.~J.}\ \bibnamefont
  {Connors}}, \bibinfo {author} {\bibfnamefont {J.}~\bibnamefont {Nelson}},
  \bibinfo {author} {\bibfnamefont {H.}~\bibnamefont {Qiao}}, \bibinfo {author}
  {\bibfnamefont {L.~F.}\ \bibnamefont {Edge}},\ and\ \bibinfo {author}
  {\bibfnamefont {J.~M.}\ \bibnamefont {Nichol}},\ }\bibfield  {title}
  {\bibinfo {title} {Low-frequency charge noise in {{Si}}/{{SiGe}} quantum
  dots},\ }\href {https://doi.org/10.1103/PhysRevB.100.165305} {\bibfield
  {journal} {\bibinfo  {journal} {Phys. Rev. B}\ }\textbf {\bibinfo {volume}
  {100}},\ \bibinfo {pages} {165305} (\bibinfo {year} {2019})}\BibitemShut
  {NoStop}%
\bibitem [{\citenamefont {Boter}\ \emph {et~al.}(2020)\citenamefont {Boter},
  \citenamefont {Xue}, \citenamefont {Kr{\"a}henmann}, \citenamefont {Watson},
  \citenamefont {Premakumar}, \citenamefont {Ward}, \citenamefont {Savage},
  \citenamefont {Lagally}, \citenamefont {Friesen}, \citenamefont
  {Coppersmith}, \citenamefont {Eriksson}, \citenamefont {Joynt},\ and\
  \citenamefont {Vandersypen}}]{boter2020}%
  \BibitemOpen
  \bibfield  {author} {\bibinfo {author} {\bibfnamefont {J.~M.}\ \bibnamefont
  {Boter}}, \bibinfo {author} {\bibfnamefont {X.}~\bibnamefont {Xue}}, \bibinfo
  {author} {\bibfnamefont {T.}~\bibnamefont {Kr{\"a}henmann}}, \bibinfo
  {author} {\bibfnamefont {T.~F.}\ \bibnamefont {Watson}}, \bibinfo {author}
  {\bibfnamefont {V.~N.}\ \bibnamefont {Premakumar}}, \bibinfo {author}
  {\bibfnamefont {D.~R.}\ \bibnamefont {Ward}}, \bibinfo {author}
  {\bibfnamefont {D.~E.}\ \bibnamefont {Savage}}, \bibinfo {author}
  {\bibfnamefont {M.~G.}\ \bibnamefont {Lagally}}, \bibinfo {author}
  {\bibfnamefont {M.}~\bibnamefont {Friesen}}, \bibinfo {author} {\bibfnamefont
  {S.~N.}\ \bibnamefont {Coppersmith}}, \bibinfo {author} {\bibfnamefont
  {M.~A.}\ \bibnamefont {Eriksson}}, \bibinfo {author} {\bibfnamefont
  {R.}~\bibnamefont {Joynt}},\ and\ \bibinfo {author} {\bibfnamefont
  {L.~M.~K.}\ \bibnamefont {Vandersypen}},\ }\bibfield  {title} {\bibinfo
  {title} {Spatial noise correlations in a {{Si}}/{{SiGe}} two-qubit device
  from {{Bell}} state coherences},\ }\href
  {https://doi.org/10.1103/PhysRevB.101.235133} {\bibfield  {journal} {\bibinfo
   {journal} {Phys. Rev. B}\ }\textbf {\bibinfo {volume} {101}},\ \bibinfo
  {pages} {235133} (\bibinfo {year} {2020})}\BibitemShut {NoStop}%
\bibitem [{\citenamefont {Wang}\ \emph {et~al.}(2012)\citenamefont {Wang},
  \citenamefont {Bishop}, \citenamefont {Kestner}, \citenamefont {Barnes},
  \citenamefont {Sun},\ and\ \citenamefont {Das~Sarma}}]{wang2012}%
  \BibitemOpen
  \bibfield  {author} {\bibinfo {author} {\bibfnamefont {X.}~\bibnamefont
  {Wang}}, \bibinfo {author} {\bibfnamefont {L.~S.}\ \bibnamefont {Bishop}},
  \bibinfo {author} {\bibfnamefont {J.}~\bibnamefont {Kestner}}, \bibinfo
  {author} {\bibfnamefont {E.}~\bibnamefont {Barnes}}, \bibinfo {author}
  {\bibfnamefont {K.}~\bibnamefont {Sun}},\ and\ \bibinfo {author}
  {\bibfnamefont {S.}~\bibnamefont {Das~Sarma}},\ }\bibfield  {title} {\bibinfo
  {title} {Composite pulses for robust universal control of singlet\textendash
  triplet qubits},\ }\bibfield  {journal} {\bibinfo  {journal} {Nature
  Communications}\ }\textbf {\bibinfo {volume} {3}},\ \href
  {https://doi.org/10.1038/ncomms2003} {10.1038/ncomms2003} (\bibinfo {year}
  {2012})\BibitemShut {NoStop}%
\bibitem [{\citenamefont {G{\"u}ng{\"o}rd{\"u}}\ and\ \citenamefont
  {Kestner}(2018)}]{gungordu2018}%
  \BibitemOpen
  \bibfield  {author} {\bibinfo {author} {\bibfnamefont {U.}~\bibnamefont
  {G{\"u}ng{\"o}rd{\"u}}}\ and\ \bibinfo {author} {\bibfnamefont {J.~P.}\
  \bibnamefont {Kestner}},\ }\bibfield  {title} {\bibinfo {title} {Indications
  of a soft cutoff frequency in the charge noise of a {{Si}}/{{SiGe}} quantum
  dot spin qubit},\ }\href {http://arxiv.org/abs/1811.06082} {\bibfield
  {journal} {\bibinfo  {journal} {arXiv:1811.06082 [cond-mat,
  physics:quant-ph]}\ } (\bibinfo {year} {2018})}\BibitemShut {NoStop}%
\bibitem [{\citenamefont {Barthel}\ \emph {et~al.}(2010)\citenamefont
  {Barthel}, \citenamefont {Kj{\ae}rgaard}, \citenamefont {Medford},
  \citenamefont {Stopa}, \citenamefont {Marcus}, \citenamefont {Hanson},\ and\
  \citenamefont {Gossard}}]{barthel2010}%
  \BibitemOpen
  \bibfield  {author} {\bibinfo {author} {\bibfnamefont {C.}~\bibnamefont
  {Barthel}}, \bibinfo {author} {\bibfnamefont {M.}~\bibnamefont
  {Kj{\ae}rgaard}}, \bibinfo {author} {\bibfnamefont {J.}~\bibnamefont
  {Medford}}, \bibinfo {author} {\bibfnamefont {M.}~\bibnamefont {Stopa}},
  \bibinfo {author} {\bibfnamefont {C.~M.}\ \bibnamefont {Marcus}}, \bibinfo
  {author} {\bibfnamefont {M.~P.}\ \bibnamefont {Hanson}},\ and\ \bibinfo
  {author} {\bibfnamefont {A.~C.}\ \bibnamefont {Gossard}},\ }\bibfield
  {title} {\bibinfo {title} {Fast sensing of double-dot charge arrangement and
  spin state with a radio-frequency sensor quantum dot},\ }\href
  {https://doi.org/10.1103/PhysRevB.81.161308} {\bibfield  {journal} {\bibinfo
  {journal} {Phys. Rev. B}\ }\textbf {\bibinfo {volume} {81}},\ \bibinfo
  {pages} {161308} (\bibinfo {year} {2010})}\BibitemShut {NoStop}%
\bibitem [{\citenamefont {Schoelkopf}\ \emph {et~al.}(1998)\citenamefont
  {Schoelkopf}, \citenamefont {Wahlgren}, \citenamefont {Kozhevnikov},
  \citenamefont {Delsing},\ and\ \citenamefont {Prober}}]{schoelkopf1998}%
  \BibitemOpen
  \bibfield  {author} {\bibinfo {author} {\bibfnamefont {R.~J.}\ \bibnamefont
  {Schoelkopf}}, \bibinfo {author} {\bibfnamefont {P.}~\bibnamefont
  {Wahlgren}}, \bibinfo {author} {\bibfnamefont {A.~A.}\ \bibnamefont
  {Kozhevnikov}}, \bibinfo {author} {\bibfnamefont {P.}~\bibnamefont
  {Delsing}},\ and\ \bibinfo {author} {\bibfnamefont {D.~E.}\ \bibnamefont
  {Prober}},\ }\bibfield  {title} {\bibinfo {title} {The {{Radio-Frequency
  Single-Electron Transistor}} ({{RF-SET}}): {{A Fast}} and {{Ultrasensitive
  Electrometer}}},\ }\href {https://doi.org/10.1126/science.280.5367.1238}
  {\bibfield  {journal} {\bibinfo  {journal} {Science}\ }\textbf {\bibinfo
  {volume} {280}},\ \bibinfo {pages} {1238} (\bibinfo {year}
  {1998})}\BibitemShut {NoStop}%
\bibitem [{\citenamefont {Reilly}\ \emph {et~al.}(2007)\citenamefont {Reilly},
  \citenamefont {Marcus}, \citenamefont {Hanson},\ and\ \citenamefont
  {Gossard}}]{reilly2007}%
  \BibitemOpen
  \bibfield  {author} {\bibinfo {author} {\bibfnamefont {D.~J.}\ \bibnamefont
  {Reilly}}, \bibinfo {author} {\bibfnamefont {C.~M.}\ \bibnamefont {Marcus}},
  \bibinfo {author} {\bibfnamefont {M.~P.}\ \bibnamefont {Hanson}},\ and\
  \bibinfo {author} {\bibfnamefont {A.~C.}\ \bibnamefont {Gossard}},\
  }\bibfield  {title} {\bibinfo {title} {Fast single-charge sensing with a rf
  quantum point contact},\ }\href {https://doi.org/10.1063/1.2794995}
  {\bibfield  {journal} {\bibinfo  {journal} {Applied Physics Letters}\
  }\textbf {\bibinfo {volume} {91}},\ \bibinfo {pages} {162101} (\bibinfo
  {year} {2007})}\BibitemShut {NoStop}%
\bibitem [{\citenamefont {Noiri}\ \emph {et~al.}(2020)\citenamefont {Noiri},
  \citenamefont {Takeda}, \citenamefont {Yoneda}, \citenamefont {Nakajima},
  \citenamefont {Kodera},\ and\ \citenamefont {Tarucha}}]{noiri2020}%
  \BibitemOpen
  \bibfield  {author} {\bibinfo {author} {\bibfnamefont {A.}~\bibnamefont
  {Noiri}}, \bibinfo {author} {\bibfnamefont {K.}~\bibnamefont {Takeda}},
  \bibinfo {author} {\bibfnamefont {J.}~\bibnamefont {Yoneda}}, \bibinfo
  {author} {\bibfnamefont {T.}~\bibnamefont {Nakajima}}, \bibinfo {author}
  {\bibfnamefont {T.}~\bibnamefont {Kodera}},\ and\ \bibinfo {author}
  {\bibfnamefont {S.}~\bibnamefont {Tarucha}},\ }\bibfield  {title} {\bibinfo
  {title} {Radio-{{Frequency-Detected Fast Charge Sensing}} in {{Undoped
  Silicon Quantum Dots}}},\ }\href
  {https://doi.org/10.1021/acs.nanolett.9b03847} {\bibfield  {journal}
  {\bibinfo  {journal} {Nano Lett.}\ }\textbf {\bibinfo {volume} {20}},\
  \bibinfo {pages} {947} (\bibinfo {year} {2020})}\BibitemShut {NoStop}%
\bibitem [{\citenamefont {Connors}\ \emph {et~al.}(2020)\citenamefont
  {Connors}, \citenamefont {Nelson},\ and\ \citenamefont
  {Nichol}}]{connors2020}%
  \BibitemOpen
  \bibfield  {author} {\bibinfo {author} {\bibfnamefont {E.~J.}\ \bibnamefont
  {Connors}}, \bibinfo {author} {\bibfnamefont {J.}~\bibnamefont {Nelson}},\
  and\ \bibinfo {author} {\bibfnamefont {J.~M.}\ \bibnamefont {Nichol}},\
  }\bibfield  {title} {\bibinfo {title} {Rapid {{High-Fidelity Spin-State
  Readout}} in {$\mathrm{Si}$}/{$\mathrm{Si}$}-{$\mathrm{Ge}$} {{Quantum Dots}}
  via rf {{Reflectometry}}},\ }\href
  {https://doi.org/10.1103/PhysRevApplied.13.024019} {\bibfield  {journal}
  {\bibinfo  {journal} {Phys. Rev. Applied}\ }\textbf {\bibinfo {volume}
  {13}},\ \bibinfo {pages} {024019} (\bibinfo {year} {2020})}\BibitemShut
  {NoStop}%
\bibitem [{\citenamefont {Liu}\ \emph {et~al.}(2021)\citenamefont {Liu},
  \citenamefont {Xu}, \citenamefont {Cao}, \citenamefont {Zhu}, \citenamefont
  {Wang},\ and\ \citenamefont {Yang}}]{liu2021}%
  \BibitemOpen
  \bibfield  {author} {\bibinfo {author} {\bibfnamefont {Z.}~\bibnamefont
  {Liu}}, \bibinfo {author} {\bibfnamefont {C.}~\bibnamefont {Xu}}, \bibinfo
  {author} {\bibfnamefont {C.}~\bibnamefont {Cao}}, \bibinfo {author}
  {\bibfnamefont {W.}~\bibnamefont {Zhu}}, \bibinfo {author} {\bibfnamefont
  {Z.~F.}\ \bibnamefont {Wang}},\ and\ \bibinfo {author} {\bibfnamefont
  {J.}~\bibnamefont {Yang}},\ }\bibfield  {title} {\bibinfo {title} {Doping
  dependence of electronic structure of infinite-layer
  {${\mathrm{NdNiO}}_{2}$}},\ }\href
  {https://doi.org/10.1103/PhysRevB.103.045103} {\bibfield  {journal} {\bibinfo
   {journal} {Phys. Rev. B}\ }\textbf {\bibinfo {volume} {103}},\ \bibinfo
  {pages} {045103} (\bibinfo {year} {2021})}\BibitemShut {NoStop}%
\bibitem [{\citenamefont {Keith}\ \emph {et~al.}(2019)\citenamefont {Keith},
  \citenamefont {House}, \citenamefont {Donnelly}, \citenamefont {Watson},
  \citenamefont {Weber},\ and\ \citenamefont {Simmons}}]{keith2019}%
  \BibitemOpen
  \bibfield  {author} {\bibinfo {author} {\bibfnamefont {D.}~\bibnamefont
  {Keith}}, \bibinfo {author} {\bibfnamefont {M.~G.}\ \bibnamefont {House}},
  \bibinfo {author} {\bibfnamefont {M.~B.}\ \bibnamefont {Donnelly}}, \bibinfo
  {author} {\bibfnamefont {T.~F.}\ \bibnamefont {Watson}}, \bibinfo {author}
  {\bibfnamefont {B.}~\bibnamefont {Weber}},\ and\ \bibinfo {author}
  {\bibfnamefont {M.~Y.}\ \bibnamefont {Simmons}},\ }\bibfield  {title}
  {\bibinfo {title} {Single-{{Shot Spin Readout}} in {{Semiconductors Near}}
  the {{Shot-Noise Sensitivity Limit}}},\ }\href
  {https://doi.org/10.1103/PhysRevX.9.041003} {\bibfield  {journal} {\bibinfo
  {journal} {Phys. Rev. X}\ }\textbf {\bibinfo {volume} {9}},\ \bibinfo {pages}
  {041003} (\bibinfo {year} {2019})}\BibitemShut {NoStop}%
\bibitem [{\citenamefont {Petersson}\ \emph
  {et~al.}(2010{\natexlab{b}})\citenamefont {Petersson}, \citenamefont {Smith},
  \citenamefont {Anderson}, \citenamefont {Atkinson}, \citenamefont {Jones},\
  and\ \citenamefont {Ritchie}}]{petersson2010}%
  \BibitemOpen
  \bibfield  {author} {\bibinfo {author} {\bibfnamefont {K.~D.}\ \bibnamefont
  {Petersson}}, \bibinfo {author} {\bibfnamefont {C.~G.}\ \bibnamefont
  {Smith}}, \bibinfo {author} {\bibfnamefont {D.}~\bibnamefont {Anderson}},
  \bibinfo {author} {\bibfnamefont {P.}~\bibnamefont {Atkinson}}, \bibinfo
  {author} {\bibfnamefont {G.~A.~C.}\ \bibnamefont {Jones}},\ and\ \bibinfo
  {author} {\bibfnamefont {D.~A.}\ \bibnamefont {Ritchie}},\ }\bibfield
  {title} {\bibinfo {title} {Charge and {{Spin State Readout}} of a {{Double
  Quantum Dot Coupled}} to a {{Resonator}}},\ }\href
  {https://doi.org/10.1021/nl100663w} {\bibfield  {journal} {\bibinfo
  {journal} {Nano Letters}\ }\textbf {\bibinfo {volume} {10}},\ \bibinfo
  {pages} {2789} (\bibinfo {year} {2010}{\natexlab{b}})}\BibitemShut {NoStop}%
\bibitem [{\citenamefont {Colless}\ \emph {et~al.}(2013)\citenamefont
  {Colless}, \citenamefont {Mahoney}, \citenamefont {Hornibrook}, \citenamefont
  {Doherty}, \citenamefont {Lu}, \citenamefont {Gossard},\ and\ \citenamefont
  {Reilly}}]{colless2013}%
  \BibitemOpen
  \bibfield  {author} {\bibinfo {author} {\bibfnamefont {J.}~\bibnamefont
  {Colless}}, \bibinfo {author} {\bibfnamefont {A.}~\bibnamefont {Mahoney}},
  \bibinfo {author} {\bibfnamefont {J.}~\bibnamefont {Hornibrook}}, \bibinfo
  {author} {\bibfnamefont {A.}~\bibnamefont {Doherty}}, \bibinfo {author}
  {\bibfnamefont {H.}~\bibnamefont {Lu}}, \bibinfo {author} {\bibfnamefont
  {A.}~\bibnamefont {Gossard}},\ and\ \bibinfo {author} {\bibfnamefont
  {D.}~\bibnamefont {Reilly}},\ }\bibfield  {title} {\bibinfo {title}
  {Dispersive {{Readout}} of a {{Few-Electron Double Quantum Dot}} with
  {{Fast}} rf {{Gate Sensors}}},\ }\bibfield  {journal} {\bibinfo  {journal}
  {Physical Review Letters}\ }\textbf {\bibinfo {volume} {110}},\ \href
  {https://doi.org/10.1103/PhysRevLett.110.046805}
  {10.1103/PhysRevLett.110.046805} (\bibinfo {year} {2013})\BibitemShut
  {NoStop}%
\bibitem [{\citenamefont {Ahmed}\ \emph {et~al.}(2018)\citenamefont {Ahmed},
  \citenamefont {Haigh}, \citenamefont {Schaal}, \citenamefont {Barraud},
  \citenamefont {Zhu}, \citenamefont {Lee}, \citenamefont {Amado},
  \citenamefont {Robinson}, \citenamefont {Rossi}, \citenamefont {Morton},\
  and\ \citenamefont {{Gonzalez-Zalba}}}]{ahmed2018}%
  \BibitemOpen
  \bibfield  {author} {\bibinfo {author} {\bibfnamefont {I.}~\bibnamefont
  {Ahmed}}, \bibinfo {author} {\bibfnamefont {J.~A.}\ \bibnamefont {Haigh}},
  \bibinfo {author} {\bibfnamefont {S.}~\bibnamefont {Schaal}}, \bibinfo
  {author} {\bibfnamefont {S.}~\bibnamefont {Barraud}}, \bibinfo {author}
  {\bibfnamefont {Y.}~\bibnamefont {Zhu}}, \bibinfo {author} {\bibfnamefont
  {C.-m.}\ \bibnamefont {Lee}}, \bibinfo {author} {\bibfnamefont
  {M.}~\bibnamefont {Amado}}, \bibinfo {author} {\bibfnamefont {J.~W.~A.}\
  \bibnamefont {Robinson}}, \bibinfo {author} {\bibfnamefont {A.}~\bibnamefont
  {Rossi}}, \bibinfo {author} {\bibfnamefont {J.~J.~L.}\ \bibnamefont
  {Morton}},\ and\ \bibinfo {author} {\bibfnamefont {M.~F.}\ \bibnamefont
  {{Gonzalez-Zalba}}},\ }\bibfield  {title} {\bibinfo {title}
  {Radio-{{Frequency Capacitive Gate-Based Sensing}}},\ }\href
  {https://doi.org/10.1103/PhysRevApplied.10.014018} {\bibfield  {journal}
  {\bibinfo  {journal} {Phys. Rev. Applied}\ }\textbf {\bibinfo {volume}
  {10}},\ \bibinfo {pages} {014018} (\bibinfo {year} {2018})}\BibitemShut
  {NoStop}%
\bibitem [{\citenamefont {West}\ \emph {et~al.}(2019)\citenamefont {West},
  \citenamefont {Hensen}, \citenamefont {Jouan}, \citenamefont {Tanttu},
  \citenamefont {Yang}, \citenamefont {Rossi}, \citenamefont
  {{Gonzalez-Zalba}}, \citenamefont {Hudson}, \citenamefont {Morello},
  \citenamefont {Reilly},\ and\ \citenamefont {Dzurak}}]{west2019}%
  \BibitemOpen
  \bibfield  {author} {\bibinfo {author} {\bibfnamefont {A.}~\bibnamefont
  {West}}, \bibinfo {author} {\bibfnamefont {B.}~\bibnamefont {Hensen}},
  \bibinfo {author} {\bibfnamefont {A.}~\bibnamefont {Jouan}}, \bibinfo
  {author} {\bibfnamefont {T.}~\bibnamefont {Tanttu}}, \bibinfo {author}
  {\bibfnamefont {C.-H.}\ \bibnamefont {Yang}}, \bibinfo {author}
  {\bibfnamefont {A.}~\bibnamefont {Rossi}}, \bibinfo {author} {\bibfnamefont
  {M.~F.}\ \bibnamefont {{Gonzalez-Zalba}}}, \bibinfo {author} {\bibfnamefont
  {F.}~\bibnamefont {Hudson}}, \bibinfo {author} {\bibfnamefont
  {A.}~\bibnamefont {Morello}}, \bibinfo {author} {\bibfnamefont {D.~J.}\
  \bibnamefont {Reilly}},\ and\ \bibinfo {author} {\bibfnamefont {A.~S.}\
  \bibnamefont {Dzurak}},\ }\bibfield  {title} {\bibinfo {title} {Gate-based
  single-shot readout of spins in silicon},\ }\href
  {https://doi.org/10.1038/s41565-019-0400-7} {\bibfield  {journal} {\bibinfo
  {journal} {Nature Nanotechnology}\ ,\ \bibinfo {pages} {1}} (\bibinfo {year}
  {2019})}\BibitemShut {NoStop}%
\bibitem [{\citenamefont {Schaal}\ \emph {et~al.}(2018)\citenamefont {Schaal},
  \citenamefont {Barraud}, \citenamefont {Morton},\ and\ \citenamefont
  {{Gonzalez-Zalba}}}]{schaal2018}%
  \BibitemOpen
  \bibfield  {author} {\bibinfo {author} {\bibfnamefont {S.}~\bibnamefont
  {Schaal}}, \bibinfo {author} {\bibfnamefont {S.}~\bibnamefont {Barraud}},
  \bibinfo {author} {\bibfnamefont {J.~J.~L.}\ \bibnamefont {Morton}},\ and\
  \bibinfo {author} {\bibfnamefont {M.~F.}\ \bibnamefont {{Gonzalez-Zalba}}},\
  }\bibfield  {title} {\bibinfo {title} {Conditional {{Dispersive Readout}} of
  a {{CMOS Single-Electron Memory Cell}}},\ }\href
  {https://doi.org/10.1103/PhysRevApplied.9.054016} {\bibfield  {journal}
  {\bibinfo  {journal} {Phys. Rev. Applied}\ }\textbf {\bibinfo {volume} {9}},\
  \bibinfo {pages} {054016} (\bibinfo {year} {2018})}\BibitemShut {NoStop}%
\bibitem [{\citenamefont {Pakkiam}\ \emph {et~al.}(2018)\citenamefont
  {Pakkiam}, \citenamefont {Timofeev}, \citenamefont {House}, \citenamefont
  {Hogg}, \citenamefont {Kobayashi}, \citenamefont {Koch}, \citenamefont
  {Rogge},\ and\ \citenamefont {Simmons}}]{pakkiam2018}%
  \BibitemOpen
  \bibfield  {author} {\bibinfo {author} {\bibfnamefont {P.}~\bibnamefont
  {Pakkiam}}, \bibinfo {author} {\bibfnamefont {A.~V.}\ \bibnamefont
  {Timofeev}}, \bibinfo {author} {\bibfnamefont {M.~G.}\ \bibnamefont {House}},
  \bibinfo {author} {\bibfnamefont {M.~R.}\ \bibnamefont {Hogg}}, \bibinfo
  {author} {\bibfnamefont {T.}~\bibnamefont {Kobayashi}}, \bibinfo {author}
  {\bibfnamefont {M.}~\bibnamefont {Koch}}, \bibinfo {author} {\bibfnamefont
  {S.}~\bibnamefont {Rogge}},\ and\ \bibinfo {author} {\bibfnamefont {M.~Y.}\
  \bibnamefont {Simmons}},\ }\bibfield  {title} {\bibinfo {title}
  {Single-{{Shot Single-Gate}} rf {{Spin Readout}} in {{Silicon}}},\ }\href
  {https://doi.org/10.1103/PhysRevX.8.041032} {\bibfield  {journal} {\bibinfo
  {journal} {Phys. Rev. X}\ }\textbf {\bibinfo {volume} {8}},\ \bibinfo {pages}
  {041032} (\bibinfo {year} {2018})}\BibitemShut {NoStop}%
\bibitem [{\citenamefont {Crippa}\ \emph {et~al.}(2019)\citenamefont {Crippa},
  \citenamefont {Ezzouch}, \citenamefont {Apr{\'a}}, \citenamefont {Amisse},
  \citenamefont {Lavi{\'e}ville}, \citenamefont {Hutin}, \citenamefont
  {Bertrand}, \citenamefont {Vinet}, \citenamefont {Urdampilleta},
  \citenamefont {Meunier}, \citenamefont {Sanquer}, \citenamefont {Jehl},
  \citenamefont {Maurand},\ and\ \citenamefont {De~Franceschi}}]{crippa2019}%
  \BibitemOpen
  \bibfield  {author} {\bibinfo {author} {\bibfnamefont {A.}~\bibnamefont
  {Crippa}}, \bibinfo {author} {\bibfnamefont {R.}~\bibnamefont {Ezzouch}},
  \bibinfo {author} {\bibfnamefont {A.}~\bibnamefont {Apr{\'a}}}, \bibinfo
  {author} {\bibfnamefont {A.}~\bibnamefont {Amisse}}, \bibinfo {author}
  {\bibfnamefont {R.}~\bibnamefont {Lavi{\'e}ville}}, \bibinfo {author}
  {\bibfnamefont {L.}~\bibnamefont {Hutin}}, \bibinfo {author} {\bibfnamefont
  {B.}~\bibnamefont {Bertrand}}, \bibinfo {author} {\bibfnamefont
  {M.}~\bibnamefont {Vinet}}, \bibinfo {author} {\bibfnamefont
  {M.}~\bibnamefont {Urdampilleta}}, \bibinfo {author} {\bibfnamefont
  {T.}~\bibnamefont {Meunier}}, \bibinfo {author} {\bibfnamefont
  {M.}~\bibnamefont {Sanquer}}, \bibinfo {author} {\bibfnamefont
  {X.}~\bibnamefont {Jehl}}, \bibinfo {author} {\bibfnamefont {R.}~\bibnamefont
  {Maurand}},\ and\ \bibinfo {author} {\bibfnamefont {S.}~\bibnamefont
  {De~Franceschi}},\ }\bibfield  {title} {\bibinfo {title} {Gate-reflectometry
  dispersive readout and coherent control of a spin qubit in silicon},\ }\href
  {https://doi.org/10.1038/s41467-019-10848-z} {\bibfield  {journal} {\bibinfo
  {journal} {Nature Communications}\ }\textbf {\bibinfo {volume} {10}},\
  \bibinfo {pages} {2776} (\bibinfo {year} {2019})}\BibitemShut {NoStop}%
\bibitem [{\citenamefont {Volk}\ \emph {et~al.}(2019)\citenamefont {Volk},
  \citenamefont {Chatterjee}, \citenamefont {Ansaloni}, \citenamefont
  {Marcus},\ and\ \citenamefont {Kuemmeth}}]{volk2019}%
  \BibitemOpen
  \bibfield  {author} {\bibinfo {author} {\bibfnamefont {C.}~\bibnamefont
  {Volk}}, \bibinfo {author} {\bibfnamefont {A.}~\bibnamefont {Chatterjee}},
  \bibinfo {author} {\bibfnamefont {F.}~\bibnamefont {Ansaloni}}, \bibinfo
  {author} {\bibfnamefont {C.~M.}\ \bibnamefont {Marcus}},\ and\ \bibinfo
  {author} {\bibfnamefont {F.}~\bibnamefont {Kuemmeth}},\ }\bibfield  {title}
  {\bibinfo {title} {Fast {{Charge Sensing}} of {{Si}} / {{SiGe Quantum Dots}}
  via a {{High-Frequency Accumulation Gate}}},\ }\href
  {https://doi.org/10.1021/acs.nanolett.9b02149} {\bibfield  {journal}
  {\bibinfo  {journal} {Nano Lett.}\ }\textbf {\bibinfo {volume} {19}},\
  \bibinfo {pages} {5628} (\bibinfo {year} {2019})}\BibitemShut {NoStop}%
\bibitem [{\citenamefont {Zheng}\ \emph {et~al.}(2019)\citenamefont {Zheng},
  \citenamefont {Samkharadze}, \citenamefont {Noordam}, \citenamefont {Kalhor},
  \citenamefont {Brousse}, \citenamefont {Sammak}, \citenamefont {Scappucci},\
  and\ \citenamefont {Vandersypen}}]{zheng2019a}%
  \BibitemOpen
  \bibfield  {author} {\bibinfo {author} {\bibfnamefont {G.}~\bibnamefont
  {Zheng}}, \bibinfo {author} {\bibfnamefont {N.}~\bibnamefont {Samkharadze}},
  \bibinfo {author} {\bibfnamefont {M.~L.}\ \bibnamefont {Noordam}}, \bibinfo
  {author} {\bibfnamefont {N.}~\bibnamefont {Kalhor}}, \bibinfo {author}
  {\bibfnamefont {D.}~\bibnamefont {Brousse}}, \bibinfo {author} {\bibfnamefont
  {A.}~\bibnamefont {Sammak}}, \bibinfo {author} {\bibfnamefont
  {G.}~\bibnamefont {Scappucci}},\ and\ \bibinfo {author} {\bibfnamefont
  {L.~M.~K.}\ \bibnamefont {Vandersypen}},\ }\bibfield  {title} {\bibinfo
  {title} {Rapid gate-based spin read-out in silicon using an on-chip
  resonator},\ }\href {https://doi.org/10.1038/s41565-019-0488-9} {\bibfield
  {journal} {\bibinfo  {journal} {Nat. Nanotechnol.}\ }\textbf {\bibinfo
  {volume} {14}},\ \bibinfo {pages} {742} (\bibinfo {year} {2019})}\BibitemShut
  {NoStop}%
\bibitem [{\citenamefont {Wallraff}\ \emph {et~al.}(2005)\citenamefont
  {Wallraff}, \citenamefont {Schuster}, \citenamefont {Blais}, \citenamefont
  {Frunzio}, \citenamefont {Majer}, \citenamefont {Devoret}, \citenamefont
  {Girvin},\ and\ \citenamefont {Schoelkopf}}]{wallraff2005}%
  \BibitemOpen
  \bibfield  {author} {\bibinfo {author} {\bibfnamefont {A.}~\bibnamefont
  {Wallraff}}, \bibinfo {author} {\bibfnamefont {D.}~\bibnamefont {Schuster}},
  \bibinfo {author} {\bibfnamefont {A.}~\bibnamefont {Blais}}, \bibinfo
  {author} {\bibfnamefont {L.}~\bibnamefont {Frunzio}}, \bibinfo {author}
  {\bibfnamefont {J.}~\bibnamefont {Majer}}, \bibinfo {author} {\bibfnamefont
  {M.}~\bibnamefont {Devoret}}, \bibinfo {author} {\bibfnamefont
  {S.}~\bibnamefont {Girvin}},\ and\ \bibinfo {author} {\bibfnamefont
  {R.}~\bibnamefont {Schoelkopf}},\ }\bibfield  {title} {\bibinfo {title}
  {Approaching {{Unit Visibility}} for {{Control}} of a {{Superconducting
  Qubit}} with {{Dispersive Readout}}},\ }\bibfield  {journal} {\bibinfo
  {journal} {Physical Review Letters}\ }\textbf {\bibinfo {volume} {95}},\
  \href {https://doi.org/10.1103/PhysRevLett.95.060501}
  {10.1103/PhysRevLett.95.060501} (\bibinfo {year} {2005})\BibitemShut
  {NoStop}%
\bibitem [{\citenamefont {Hughes}\ \emph {et~al.}(2004)\citenamefont {Hughes},
  \citenamefont {Awschalom}, \citenamefont {Caves}, \citenamefont {Chapman},
  \citenamefont {Clark}, \citenamefont {Cory}, \citenamefont {DiVincenzo},
  \citenamefont {Ekert}, \citenamefont {Hammel}, \citenamefont {Kwiat},
  \citenamefont {Lloyd}, \citenamefont {Milburn}, \citenamefont {Orlando},
  \citenamefont {Steel}, \citenamefont {Vazirani}, \citenamefont {Whaley},\
  and\ \citenamefont {Wineland}}]{hughes2004}%
  \BibitemOpen
  \bibfield  {author} {\bibinfo {author} {\bibfnamefont {D.~R.}\ \bibnamefont
  {Hughes}}, \bibinfo {author} {\bibfnamefont {D.}~\bibnamefont {Awschalom}},
  \bibinfo {author} {\bibfnamefont {C.}~\bibnamefont {Caves}}, \bibinfo
  {author} {\bibfnamefont {M.}~\bibnamefont {Chapman}}, \bibinfo {author}
  {\bibfnamefont {R.}~\bibnamefont {Clark}}, \bibinfo {author} {\bibfnamefont
  {D.}~\bibnamefont {Cory}}, \bibinfo {author} {\bibfnamefont {D.~D.}\
  \bibnamefont {DiVincenzo}}, \bibinfo {author} {\bibfnamefont
  {A.}~\bibnamefont {Ekert}}, \bibinfo {author} {\bibfnamefont {P.~C.}\
  \bibnamefont {Hammel}}, \bibinfo {author} {\bibfnamefont {P.}~\bibnamefont
  {Kwiat}}, \bibinfo {author} {\bibfnamefont {S.}~\bibnamefont {Lloyd}},
  \bibinfo {author} {\bibfnamefont {G.}~\bibnamefont {Milburn}}, \bibinfo
  {author} {\bibfnamefont {T.}~\bibnamefont {Orlando}}, \bibinfo {author}
  {\bibfnamefont {D.}~\bibnamefont {Steel}}, \bibinfo {author} {\bibfnamefont
  {U.}~\bibnamefont {Vazirani}}, \bibinfo {author} {\bibfnamefont {K.~B.}\
  \bibnamefont {Whaley}},\ and\ \bibinfo {author} {\bibfnamefont {D.~D.}\
  \bibnamefont {Wineland}},\ }\bibfield  {title} {\bibinfo {title} {A {{Quantum
  Information Science}} and {{Technology Roadmap}}},\ }\href@noop {} {\ ,\
  \bibinfo {pages} {268} (\bibinfo {year} {2004})}\BibitemShut {NoStop}%
\bibitem [{\citenamefont {Devoret}\ and\ \citenamefont
  {Schoelkopf}(2013)}]{devoret2013}%
  \BibitemOpen
  \bibfield  {author} {\bibinfo {author} {\bibfnamefont {M.~H.}\ \bibnamefont
  {Devoret}}\ and\ \bibinfo {author} {\bibfnamefont {R.~J.}\ \bibnamefont
  {Schoelkopf}},\ }\bibfield  {title} {\bibinfo {title} {Superconducting
  {{Circuits}} for {{Quantum Information}}: {{An Outlook}}},\ }\href
  {https://doi.org/10.1126/science.1231930} {\bibfield  {journal} {\bibinfo
  {journal} {Science}\ }\textbf {\bibinfo {volume} {339}},\ \bibinfo {pages}
  {1169} (\bibinfo {year} {2013})}\BibitemShut {NoStop}%
\bibitem [{\citenamefont {Mi}\ \emph {et~al.}(2017{\natexlab{a}})\citenamefont
  {Mi}, \citenamefont {Cady}, \citenamefont {Zajac}, \citenamefont {Deelman},\
  and\ \citenamefont {Petta}}]{mi2017}%
  \BibitemOpen
  \bibfield  {author} {\bibinfo {author} {\bibfnamefont {X.}~\bibnamefont
  {Mi}}, \bibinfo {author} {\bibfnamefont {J.~V.}\ \bibnamefont {Cady}},
  \bibinfo {author} {\bibfnamefont {D.~M.}\ \bibnamefont {Zajac}}, \bibinfo
  {author} {\bibfnamefont {P.~W.}\ \bibnamefont {Deelman}},\ and\ \bibinfo
  {author} {\bibfnamefont {J.~R.}\ \bibnamefont {Petta}},\ }\bibfield  {title}
  {\bibinfo {title} {Strong coupling of a single electron in silicon to a
  microwave photon},\ }\href {https://doi.org/10.1126/science.aal2469}
  {\bibfield  {journal} {\bibinfo  {journal} {Science}\ }\textbf {\bibinfo
  {volume} {355}},\ \bibinfo {pages} {156} (\bibinfo {year}
  {2017}{\natexlab{a}})}\BibitemShut {NoStop}%
\bibitem [{\citenamefont {Stockklauser}\ \emph {et~al.}(2017)\citenamefont
  {Stockklauser}, \citenamefont {Scarlino}, \citenamefont {Koski},
  \citenamefont {Gasparinetti}, \citenamefont {Andersen}, \citenamefont
  {Reichl}, \citenamefont {Wegscheider}, \citenamefont {Ihn}, \citenamefont
  {Ensslin},\ and\ \citenamefont {Wallraff}}]{stockklauser2017}%
  \BibitemOpen
  \bibfield  {author} {\bibinfo {author} {\bibfnamefont {A.}~\bibnamefont
  {Stockklauser}}, \bibinfo {author} {\bibfnamefont {P.}~\bibnamefont
  {Scarlino}}, \bibinfo {author} {\bibfnamefont {J.}~\bibnamefont {Koski}},
  \bibinfo {author} {\bibfnamefont {S.}~\bibnamefont {Gasparinetti}}, \bibinfo
  {author} {\bibfnamefont {C.~K.}\ \bibnamefont {Andersen}}, \bibinfo {author}
  {\bibfnamefont {C.}~\bibnamefont {Reichl}}, \bibinfo {author} {\bibfnamefont
  {W.}~\bibnamefont {Wegscheider}}, \bibinfo {author} {\bibfnamefont
  {T.}~\bibnamefont {Ihn}}, \bibinfo {author} {\bibfnamefont {K.}~\bibnamefont
  {Ensslin}},\ and\ \bibinfo {author} {\bibfnamefont {A.}~\bibnamefont
  {Wallraff}},\ }\bibfield  {title} {\bibinfo {title} {Strong {{Coupling Cavity
  QED}} with {{Gate-Defined Double Quantum Dots Enabled}} by a {{High Impedance
  Resonator}}},\ }\bibfield  {journal} {\bibinfo  {journal} {Physical Review
  X}\ }\textbf {\bibinfo {volume} {7}},\ \href
  {https://doi.org/10.1103/PhysRevX.7.011030} {10.1103/PhysRevX.7.011030}
  (\bibinfo {year} {2017})\BibitemShut {NoStop}%
\bibitem [{\citenamefont {Mi}\ \emph {et~al.}(2018)\citenamefont {Mi},
  \citenamefont {Benito}, \citenamefont {Putz}, \citenamefont {Zajac},
  \citenamefont {Taylor}, \citenamefont {Burkard},\ and\ \citenamefont
  {Petta}}]{mi2018}%
  \BibitemOpen
  \bibfield  {author} {\bibinfo {author} {\bibfnamefont {X.}~\bibnamefont
  {Mi}}, \bibinfo {author} {\bibfnamefont {M.}~\bibnamefont {Benito}}, \bibinfo
  {author} {\bibfnamefont {S.}~\bibnamefont {Putz}}, \bibinfo {author}
  {\bibfnamefont {D.~M.}\ \bibnamefont {Zajac}}, \bibinfo {author}
  {\bibfnamefont {J.~M.}\ \bibnamefont {Taylor}}, \bibinfo {author}
  {\bibfnamefont {G.}~\bibnamefont {Burkard}},\ and\ \bibinfo {author}
  {\bibfnamefont {J.~R.}\ \bibnamefont {Petta}},\ }\bibfield  {title} {\bibinfo
  {title} {A coherent spin\textendash photon interface in silicon},\ }\href
  {https://doi.org/10.1038/nature25769} {\bibfield  {journal} {\bibinfo
  {journal} {Nature}\ }\textbf {\bibinfo {volume} {555}},\ \bibinfo {pages}
  {599} (\bibinfo {year} {2018})}\BibitemShut {NoStop}%
\bibitem [{\citenamefont {Samkharadze}\ \emph {et~al.}(2018)\citenamefont
  {Samkharadze}, \citenamefont {Zheng}, \citenamefont {Kalhor}, \citenamefont
  {Brousse}, \citenamefont {Sammak}, \citenamefont {Mendes}, \citenamefont
  {Blais}, \citenamefont {Scappucci},\ and\ \citenamefont
  {Vandersypen}}]{samkharadze2018}%
  \BibitemOpen
  \bibfield  {author} {\bibinfo {author} {\bibfnamefont {N.}~\bibnamefont
  {Samkharadze}}, \bibinfo {author} {\bibfnamefont {G.}~\bibnamefont {Zheng}},
  \bibinfo {author} {\bibfnamefont {N.}~\bibnamefont {Kalhor}}, \bibinfo
  {author} {\bibfnamefont {D.}~\bibnamefont {Brousse}}, \bibinfo {author}
  {\bibfnamefont {A.}~\bibnamefont {Sammak}}, \bibinfo {author} {\bibfnamefont
  {U.~C.}\ \bibnamefont {Mendes}}, \bibinfo {author} {\bibfnamefont
  {A.}~\bibnamefont {Blais}}, \bibinfo {author} {\bibfnamefont
  {G.}~\bibnamefont {Scappucci}},\ and\ \bibinfo {author} {\bibfnamefont
  {L.~M.~K.}\ \bibnamefont {Vandersypen}},\ }\bibfield  {title} {\bibinfo
  {title} {Strong spin-photon coupling in silicon},\ }\href
  {https://doi.org/10.1126/science.aar4054} {\bibfield  {journal} {\bibinfo
  {journal} {Science}\ }\textbf {\bibinfo {volume} {359}},\ \bibinfo {pages}
  {1123} (\bibinfo {year} {2018})}\BibitemShut {NoStop}%
\bibitem [{\citenamefont {Landig}\ \emph {et~al.}(2018)\citenamefont {Landig},
  \citenamefont {Koski}, \citenamefont {Scarlino}, \citenamefont {Mendes},
  \citenamefont {Blais}, \citenamefont {Reichl}, \citenamefont {Wegscheider},
  \citenamefont {Wallraff}, \citenamefont {Ensslin},\ and\ \citenamefont
  {Ihn}}]{landig2018a}%
  \BibitemOpen
  \bibfield  {author} {\bibinfo {author} {\bibfnamefont {A.~J.}\ \bibnamefont
  {Landig}}, \bibinfo {author} {\bibfnamefont {J.~V.}\ \bibnamefont {Koski}},
  \bibinfo {author} {\bibfnamefont {P.}~\bibnamefont {Scarlino}}, \bibinfo
  {author} {\bibfnamefont {U.~C.}\ \bibnamefont {Mendes}}, \bibinfo {author}
  {\bibfnamefont {A.}~\bibnamefont {Blais}}, \bibinfo {author} {\bibfnamefont
  {C.}~\bibnamefont {Reichl}}, \bibinfo {author} {\bibfnamefont
  {W.}~\bibnamefont {Wegscheider}}, \bibinfo {author} {\bibfnamefont
  {A.}~\bibnamefont {Wallraff}}, \bibinfo {author} {\bibfnamefont
  {K.}~\bibnamefont {Ensslin}},\ and\ \bibinfo {author} {\bibfnamefont
  {T.}~\bibnamefont {Ihn}},\ }\bibfield  {title} {\bibinfo {title} {Coherent
  spin\textendash photon coupling using a resonant exchange qubit},\ }\href
  {https://doi.org/10.1038/s41586-018-0365-y} {\bibfield  {journal} {\bibinfo
  {journal} {Nature}\ }\textbf {\bibinfo {volume} {560}},\ \bibinfo {pages}
  {179} (\bibinfo {year} {2018})}\BibitemShut {NoStop}%
\bibitem [{\citenamefont {Mi}\ \emph {et~al.}(2017{\natexlab{b}})\citenamefont
  {Mi}, \citenamefont {Cady}, \citenamefont {Zajac}, \citenamefont {Stehlik},
  \citenamefont {Edge},\ and\ \citenamefont {Petta}}]{mi2017a}%
  \BibitemOpen
  \bibfield  {author} {\bibinfo {author} {\bibfnamefont {X.}~\bibnamefont
  {Mi}}, \bibinfo {author} {\bibfnamefont {J.~V.}\ \bibnamefont {Cady}},
  \bibinfo {author} {\bibfnamefont {D.~M.}\ \bibnamefont {Zajac}}, \bibinfo
  {author} {\bibfnamefont {J.}~\bibnamefont {Stehlik}}, \bibinfo {author}
  {\bibfnamefont {L.~F.}\ \bibnamefont {Edge}},\ and\ \bibinfo {author}
  {\bibfnamefont {J.~R.}\ \bibnamefont {Petta}},\ }\bibfield  {title} {\bibinfo
  {title} {Circuit quantum electrodynamics architecture for gate-defined
  quantum dots in silicon},\ }\href {https://doi.org/10.1063/1.4974536}
  {\bibfield  {journal} {\bibinfo  {journal} {Appl. Phys. Lett.}\ }\textbf
  {\bibinfo {volume} {110}},\ \bibinfo {pages} {043502} (\bibinfo {year}
  {2017}{\natexlab{b}})}\BibitemShut {NoStop}%
\bibitem [{\citenamefont {{Harvey-Collard}}\ \emph {et~al.}(2020)\citenamefont
  {{Harvey-Collard}}, \citenamefont {Zheng}, \citenamefont {Dijkema},
  \citenamefont {Samkharadze}, \citenamefont {Sammak}, \citenamefont
  {Scappucci},\ and\ \citenamefont {Vandersypen}}]{harvey-collard2020a}%
  \BibitemOpen
  \bibfield  {author} {\bibinfo {author} {\bibfnamefont {P.}~\bibnamefont
  {{Harvey-Collard}}}, \bibinfo {author} {\bibfnamefont {G.}~\bibnamefont
  {Zheng}}, \bibinfo {author} {\bibfnamefont {J.}~\bibnamefont {Dijkema}},
  \bibinfo {author} {\bibfnamefont {N.}~\bibnamefont {Samkharadze}}, \bibinfo
  {author} {\bibfnamefont {A.}~\bibnamefont {Sammak}}, \bibinfo {author}
  {\bibfnamefont {G.}~\bibnamefont {Scappucci}},\ and\ \bibinfo {author}
  {\bibfnamefont {L.~M.~K.}\ \bibnamefont {Vandersypen}},\ }\bibfield  {title}
  {\bibinfo {title} {On-{{Chip Microwave Filters}} for {{High-Impedance
  Resonators}} with {{Gate-Defined Quantum Dots}}},\ }\href
  {https://doi.org/10.1103/PhysRevApplied.14.034025} {\bibfield  {journal}
  {\bibinfo  {journal} {Phys. Rev. Applied}\ }\textbf {\bibinfo {volume}
  {14}},\ \bibinfo {pages} {034025} (\bibinfo {year} {2020})}\BibitemShut
  {NoStop}%
\bibitem [{\citenamefont {Holman}\ \emph {et~al.}(2020)\citenamefont {Holman},
  \citenamefont {Dodson}, \citenamefont {Edge}, \citenamefont {Coppersmith},
  \citenamefont {Friesen}, \citenamefont {McDermott},\ and\ \citenamefont
  {Eriksson}}]{holman2020}%
  \BibitemOpen
  \bibfield  {author} {\bibinfo {author} {\bibfnamefont {N.}~\bibnamefont
  {Holman}}, \bibinfo {author} {\bibfnamefont {J.~P.}\ \bibnamefont {Dodson}},
  \bibinfo {author} {\bibfnamefont {L.~F.}\ \bibnamefont {Edge}}, \bibinfo
  {author} {\bibfnamefont {S.~N.}\ \bibnamefont {Coppersmith}}, \bibinfo
  {author} {\bibfnamefont {M.}~\bibnamefont {Friesen}}, \bibinfo {author}
  {\bibfnamefont {R.}~\bibnamefont {McDermott}},\ and\ \bibinfo {author}
  {\bibfnamefont {M.~A.}\ \bibnamefont {Eriksson}},\ }\bibfield  {title}
  {\bibinfo {title} {Microwave engineering for semiconductor quantum dots in a
  {{cQED}} architecture},\ }\href {https://doi.org/10.1063/5.0016248}
  {\bibfield  {journal} {\bibinfo  {journal} {Appl. Phys. Lett.}\ }\textbf
  {\bibinfo {volume} {117}},\ \bibinfo {pages} {083502} (\bibinfo {year}
  {2020})}\BibitemShut {NoStop}%
\bibitem [{\citenamefont {Borjans}\ \emph {et~al.}(2019)\citenamefont
  {Borjans}, \citenamefont {Croot}, \citenamefont {Mi}, \citenamefont
  {Gullans},\ and\ \citenamefont {Petta}}]{borjans2019}%
  \BibitemOpen
  \bibfield  {author} {\bibinfo {author} {\bibfnamefont {F.}~\bibnamefont
  {Borjans}}, \bibinfo {author} {\bibfnamefont {X.~G.}\ \bibnamefont {Croot}},
  \bibinfo {author} {\bibfnamefont {X.}~\bibnamefont {Mi}}, \bibinfo {author}
  {\bibfnamefont {M.~J.}\ \bibnamefont {Gullans}},\ and\ \bibinfo {author}
  {\bibfnamefont {J.~R.}\ \bibnamefont {Petta}},\ }\bibfield  {title} {\bibinfo
  {title} {Resonant microwave-mediated interactions between distant electron
  spins},\ }\href {https://doi.org/10.1038/s41586-019-1867-y} {\bibfield
  {journal} {\bibinfo  {journal} {Nature}\ ,\ \bibinfo {pages} {1}} (\bibinfo
  {year} {2019})}\BibitemShut {NoStop}%
\bibitem [{\citenamefont {Borjans}\ \emph {et~al.}(2020)\citenamefont
  {Borjans}, \citenamefont {Croot}, \citenamefont {Putz}, \citenamefont {Mi},
  \citenamefont {Quinn}, \citenamefont {Pan}, \citenamefont {Kerckhoff},
  \citenamefont {Pritchett}, \citenamefont {Jackson}, \citenamefont {Edge},
  \citenamefont {Ross}, \citenamefont {Ladd}, \citenamefont {Borselli},
  \citenamefont {Gyure},\ and\ \citenamefont {Petta}}]{borjans2020}%
  \BibitemOpen
  \bibfield  {author} {\bibinfo {author} {\bibfnamefont {F.}~\bibnamefont
  {Borjans}}, \bibinfo {author} {\bibfnamefont {X.}~\bibnamefont {Croot}},
  \bibinfo {author} {\bibfnamefont {S.}~\bibnamefont {Putz}}, \bibinfo {author}
  {\bibfnamefont {X.}~\bibnamefont {Mi}}, \bibinfo {author} {\bibfnamefont
  {S.~M.}\ \bibnamefont {Quinn}}, \bibinfo {author} {\bibfnamefont
  {A.}~\bibnamefont {Pan}}, \bibinfo {author} {\bibfnamefont {J.}~\bibnamefont
  {Kerckhoff}}, \bibinfo {author} {\bibfnamefont {E.~J.}\ \bibnamefont
  {Pritchett}}, \bibinfo {author} {\bibfnamefont {C.~A.}\ \bibnamefont
  {Jackson}}, \bibinfo {author} {\bibfnamefont {L.~F.}\ \bibnamefont {Edge}},
  \bibinfo {author} {\bibfnamefont {R.~S.}\ \bibnamefont {Ross}}, \bibinfo
  {author} {\bibfnamefont {T.~D.}\ \bibnamefont {Ladd}}, \bibinfo {author}
  {\bibfnamefont {M.~G.}\ \bibnamefont {Borselli}}, \bibinfo {author}
  {\bibfnamefont {M.~F.}\ \bibnamefont {Gyure}},\ and\ \bibinfo {author}
  {\bibfnamefont {J.~R.}\ \bibnamefont {Petta}},\ }\bibfield  {title} {\bibinfo
  {title} {Split-gate cavity coupler for silicon circuit quantum
  electrodynamics},\ }\href {https://doi.org/10.1063/5.0006442} {\bibfield
  {journal} {\bibinfo  {journal} {Appl. Phys. Lett.}\ }\textbf {\bibinfo
  {volume} {116}},\ \bibinfo {pages} {234001} (\bibinfo {year}
  {2020})}\BibitemShut {NoStop}%
\bibitem [{\citenamefont {Burkard}\ \emph {et~al.}(2020)\citenamefont
  {Burkard}, \citenamefont {Gullans}, \citenamefont {Mi},\ and\ \citenamefont
  {Petta}}]{burkard2020}%
  \BibitemOpen
  \bibfield  {author} {\bibinfo {author} {\bibfnamefont {G.}~\bibnamefont
  {Burkard}}, \bibinfo {author} {\bibfnamefont {M.~J.}\ \bibnamefont
  {Gullans}}, \bibinfo {author} {\bibfnamefont {X.}~\bibnamefont {Mi}},\ and\
  \bibinfo {author} {\bibfnamefont {J.~R.}\ \bibnamefont {Petta}},\ }\bibfield
  {title} {\bibinfo {title} {Superconductor\textendash semiconductor
  hybrid-circuit quantum electrodynamics},\ }\href
  {https://doi.org/10.1038/s42254-019-0135-2} {\bibfield  {journal} {\bibinfo
  {journal} {Nature Reviews Physics}\ }\textbf {\bibinfo {volume} {2}},\
  \bibinfo {pages} {129} (\bibinfo {year} {2020})}\BibitemShut {NoStop}%
\bibitem [{\citenamefont {Brun}(2020)}]{brun2020}%
  \BibitemOpen
  \bibfield  {author} {\bibinfo {author} {\bibfnamefont {T.~A.}\ \bibnamefont
  {Brun}},\ }\href {https://doi.org/10.1093/acrefore/9780190871994.013.35}
  {\bibinfo {title} {Quantum {{Error Correction}}}} (\bibinfo {year}
  {2020})\BibitemShut {NoStop}%
\bibitem [{\citenamefont {Baart}\ \emph {et~al.}(2016)\citenamefont {Baart},
  \citenamefont {Eendebak}, \citenamefont {Reichl}, \citenamefont
  {Wegscheider},\ and\ \citenamefont {Vandersypen}}]{baart2016}%
  \BibitemOpen
  \bibfield  {author} {\bibinfo {author} {\bibfnamefont {T.~A.}\ \bibnamefont
  {Baart}}, \bibinfo {author} {\bibfnamefont {P.~T.}\ \bibnamefont {Eendebak}},
  \bibinfo {author} {\bibfnamefont {C.}~\bibnamefont {Reichl}}, \bibinfo
  {author} {\bibfnamefont {W.}~\bibnamefont {Wegscheider}},\ and\ \bibinfo
  {author} {\bibfnamefont {L.~M.~K.}\ \bibnamefont {Vandersypen}},\ }\bibfield
  {title} {\bibinfo {title} {Computer-automated tuning of semiconductor double
  quantum dots into the single-electron regime},\ }\href
  {https://doi.org/10.1063/1.4952624} {\bibfield  {journal} {\bibinfo
  {journal} {Appl. Phys. Lett.}\ }\textbf {\bibinfo {volume} {108}},\ \bibinfo
  {pages} {213104} (\bibinfo {year} {2016})}\BibitemShut {NoStop}%
\bibitem [{\citenamefont {Lennon}\ \emph {et~al.}(2019)\citenamefont {Lennon},
  \citenamefont {Moon}, \citenamefont {Camenzind}, \citenamefont {Yu},
  \citenamefont {Zumb{\"u}hl}, \citenamefont {Briggs}, \citenamefont {Osborne},
  \citenamefont {Laird},\ and\ \citenamefont {Ares}}]{lennon2019}%
  \BibitemOpen
  \bibfield  {author} {\bibinfo {author} {\bibfnamefont {D.~T.}\ \bibnamefont
  {Lennon}}, \bibinfo {author} {\bibfnamefont {H.}~\bibnamefont {Moon}},
  \bibinfo {author} {\bibfnamefont {L.~C.}\ \bibnamefont {Camenzind}}, \bibinfo
  {author} {\bibfnamefont {L.}~\bibnamefont {Yu}}, \bibinfo {author}
  {\bibfnamefont {D.~M.}\ \bibnamefont {Zumb{\"u}hl}}, \bibinfo {author}
  {\bibfnamefont {G.~a.~D.}\ \bibnamefont {Briggs}}, \bibinfo {author}
  {\bibfnamefont {M.~A.}\ \bibnamefont {Osborne}}, \bibinfo {author}
  {\bibfnamefont {E.~A.}\ \bibnamefont {Laird}},\ and\ \bibinfo {author}
  {\bibfnamefont {N.}~\bibnamefont {Ares}},\ }\bibfield  {title} {\bibinfo
  {title} {Efficiently measuring a quantum device using machine learning},\
  }\href {https://doi.org/10.1038/s41534-019-0193-4} {\bibfield  {journal}
  {\bibinfo  {journal} {npj Quantum Information}\ }\textbf {\bibinfo {volume}
  {5}},\ \bibinfo {pages} {1} (\bibinfo {year} {2019})}\BibitemShut {NoStop}%
\bibitem [{\citenamefont {Moon}\ \emph {et~al.}(2020)\citenamefont {Moon},
  \citenamefont {Lennon}, \citenamefont {Kirkpatrick}, \citenamefont {{van
  Esbroeck}}, \citenamefont {Camenzind}, \citenamefont {Yu}, \citenamefont
  {Vigneau}, \citenamefont {Zumb{\"u}hl}, \citenamefont {Briggs}, \citenamefont
  {Osborne}, \citenamefont {Sejdinovic}, \citenamefont {Laird},\ and\
  \citenamefont {Ares}}]{moon2020}%
  \BibitemOpen
  \bibfield  {author} {\bibinfo {author} {\bibfnamefont {H.}~\bibnamefont
  {Moon}}, \bibinfo {author} {\bibfnamefont {D.~T.}\ \bibnamefont {Lennon}},
  \bibinfo {author} {\bibfnamefont {J.}~\bibnamefont {Kirkpatrick}}, \bibinfo
  {author} {\bibfnamefont {N.~M.}\ \bibnamefont {{van Esbroeck}}}, \bibinfo
  {author} {\bibfnamefont {L.~C.}\ \bibnamefont {Camenzind}}, \bibinfo {author}
  {\bibfnamefont {L.}~\bibnamefont {Yu}}, \bibinfo {author} {\bibfnamefont
  {F.}~\bibnamefont {Vigneau}}, \bibinfo {author} {\bibfnamefont {D.~M.}\
  \bibnamefont {Zumb{\"u}hl}}, \bibinfo {author} {\bibfnamefont {G.~a.~D.}\
  \bibnamefont {Briggs}}, \bibinfo {author} {\bibfnamefont {M.~A.}\
  \bibnamefont {Osborne}}, \bibinfo {author} {\bibfnamefont {D.}~\bibnamefont
  {Sejdinovic}}, \bibinfo {author} {\bibfnamefont {E.~A.}\ \bibnamefont
  {Laird}},\ and\ \bibinfo {author} {\bibfnamefont {N.}~\bibnamefont {Ares}},\
  }\bibfield  {title} {\bibinfo {title} {Machine learning enables completely
  automatic tuning of a quantum device faster than human experts},\ }\href
  {https://doi.org/10.1038/s41467-020-17835-9} {\bibfield  {journal} {\bibinfo
  {journal} {Nature Communications}\ }\textbf {\bibinfo {volume} {11}},\
  \bibinfo {pages} {4161} (\bibinfo {year} {2020})}\BibitemShut {NoStop}%
\bibitem [{\citenamefont {Zwolak}\ \emph {et~al.}(2020)\citenamefont {Zwolak},
  \citenamefont {McJunkin}, \citenamefont {Kalantre}, \citenamefont {Dodson},
  \citenamefont {MacQuarrie}, \citenamefont {Savage}, \citenamefont {Lagally},
  \citenamefont {Coppersmith}, \citenamefont {Eriksson},\ and\ \citenamefont
  {Taylor}}]{zwolak2020a}%
  \BibitemOpen
  \bibfield  {author} {\bibinfo {author} {\bibfnamefont {J.~P.}\ \bibnamefont
  {Zwolak}}, \bibinfo {author} {\bibfnamefont {T.}~\bibnamefont {McJunkin}},
  \bibinfo {author} {\bibfnamefont {S.~S.}\ \bibnamefont {Kalantre}}, \bibinfo
  {author} {\bibfnamefont {J.}~\bibnamefont {Dodson}}, \bibinfo {author}
  {\bibfnamefont {E.}~\bibnamefont {MacQuarrie}}, \bibinfo {author}
  {\bibfnamefont {D.}~\bibnamefont {Savage}}, \bibinfo {author} {\bibfnamefont
  {M.}~\bibnamefont {Lagally}}, \bibinfo {author} {\bibfnamefont
  {S.}~\bibnamefont {Coppersmith}}, \bibinfo {author} {\bibfnamefont {M.~A.}\
  \bibnamefont {Eriksson}},\ and\ \bibinfo {author} {\bibfnamefont {J.~M.}\
  \bibnamefont {Taylor}},\ }\bibfield  {title} {\bibinfo {title} {Autotuning of
  {{Double-Dot Devices In Situ}} with {{Machine Learning}}},\ }\href
  {https://doi.org/10.1103/PhysRevApplied.13.034075} {\bibfield  {journal}
  {\bibinfo  {journal} {Phys. Rev. Applied}\ }\textbf {\bibinfo {volume}
  {13}},\ \bibinfo {pages} {034075} (\bibinfo {year} {2020})}\BibitemShut
  {NoStop}%
\bibitem [{\citenamefont {{van Diepen}}\ \emph {et~al.}(2018)\citenamefont
  {{van Diepen}}, \citenamefont {Eendebak}, \citenamefont {Buijtendorp},
  \citenamefont {Mukhopadhyay}, \citenamefont {Fujita}, \citenamefont {Reichl},
  \citenamefont {Wegscheider},\ and\ \citenamefont
  {Vandersypen}}]{vandiepen2018}%
  \BibitemOpen
  \bibfield  {author} {\bibinfo {author} {\bibfnamefont {C.~J.}\ \bibnamefont
  {{van Diepen}}}, \bibinfo {author} {\bibfnamefont {P.~T.}\ \bibnamefont
  {Eendebak}}, \bibinfo {author} {\bibfnamefont {B.~T.}\ \bibnamefont
  {Buijtendorp}}, \bibinfo {author} {\bibfnamefont {U.}~\bibnamefont
  {Mukhopadhyay}}, \bibinfo {author} {\bibfnamefont {T.}~\bibnamefont
  {Fujita}}, \bibinfo {author} {\bibfnamefont {C.}~\bibnamefont {Reichl}},
  \bibinfo {author} {\bibfnamefont {W.}~\bibnamefont {Wegscheider}},\ and\
  \bibinfo {author} {\bibfnamefont {L.~M.~K.}\ \bibnamefont {Vandersypen}},\
  }\bibfield  {title} {\bibinfo {title} {Automated tuning of inter-dot tunnel
  coupling in double quantum dots},\ }\href {https://doi.org/10.1063/1.5031034}
  {\bibfield  {journal} {\bibinfo  {journal} {Appl. Phys. Lett.}\ }\textbf
  {\bibinfo {volume} {113}},\ \bibinfo {pages} {033101} (\bibinfo {year}
  {2018})}\BibitemShut {NoStop}%
\bibitem [{\citenamefont {Teske}\ \emph {et~al.}(2019)\citenamefont {Teske},
  \citenamefont {Humpohl}, \citenamefont {Otten}, \citenamefont {Bethke},
  \citenamefont {Cerfontaine}, \citenamefont {Dedden}, \citenamefont {Ludwig},
  \citenamefont {Wieck},\ and\ \citenamefont {Bluhm}}]{teske2019}%
  \BibitemOpen
  \bibfield  {author} {\bibinfo {author} {\bibfnamefont {J.~D.}\ \bibnamefont
  {Teske}}, \bibinfo {author} {\bibfnamefont {S.~S.}\ \bibnamefont {Humpohl}},
  \bibinfo {author} {\bibfnamefont {R.}~\bibnamefont {Otten}}, \bibinfo
  {author} {\bibfnamefont {P.}~\bibnamefont {Bethke}}, \bibinfo {author}
  {\bibfnamefont {P.}~\bibnamefont {Cerfontaine}}, \bibinfo {author}
  {\bibfnamefont {J.}~\bibnamefont {Dedden}}, \bibinfo {author} {\bibfnamefont
  {A.}~\bibnamefont {Ludwig}}, \bibinfo {author} {\bibfnamefont {A.~D.}\
  \bibnamefont {Wieck}},\ and\ \bibinfo {author} {\bibfnamefont
  {H.}~\bibnamefont {Bluhm}},\ }\bibfield  {title} {\bibinfo {title} {A machine
  learning approach for automated fine-tuning of semiconductor spin qubits},\
  }\href {https://doi.org/10.1063/1.5088412} {\bibfield  {journal} {\bibinfo
  {journal} {Appl. Phys. Lett.}\ }\textbf {\bibinfo {volume} {114}},\ \bibinfo
  {pages} {133102} (\bibinfo {year} {2019})}\BibitemShut {NoStop}%
\bibitem [{\citenamefont {Hsiao}\ \emph {et~al.}(2020)\citenamefont {Hsiao},
  \citenamefont {{van Diepen}}, \citenamefont {Mukhopadhyay}, \citenamefont
  {Reichl}, \citenamefont {Wegscheider},\ and\ \citenamefont
  {Vandersypen}}]{hsiao2020}%
  \BibitemOpen
  \bibfield  {author} {\bibinfo {author} {\bibfnamefont {T.-K.}\ \bibnamefont
  {Hsiao}}, \bibinfo {author} {\bibfnamefont {C.}~\bibnamefont {{van Diepen}}},
  \bibinfo {author} {\bibfnamefont {U.}~\bibnamefont {Mukhopadhyay}}, \bibinfo
  {author} {\bibfnamefont {C.}~\bibnamefont {Reichl}}, \bibinfo {author}
  {\bibfnamefont {W.}~\bibnamefont {Wegscheider}},\ and\ \bibinfo {author}
  {\bibfnamefont {L.}~\bibnamefont {Vandersypen}},\ }\bibfield  {title}
  {\bibinfo {title} {Efficient {{Orthogonal Control}} of {{Tunnel Couplings}}
  in a {{Quantum Dot Array}}},\ }\href
  {https://doi.org/10.1103/PhysRevApplied.13.054018} {\bibfield  {journal}
  {\bibinfo  {journal} {Phys. Rev. Applied}\ }\textbf {\bibinfo {volume}
  {13}},\ \bibinfo {pages} {054018} (\bibinfo {year} {2020})}\BibitemShut
  {NoStop}%
\bibitem [{\citenamefont {Botzem}\ \emph {et~al.}(2018)\citenamefont {Botzem},
  \citenamefont {Shulman}, \citenamefont {Foletti}, \citenamefont {Harvey},
  \citenamefont {Dial}, \citenamefont {Bethke}, \citenamefont {Cerfontaine},
  \citenamefont {McNeil}, \citenamefont {Mahalu}, \citenamefont {Umansky},
  \citenamefont {Ludwig}, \citenamefont {Wieck}, \citenamefont {Schuh},
  \citenamefont {Bougeard}, \citenamefont {Yacoby},\ and\ \citenamefont
  {Bluhm}}]{botzem2018}%
  \BibitemOpen
  \bibfield  {author} {\bibinfo {author} {\bibfnamefont {T.}~\bibnamefont
  {Botzem}}, \bibinfo {author} {\bibfnamefont {M.~D.}\ \bibnamefont {Shulman}},
  \bibinfo {author} {\bibfnamefont {S.}~\bibnamefont {Foletti}}, \bibinfo
  {author} {\bibfnamefont {S.~P.}\ \bibnamefont {Harvey}}, \bibinfo {author}
  {\bibfnamefont {O.~E.}\ \bibnamefont {Dial}}, \bibinfo {author}
  {\bibfnamefont {P.}~\bibnamefont {Bethke}}, \bibinfo {author} {\bibfnamefont
  {P.}~\bibnamefont {Cerfontaine}}, \bibinfo {author} {\bibfnamefont
  {R.~P.~G.}\ \bibnamefont {McNeil}}, \bibinfo {author} {\bibfnamefont
  {D.}~\bibnamefont {Mahalu}}, \bibinfo {author} {\bibfnamefont
  {V.}~\bibnamefont {Umansky}}, \bibinfo {author} {\bibfnamefont
  {A.}~\bibnamefont {Ludwig}}, \bibinfo {author} {\bibfnamefont
  {A.}~\bibnamefont {Wieck}}, \bibinfo {author} {\bibfnamefont
  {D.}~\bibnamefont {Schuh}}, \bibinfo {author} {\bibfnamefont
  {D.}~\bibnamefont {Bougeard}}, \bibinfo {author} {\bibfnamefont
  {A.}~\bibnamefont {Yacoby}},\ and\ \bibinfo {author} {\bibfnamefont
  {H.}~\bibnamefont {Bluhm}},\ }\bibfield  {title} {\bibinfo {title} {Tuning
  {{Methods}} for {{Semiconductor Spin Qubits}}},\ }\href
  {https://doi.org/10.1103/PhysRevApplied.10.054026} {\bibfield  {journal}
  {\bibinfo  {journal} {Phys. Rev. Applied}\ }\textbf {\bibinfo {volume}
  {10}},\ \bibinfo {pages} {054026} (\bibinfo {year} {2018})}\BibitemShut
  {NoStop}%
\bibitem [{\citenamefont {Zajac}\ \emph {et~al.}(2016)\citenamefont {Zajac},
  \citenamefont {Hazard}, \citenamefont {Mi}, \citenamefont {Nielsen},\ and\
  \citenamefont {Petta}}]{zajac2016}%
  \BibitemOpen
  \bibfield  {author} {\bibinfo {author} {\bibfnamefont {D.~M.}\ \bibnamefont
  {Zajac}}, \bibinfo {author} {\bibfnamefont {T.~M.}\ \bibnamefont {Hazard}},
  \bibinfo {author} {\bibfnamefont {X.}~\bibnamefont {Mi}}, \bibinfo {author}
  {\bibfnamefont {E.}~\bibnamefont {Nielsen}},\ and\ \bibinfo {author}
  {\bibfnamefont {J.~R.}\ \bibnamefont {Petta}},\ }\bibfield  {title} {\bibinfo
  {title} {Scalable {{Gate Architecture}} for a {{One-Dimensional Array}} of
  {{Semiconductor Spin Qubits}}},\ }\href
  {https://doi.org/10.1103/PhysRevApplied.6.054013} {\bibfield  {journal}
  {\bibinfo  {journal} {Phys. Rev. Applied}\ }\textbf {\bibinfo {volume} {6}},\
  \bibinfo {pages} {054013} (\bibinfo {year} {2016})}\BibitemShut {NoStop}%
\bibitem [{\citenamefont {Borselli}\ \emph {et~al.}(2015)\citenamefont
  {Borselli}, \citenamefont {Eng}, \citenamefont {Ross}, \citenamefont
  {Hazard}, \citenamefont {Holabird}, \citenamefont {Huang}, \citenamefont
  {Kiselev}, \citenamefont {Deelman}, \citenamefont {Warren}, \citenamefont {{I
  Milosavljevic}}, \citenamefont {Schmitz}, \citenamefont {Sokolich},
  \citenamefont {Gyure},\ and\ \citenamefont {Hunter}}]{borselli2015}%
  \BibitemOpen
  \bibfield  {author} {\bibinfo {author} {\bibfnamefont {M.~G.}\ \bibnamefont
  {Borselli}}, \bibinfo {author} {\bibfnamefont {K.}~\bibnamefont {Eng}},
  \bibinfo {author} {\bibfnamefont {R.~S.}\ \bibnamefont {Ross}}, \bibinfo
  {author} {\bibfnamefont {T.~M.}\ \bibnamefont {Hazard}}, \bibinfo {author}
  {\bibfnamefont {K.~S.}\ \bibnamefont {Holabird}}, \bibinfo {author}
  {\bibfnamefont {B.}~\bibnamefont {Huang}}, \bibinfo {author} {\bibfnamefont
  {A.~A.}\ \bibnamefont {Kiselev}}, \bibinfo {author} {\bibfnamefont {P.~W.}\
  \bibnamefont {Deelman}}, \bibinfo {author} {\bibfnamefont {L.~D.}\
  \bibnamefont {Warren}}, \bibinfo {author} {\bibnamefont {{I Milosavljevic}}},
  \bibinfo {author} {\bibfnamefont {A.~E.}\ \bibnamefont {Schmitz}}, \bibinfo
  {author} {\bibfnamefont {M.}~\bibnamefont {Sokolich}}, \bibinfo {author}
  {\bibfnamefont {M.~F.}\ \bibnamefont {Gyure}},\ and\ \bibinfo {author}
  {\bibfnamefont {A.~T.}\ \bibnamefont {Hunter}},\ }\bibfield  {title}
  {\bibinfo {title} {Undoped accumulation-mode {{Si}}/{{SiGe}} quantum dots},\
  }\href {https://doi.org/10.1088/0957-4484/26/37/375202} {\bibfield  {journal}
  {\bibinfo  {journal} {Nanotechnology}\ }\textbf {\bibinfo {volume} {26}},\
  \bibinfo {pages} {375202} (\bibinfo {year} {2015})}\BibitemShut {NoStop}%
\bibitem [{\citenamefont {Frees}\ \emph {et~al.}(2019)\citenamefont {Frees},
  \citenamefont {Gamble}, \citenamefont {Ward}, \citenamefont {{Blume-Kohout}},
  \citenamefont {Eriksson}, \citenamefont {Friesen},\ and\ \citenamefont
  {Coppersmith}}]{frees2019}%
  \BibitemOpen
  \bibfield  {author} {\bibinfo {author} {\bibfnamefont {A.}~\bibnamefont
  {Frees}}, \bibinfo {author} {\bibfnamefont {J.~K.}\ \bibnamefont {Gamble}},
  \bibinfo {author} {\bibfnamefont {D.~R.}\ \bibnamefont {Ward}}, \bibinfo
  {author} {\bibfnamefont {R.}~\bibnamefont {{Blume-Kohout}}}, \bibinfo
  {author} {\bibfnamefont {M.}~\bibnamefont {Eriksson}}, \bibinfo {author}
  {\bibfnamefont {M.}~\bibnamefont {Friesen}},\ and\ \bibinfo {author}
  {\bibfnamefont {S.}~\bibnamefont {Coppersmith}},\ }\bibfield  {title}
  {\bibinfo {title} {Compressed {{Optimization}} of {{Device Architectures}}
  for {{Semiconductor Quantum Devices}}},\ }\href
  {https://doi.org/10.1103/PhysRevApplied.11.024063} {\bibfield  {journal}
  {\bibinfo  {journal} {Phys. Rev. Applied}\ }\textbf {\bibinfo {volume}
  {11}},\ \bibinfo {pages} {024063} (\bibinfo {year} {2019})}\BibitemShut
  {NoStop}%
\bibitem [{\citenamefont {Jones}\ \emph {et~al.}(2012)\citenamefont {Jones},
  \citenamefont {Van~Meter}, \citenamefont {Fowler}, \citenamefont {McMahon},
  \citenamefont {Kim}, \citenamefont {Ladd},\ and\ \citenamefont
  {Yamamoto}}]{jones2012}%
  \BibitemOpen
  \bibfield  {author} {\bibinfo {author} {\bibfnamefont {N.~C.}\ \bibnamefont
  {Jones}}, \bibinfo {author} {\bibfnamefont {R.}~\bibnamefont {Van~Meter}},
  \bibinfo {author} {\bibfnamefont {A.~G.}\ \bibnamefont {Fowler}}, \bibinfo
  {author} {\bibfnamefont {P.~L.}\ \bibnamefont {McMahon}}, \bibinfo {author}
  {\bibfnamefont {J.}~\bibnamefont {Kim}}, \bibinfo {author} {\bibfnamefont
  {T.~D.}\ \bibnamefont {Ladd}},\ and\ \bibinfo {author} {\bibfnamefont
  {Y.}~\bibnamefont {Yamamoto}},\ }\bibfield  {title} {\bibinfo {title}
  {Layered {{Architecture}} for {{Quantum Computing}}},\ }\bibfield  {journal}
  {\bibinfo  {journal} {Physical Review X}\ }\textbf {\bibinfo {volume} {2}},\
  \href {https://doi.org/10.1103/PhysRevX.2.031007} {10.1103/PhysRevX.2.031007}
  (\bibinfo {year} {2012})\BibitemShut {NoStop}%
\bibitem [{\citenamefont {Jones}\ \emph {et~al.}(2018)\citenamefont {Jones},
  \citenamefont {Fogarty}, \citenamefont {Morello}, \citenamefont {Gyure},
  \citenamefont {Dzurak},\ and\ \citenamefont {Ladd}}]{jones2018}%
  \BibitemOpen
  \bibfield  {author} {\bibinfo {author} {\bibfnamefont {C.}~\bibnamefont
  {Jones}}, \bibinfo {author} {\bibfnamefont {M.~A.}\ \bibnamefont {Fogarty}},
  \bibinfo {author} {\bibfnamefont {A.}~\bibnamefont {Morello}}, \bibinfo
  {author} {\bibfnamefont {M.~F.}\ \bibnamefont {Gyure}}, \bibinfo {author}
  {\bibfnamefont {A.~S.}\ \bibnamefont {Dzurak}},\ and\ \bibinfo {author}
  {\bibfnamefont {T.~D.}\ \bibnamefont {Ladd}},\ }\bibfield  {title} {\bibinfo
  {title} {Logical {{Qubit}} in a {{Linear Array}} of {{Semiconductor Quantum
  Dots}}},\ }\href {https://doi.org/10.1103/PhysRevX.8.021058} {\bibfield
  {journal} {\bibinfo  {journal} {Phys. Rev. X}\ }\textbf {\bibinfo {volume}
  {8}},\ \bibinfo {pages} {021058} (\bibinfo {year} {2018})}\BibitemShut
  {NoStop}%
\bibitem [{\citenamefont {Tosi}\ \emph {et~al.}(2017)\citenamefont {Tosi},
  \citenamefont {Mohiyaddin}, \citenamefont {Schmitt}, \citenamefont {Tenberg},
  \citenamefont {Rahman}, \citenamefont {Klimeck},\ and\ \citenamefont
  {Morello}}]{tosi2017}%
  \BibitemOpen
  \bibfield  {author} {\bibinfo {author} {\bibfnamefont {G.}~\bibnamefont
  {Tosi}}, \bibinfo {author} {\bibfnamefont {F.~A.}\ \bibnamefont
  {Mohiyaddin}}, \bibinfo {author} {\bibfnamefont {V.}~\bibnamefont {Schmitt}},
  \bibinfo {author} {\bibfnamefont {S.}~\bibnamefont {Tenberg}}, \bibinfo
  {author} {\bibfnamefont {R.}~\bibnamefont {Rahman}}, \bibinfo {author}
  {\bibfnamefont {G.}~\bibnamefont {Klimeck}},\ and\ \bibinfo {author}
  {\bibfnamefont {A.}~\bibnamefont {Morello}},\ }\bibfield  {title} {\bibinfo
  {title} {Silicon quantum processor with robust long-distance qubit
  couplings},\ }\href {https://doi.org/10.1038/s41467-017-00378-x} {\bibfield
  {journal} {\bibinfo  {journal} {Nature Communications}\ }\textbf {\bibinfo
  {volume} {8}},\ \bibinfo {pages} {450} (\bibinfo {year} {2017})}\BibitemShut
  {NoStop}%
\bibitem [{\citenamefont {Veldhorst}\ \emph {et~al.}(2017)\citenamefont
  {Veldhorst}, \citenamefont {Eenink}, \citenamefont {Yang},\ and\
  \citenamefont {Dzurak}}]{veldhorst2017}%
  \BibitemOpen
  \bibfield  {author} {\bibinfo {author} {\bibfnamefont {M.}~\bibnamefont
  {Veldhorst}}, \bibinfo {author} {\bibfnamefont {H.~G.~J.}\ \bibnamefont
  {Eenink}}, \bibinfo {author} {\bibfnamefont {C.~H.}\ \bibnamefont {Yang}},\
  and\ \bibinfo {author} {\bibfnamefont {A.~S.}\ \bibnamefont {Dzurak}},\
  }\bibfield  {title} {\bibinfo {title} {Silicon {{CMOS}} architecture for a
  spin-based quantum computer},\ }\href
  {https://doi.org/10.1038/s41467-017-01905-6} {\bibfield  {journal} {\bibinfo
  {journal} {Nature Communications}\ }\textbf {\bibinfo {volume} {8}},\
  \bibinfo {pages} {1766} (\bibinfo {year} {2017})}\BibitemShut {NoStop}%
\bibitem [{\citenamefont {Li}\ \emph {et~al.}(2018)\citenamefont {Li},
  \citenamefont {Petit}, \citenamefont {Franke}, \citenamefont {Dehollain},
  \citenamefont {Helsen}, \citenamefont {Steudtner}, \citenamefont {Thomas},
  \citenamefont {Yoscovits}, \citenamefont {Singh}, \citenamefont {Wehner},
  \citenamefont {Vandersypen}, \citenamefont {Clarke},\ and\ \citenamefont
  {Veldhorst}}]{li2018}%
  \BibitemOpen
  \bibfield  {author} {\bibinfo {author} {\bibfnamefont {R.}~\bibnamefont
  {Li}}, \bibinfo {author} {\bibfnamefont {L.}~\bibnamefont {Petit}}, \bibinfo
  {author} {\bibfnamefont {D.~P.}\ \bibnamefont {Franke}}, \bibinfo {author}
  {\bibfnamefont {J.~P.}\ \bibnamefont {Dehollain}}, \bibinfo {author}
  {\bibfnamefont {J.}~\bibnamefont {Helsen}}, \bibinfo {author} {\bibfnamefont
  {M.}~\bibnamefont {Steudtner}}, \bibinfo {author} {\bibfnamefont {N.~K.}\
  \bibnamefont {Thomas}}, \bibinfo {author} {\bibfnamefont {Z.~R.}\
  \bibnamefont {Yoscovits}}, \bibinfo {author} {\bibfnamefont {K.~J.}\
  \bibnamefont {Singh}}, \bibinfo {author} {\bibfnamefont {S.}~\bibnamefont
  {Wehner}}, \bibinfo {author} {\bibfnamefont {L.~M.~K.}\ \bibnamefont
  {Vandersypen}}, \bibinfo {author} {\bibfnamefont {J.~S.}\ \bibnamefont
  {Clarke}},\ and\ \bibinfo {author} {\bibfnamefont {M.}~\bibnamefont
  {Veldhorst}},\ }\bibfield  {title} {\bibinfo {title} {A crossbar network for
  silicon quantum dot qubits},\ }\href {https://doi.org/10.1126/sciadv.aar3960}
  {\bibfield  {journal} {\bibinfo  {journal} {Science Advances}\ }\textbf
  {\bibinfo {volume} {4}},\ \bibinfo {pages} {eaar3960} (\bibinfo {year}
  {2018})}\BibitemShut {NoStop}%
\bibitem [{\citenamefont {Reilly}(2015)}]{reilly2015}%
  \BibitemOpen
  \bibfield  {author} {\bibinfo {author} {\bibfnamefont {D.~J.}\ \bibnamefont
  {Reilly}},\ }\bibfield  {title} {\bibinfo {title} {Engineering the
  quantum-classical interface of solid-state qubits},\ }\href
  {https://doi.org/10.1038/npjqi.2015.11} {\bibfield  {journal} {\bibinfo
  {journal} {npj Quantum Information}\ }\textbf {\bibinfo {volume} {1}},\
  \bibinfo {pages} {15011} (\bibinfo {year} {2015})}\BibitemShut {NoStop}%
\bibitem [{\citenamefont {{van Dijk}}\ \emph {et~al.}(2019)\citenamefont {{van
  Dijk}}, \citenamefont {Kawakami}, \citenamefont {Schouten}, \citenamefont
  {Veldhorst}, \citenamefont {Vandersypen}, \citenamefont {Babaie},
  \citenamefont {Charbon},\ and\ \citenamefont {Sebastiano}}]{vandijk2019}%
  \BibitemOpen
  \bibfield  {author} {\bibinfo {author} {\bibfnamefont {J.}~\bibnamefont {{van
  Dijk}}}, \bibinfo {author} {\bibfnamefont {E.}~\bibnamefont {Kawakami}},
  \bibinfo {author} {\bibfnamefont {R.}~\bibnamefont {Schouten}}, \bibinfo
  {author} {\bibfnamefont {M.}~\bibnamefont {Veldhorst}}, \bibinfo {author}
  {\bibfnamefont {L.}~\bibnamefont {Vandersypen}}, \bibinfo {author}
  {\bibfnamefont {M.}~\bibnamefont {Babaie}}, \bibinfo {author} {\bibfnamefont
  {E.}~\bibnamefont {Charbon}},\ and\ \bibinfo {author} {\bibfnamefont
  {F.}~\bibnamefont {Sebastiano}},\ }\bibfield  {title} {\bibinfo {title}
  {Impact of {{Classical Control Electronics}} on {{Qubit Fidelity}}},\ }\href
  {https://doi.org/10.1103/PhysRevApplied.12.044054} {\bibfield  {journal}
  {\bibinfo  {journal} {Phys. Rev. Applied}\ }\textbf {\bibinfo {volume}
  {12}},\ \bibinfo {pages} {044054} (\bibinfo {year} {2019})}\BibitemShut
  {NoStop}%
\bibitem [{\citenamefont {Bluhm}\ and\ \citenamefont
  {Schreiber}(2019)}]{bluhm2019}%
  \BibitemOpen
  \bibfield  {author} {\bibinfo {author} {\bibfnamefont {H.}~\bibnamefont
  {Bluhm}}\ and\ \bibinfo {author} {\bibfnamefont {L.~R.}\ \bibnamefont
  {Schreiber}},\ }\bibfield  {title} {\bibinfo {title} {Semiconductor {{Spin
  Qubits}} \textemdash{} {{A Scalable Platform}} for {{Quantum Computing}}?},\
  }in\ \href {https://doi.org/10.1109/ISCAS.2019.8702477} {\emph {\bibinfo
  {booktitle} {2019 {{IEEE International Symposium}} on {{Circuits}} and
  {{Systems}} ({{ISCAS}})}}}\ (\bibinfo {year} {2019})\ pp.\ \bibinfo {pages}
  {1--5}\BibitemShut {NoStop}%
\bibitem [{\citenamefont {Geck}\ \emph {et~al.}(2019)\citenamefont {Geck},
  \citenamefont {Kruth}, \citenamefont {Bluhm}, \citenamefont {van Waasen},\
  and\ \citenamefont {Heinen}}]{geck2019}%
  \BibitemOpen
  \bibfield  {author} {\bibinfo {author} {\bibfnamefont {L.}~\bibnamefont
  {Geck}}, \bibinfo {author} {\bibfnamefont {A.}~\bibnamefont {Kruth}},
  \bibinfo {author} {\bibfnamefont {H.}~\bibnamefont {Bluhm}}, \bibinfo
  {author} {\bibfnamefont {S.}~\bibnamefont {van Waasen}},\ and\ \bibinfo
  {author} {\bibfnamefont {S.}~\bibnamefont {Heinen}},\ }\bibfield  {title}
  {\bibinfo {title} {Control electronics for semiconductor spin qubits},\
  }\href {https://doi.org/10.1088/2058-9565/ab5e07} {\bibfield  {journal}
  {\bibinfo  {journal} {Quantum Sci. Technol.}\ }\textbf {\bibinfo {volume}
  {5}},\ \bibinfo {pages} {015004} (\bibinfo {year} {2019})}\BibitemShut
  {NoStop}%
\bibitem [{\citenamefont {{Gonzalez-Zalba}}\ \emph {et~al.}(2020)\citenamefont
  {{Gonzalez-Zalba}}, \citenamefont {{de Franceschi}}, \citenamefont {Charbon},
  \citenamefont {Meunier}, \citenamefont {Vinet},\ and\ \citenamefont
  {Dzurak}}]{gonzalez-zalba2020}%
  \BibitemOpen
  \bibfield  {author} {\bibinfo {author} {\bibfnamefont {M.~F.}\ \bibnamefont
  {{Gonzalez-Zalba}}}, \bibinfo {author} {\bibfnamefont {S.}~\bibnamefont {{de
  Franceschi}}}, \bibinfo {author} {\bibfnamefont {E.}~\bibnamefont {Charbon}},
  \bibinfo {author} {\bibfnamefont {T.}~\bibnamefont {Meunier}}, \bibinfo
  {author} {\bibfnamefont {M.}~\bibnamefont {Vinet}},\ and\ \bibinfo {author}
  {\bibfnamefont {A.~S.}\ \bibnamefont {Dzurak}},\ }\bibfield  {title}
  {\bibinfo {title} {Scaling silicon-based quantum computing using {{CMOS}}
  technology: {{State-of-the-art}}, {{Challenges}} and {{Perspectives}}},\
  }\href {http://arxiv.org/abs/2011.11753} {\bibfield  {journal} {\bibinfo
  {journal} {arXiv:2011.11753 [cond-mat, physics:quant-ph]}\ } (\bibinfo {year}
  {2020})}\BibitemShut {NoStop}%
\bibitem [{\citenamefont {Hornibrook}\ \emph {et~al.}(2015)\citenamefont
  {Hornibrook}, \citenamefont {Colless}, \citenamefont {Conway~Lamb},
  \citenamefont {Pauka}, \citenamefont {Lu}, \citenamefont {Gossard},
  \citenamefont {Watson}, \citenamefont {Gardner}, \citenamefont {Fallahi},
  \citenamefont {Manfra},\ and\ \citenamefont {Reilly}}]{hornibrook2015}%
  \BibitemOpen
  \bibfield  {author} {\bibinfo {author} {\bibfnamefont {J.~M.}\ \bibnamefont
  {Hornibrook}}, \bibinfo {author} {\bibfnamefont {J.~I.}\ \bibnamefont
  {Colless}}, \bibinfo {author} {\bibfnamefont {I.~D.}\ \bibnamefont
  {Conway~Lamb}}, \bibinfo {author} {\bibfnamefont {S.~J.}\ \bibnamefont
  {Pauka}}, \bibinfo {author} {\bibfnamefont {H.}~\bibnamefont {Lu}}, \bibinfo
  {author} {\bibfnamefont {A.~C.}\ \bibnamefont {Gossard}}, \bibinfo {author}
  {\bibfnamefont {J.~D.}\ \bibnamefont {Watson}}, \bibinfo {author}
  {\bibfnamefont {G.~C.}\ \bibnamefont {Gardner}}, \bibinfo {author}
  {\bibfnamefont {S.}~\bibnamefont {Fallahi}}, \bibinfo {author} {\bibfnamefont
  {M.~J.}\ \bibnamefont {Manfra}},\ and\ \bibinfo {author} {\bibfnamefont
  {D.~J.}\ \bibnamefont {Reilly}},\ }\bibfield  {title} {\bibinfo {title}
  {Cryogenic {{Control Architecture}} for {{Large-Scale Quantum Computing}}},\
  }\href {https://doi.org/10.1103/PhysRevApplied.3.024010} {\bibfield
  {journal} {\bibinfo  {journal} {Phys. Rev. Applied}\ }\textbf {\bibinfo
  {volume} {3}},\ \bibinfo {pages} {024010} (\bibinfo {year}
  {2015})}\BibitemShut {NoStop}%
\bibitem [{\citenamefont {Xue}\ \emph {et~al.}(2021)\citenamefont {Xue},
  \citenamefont {Patra}, \citenamefont {{van Dijk}}, \citenamefont
  {Samkharadze}, \citenamefont {Subramanian}, \citenamefont {Corna},
  \citenamefont {Paquelet~Wuetz}, \citenamefont {Jeon}, \citenamefont {Sheikh},
  \citenamefont {{Juarez-Hernandez}}, \citenamefont {Esparza}, \citenamefont
  {Rampurawala}, \citenamefont {Carlton}, \citenamefont {Ravikumar},
  \citenamefont {Nieva}, \citenamefont {Kim}, \citenamefont {Lee},
  \citenamefont {Sammak}, \citenamefont {Scappucci}, \citenamefont {Veldhorst},
  \citenamefont {Sebastiano}, \citenamefont {Babaie}, \citenamefont
  {Pellerano}, \citenamefont {Charbon},\ and\ \citenamefont
  {Vandersypen}}]{xue2021}%
  \BibitemOpen
  \bibfield  {author} {\bibinfo {author} {\bibfnamefont {X.}~\bibnamefont
  {Xue}}, \bibinfo {author} {\bibfnamefont {B.}~\bibnamefont {Patra}}, \bibinfo
  {author} {\bibfnamefont {J.~P.~G.}\ \bibnamefont {{van Dijk}}}, \bibinfo
  {author} {\bibfnamefont {N.}~\bibnamefont {Samkharadze}}, \bibinfo {author}
  {\bibfnamefont {S.}~\bibnamefont {Subramanian}}, \bibinfo {author}
  {\bibfnamefont {A.}~\bibnamefont {Corna}}, \bibinfo {author} {\bibfnamefont
  {B.}~\bibnamefont {Paquelet~Wuetz}}, \bibinfo {author} {\bibfnamefont
  {C.}~\bibnamefont {Jeon}}, \bibinfo {author} {\bibfnamefont {F.}~\bibnamefont
  {Sheikh}}, \bibinfo {author} {\bibfnamefont {E.}~\bibnamefont
  {{Juarez-Hernandez}}}, \bibinfo {author} {\bibfnamefont {B.~P.}\ \bibnamefont
  {Esparza}}, \bibinfo {author} {\bibfnamefont {H.}~\bibnamefont
  {Rampurawala}}, \bibinfo {author} {\bibfnamefont {B.}~\bibnamefont
  {Carlton}}, \bibinfo {author} {\bibfnamefont {S.}~\bibnamefont {Ravikumar}},
  \bibinfo {author} {\bibfnamefont {C.}~\bibnamefont {Nieva}}, \bibinfo
  {author} {\bibfnamefont {S.}~\bibnamefont {Kim}}, \bibinfo {author}
  {\bibfnamefont {H.-J.}\ \bibnamefont {Lee}}, \bibinfo {author} {\bibfnamefont
  {A.}~\bibnamefont {Sammak}}, \bibinfo {author} {\bibfnamefont
  {G.}~\bibnamefont {Scappucci}}, \bibinfo {author} {\bibfnamefont
  {M.}~\bibnamefont {Veldhorst}}, \bibinfo {author} {\bibfnamefont
  {F.}~\bibnamefont {Sebastiano}}, \bibinfo {author} {\bibfnamefont
  {M.}~\bibnamefont {Babaie}}, \bibinfo {author} {\bibfnamefont
  {S.}~\bibnamefont {Pellerano}}, \bibinfo {author} {\bibfnamefont
  {E.}~\bibnamefont {Charbon}},\ and\ \bibinfo {author} {\bibfnamefont
  {L.~M.~K.}\ \bibnamefont {Vandersypen}},\ }\bibfield  {title} {\bibinfo
  {title} {{{CMOS-based}} cryogenic control of silicon quantum circuits},\
  }\href {https://doi.org/10.1038/s41586-021-03469-4} {\bibfield  {journal}
  {\bibinfo  {journal} {Nature}\ }\textbf {\bibinfo {volume} {593}},\ \bibinfo
  {pages} {205} (\bibinfo {year} {2021})}\BibitemShut {NoStop}%
\bibitem [{\citenamefont {Pauka}\ \emph {et~al.}(2021)\citenamefont {Pauka},
  \citenamefont {Das}, \citenamefont {Kalra}, \citenamefont {Moini},
  \citenamefont {Yang}, \citenamefont {Trainer}, \citenamefont {Bousquet},
  \citenamefont {Cantaloube}, \citenamefont {Dick}, \citenamefont {Gardner},
  \citenamefont {Manfra},\ and\ \citenamefont {Reilly}}]{pauka2021}%
  \BibitemOpen
  \bibfield  {author} {\bibinfo {author} {\bibfnamefont {S.~J.}\ \bibnamefont
  {Pauka}}, \bibinfo {author} {\bibfnamefont {K.}~\bibnamefont {Das}}, \bibinfo
  {author} {\bibfnamefont {R.}~\bibnamefont {Kalra}}, \bibinfo {author}
  {\bibfnamefont {A.}~\bibnamefont {Moini}}, \bibinfo {author} {\bibfnamefont
  {Y.}~\bibnamefont {Yang}}, \bibinfo {author} {\bibfnamefont {M.}~\bibnamefont
  {Trainer}}, \bibinfo {author} {\bibfnamefont {A.}~\bibnamefont {Bousquet}},
  \bibinfo {author} {\bibfnamefont {C.}~\bibnamefont {Cantaloube}}, \bibinfo
  {author} {\bibfnamefont {N.}~\bibnamefont {Dick}}, \bibinfo {author}
  {\bibfnamefont {G.~C.}\ \bibnamefont {Gardner}}, \bibinfo {author}
  {\bibfnamefont {M.~J.}\ \bibnamefont {Manfra}},\ and\ \bibinfo {author}
  {\bibfnamefont {D.~J.}\ \bibnamefont {Reilly}},\ }\bibfield  {title}
  {\bibinfo {title} {A cryogenic {{CMOS}} chip for generating control signals
  for multiple qubits},\ }\href {https://doi.org/10.1038/s41928-020-00528-y}
  {\bibfield  {journal} {\bibinfo  {journal} {Nature Electronics}\ }\textbf
  {\bibinfo {volume} {4}},\ \bibinfo {pages} {64} (\bibinfo {year}
  {2021})}\BibitemShut {NoStop}%
\bibitem [{\citenamefont {Yang}\ \emph {et~al.}(2020)\citenamefont {Yang},
  \citenamefont {Wan}, \citenamefont {Yan}, \citenamefont {Zhu}, \citenamefont
  {Wang}, \citenamefont {Peng}, \citenamefont {Huang}, \citenamefont {Yu},
  \citenamefont {Hu}, \citenamefont {Mao}, \citenamefont {Li}, \citenamefont
  {Yang}, \citenamefont {Zheng}, \citenamefont {Jia}, \citenamefont {Shi},\
  and\ \citenamefont {Xu}}]{yang2020}%
  \BibitemOpen
  \bibfield  {author} {\bibinfo {author} {\bibfnamefont {T.~Y.}\ \bibnamefont
  {Yang}}, \bibinfo {author} {\bibfnamefont {Q.}~\bibnamefont {Wan}}, \bibinfo
  {author} {\bibfnamefont {D.~Y.}\ \bibnamefont {Yan}}, \bibinfo {author}
  {\bibfnamefont {Z.}~\bibnamefont {Zhu}}, \bibinfo {author} {\bibfnamefont
  {Z.~W.}\ \bibnamefont {Wang}}, \bibinfo {author} {\bibfnamefont
  {C.}~\bibnamefont {Peng}}, \bibinfo {author} {\bibfnamefont {Y.~B.}\
  \bibnamefont {Huang}}, \bibinfo {author} {\bibfnamefont {R.}~\bibnamefont
  {Yu}}, \bibinfo {author} {\bibfnamefont {J.}~\bibnamefont {Hu}}, \bibinfo
  {author} {\bibfnamefont {Z.~Q.}\ \bibnamefont {Mao}}, \bibinfo {author}
  {\bibfnamefont {S.}~\bibnamefont {Li}}, \bibinfo {author} {\bibfnamefont
  {S.~A.}\ \bibnamefont {Yang}}, \bibinfo {author} {\bibfnamefont
  {H.}~\bibnamefont {Zheng}}, \bibinfo {author} {\bibfnamefont {J.-F.}\
  \bibnamefont {Jia}}, \bibinfo {author} {\bibfnamefont {Y.~G.}\ \bibnamefont
  {Shi}},\ and\ \bibinfo {author} {\bibfnamefont {N.}~\bibnamefont {Xu}},\
  }\bibfield  {title} {\bibinfo {title} {Directional massless {{Dirac}}
  fermions in a layered van der {{Waals}} material with one-dimensional
  long-range order},\ }\href {https://doi.org/10.1038/s41563-019-0494-1}
  {\bibfield  {journal} {\bibinfo  {journal} {Nature Materials}\ }\textbf
  {\bibinfo {volume} {19}},\ \bibinfo {pages} {27} (\bibinfo {year}
  {2020})}\BibitemShut {NoStop}%
\bibitem [{\citenamefont {Petit}\ \emph {et~al.}(2020)\citenamefont {Petit},
  \citenamefont {Eenink}, \citenamefont {Russ}, \citenamefont {Lawrie},
  \citenamefont {Hendrickx}, \citenamefont {Philips}, \citenamefont {Clarke},
  \citenamefont {Vandersypen},\ and\ \citenamefont {Veldhorst}}]{petit2020a}%
  \BibitemOpen
  \bibfield  {author} {\bibinfo {author} {\bibfnamefont {L.}~\bibnamefont
  {Petit}}, \bibinfo {author} {\bibfnamefont {H.~G.~J.}\ \bibnamefont
  {Eenink}}, \bibinfo {author} {\bibfnamefont {M.}~\bibnamefont {Russ}},
  \bibinfo {author} {\bibfnamefont {W.~I.~L.}\ \bibnamefont {Lawrie}}, \bibinfo
  {author} {\bibfnamefont {N.~W.}\ \bibnamefont {Hendrickx}}, \bibinfo {author}
  {\bibfnamefont {S.~G.~J.}\ \bibnamefont {Philips}}, \bibinfo {author}
  {\bibfnamefont {J.~S.}\ \bibnamefont {Clarke}}, \bibinfo {author}
  {\bibfnamefont {L.~M.~K.}\ \bibnamefont {Vandersypen}},\ and\ \bibinfo
  {author} {\bibfnamefont {M.}~\bibnamefont {Veldhorst}},\ }\bibfield  {title}
  {\bibinfo {title} {Universal quantum logic in hot silicon qubits},\ }\href
  {https://doi.org/10.1038/s41586-020-2170-7} {\bibfield  {journal} {\bibinfo
  {journal} {Nature}\ }\textbf {\bibinfo {volume} {580}},\ \bibinfo {pages}
  {355} (\bibinfo {year} {2020})}\BibitemShut {NoStop}%
\bibitem [{\citenamefont {Vandersypen}\ and\ \citenamefont {van
  Leeuwenhoek}(2017)}]{vandersypen2017}%
  \BibitemOpen
  \bibfield  {author} {\bibinfo {author} {\bibfnamefont {L.}~\bibnamefont
  {Vandersypen}}\ and\ \bibinfo {author} {\bibfnamefont {A.}~\bibnamefont {van
  Leeuwenhoek}},\ }\bibfield  {title} {\bibinfo {title} {1.4 {{Quantum}}
  computing - the next challenge in circuit and system design},\ }in\ \href
  {https://doi.org/10.1109/ISSCC.2017.7870244} {\emph {\bibinfo {booktitle}
  {2017 {{IEEE International Solid-State Circuits Conference}} ({{ISSCC}})}}}\
  (\bibinfo {year} {2017})\ pp.\ \bibinfo {pages} {24--29}\BibitemShut
  {NoStop}%
\bibitem [{\citenamefont {Maurand}\ \emph {et~al.}(2016)\citenamefont
  {Maurand}, \citenamefont {Jehl}, \citenamefont {{Kotekar-Patil}},
  \citenamefont {Corna}, \citenamefont {Bohuslavskyi}, \citenamefont
  {Lavi{\'e}ville}, \citenamefont {Hutin}, \citenamefont {Barraud},
  \citenamefont {Vinet}, \citenamefont {Sanquer},\ and\ \citenamefont
  {De~Franceschi}}]{maurand2016}%
  \BibitemOpen
  \bibfield  {author} {\bibinfo {author} {\bibfnamefont {R.}~\bibnamefont
  {Maurand}}, \bibinfo {author} {\bibfnamefont {X.}~\bibnamefont {Jehl}},
  \bibinfo {author} {\bibfnamefont {D.}~\bibnamefont {{Kotekar-Patil}}},
  \bibinfo {author} {\bibfnamefont {A.}~\bibnamefont {Corna}}, \bibinfo
  {author} {\bibfnamefont {H.}~\bibnamefont {Bohuslavskyi}}, \bibinfo {author}
  {\bibfnamefont {R.}~\bibnamefont {Lavi{\'e}ville}}, \bibinfo {author}
  {\bibfnamefont {L.}~\bibnamefont {Hutin}}, \bibinfo {author} {\bibfnamefont
  {S.}~\bibnamefont {Barraud}}, \bibinfo {author} {\bibfnamefont
  {M.}~\bibnamefont {Vinet}}, \bibinfo {author} {\bibfnamefont
  {M.}~\bibnamefont {Sanquer}},\ and\ \bibinfo {author} {\bibfnamefont
  {S.}~\bibnamefont {De~Franceschi}},\ }\bibfield  {title} {\bibinfo {title} {A
  {{CMOS}} silicon spin qubit},\ }\href {https://doi.org/10.1038/ncomms13575}
  {\bibfield  {journal} {\bibinfo  {journal} {Nature Communications}\ }\textbf
  {\bibinfo {volume} {7}},\ \bibinfo {pages} {13575} (\bibinfo {year}
  {2016})}\BibitemShut {NoStop}%
\bibitem [{\citenamefont {{Ciriano-Tejel}}\ \emph {et~al.}(2020)\citenamefont
  {{Ciriano-Tejel}}, \citenamefont {Fogarty}, \citenamefont {Schaal},
  \citenamefont {Hutin}, \citenamefont {Bertrand}, \citenamefont {Ibberson},
  \citenamefont {{Gonzalez-Zalba}}, \citenamefont {Li}, \citenamefont {Niquet},
  \citenamefont {Vinet},\ and\ \citenamefont {Morton}}]{ciriano-tejel2020}%
  \BibitemOpen
  \bibfield  {author} {\bibinfo {author} {\bibfnamefont {V.~N.}\ \bibnamefont
  {{Ciriano-Tejel}}}, \bibinfo {author} {\bibfnamefont {M.~A.}\ \bibnamefont
  {Fogarty}}, \bibinfo {author} {\bibfnamefont {S.}~\bibnamefont {Schaal}},
  \bibinfo {author} {\bibfnamefont {L.}~\bibnamefont {Hutin}}, \bibinfo
  {author} {\bibfnamefont {B.}~\bibnamefont {Bertrand}}, \bibinfo {author}
  {\bibfnamefont {L.}~\bibnamefont {Ibberson}}, \bibinfo {author}
  {\bibfnamefont {M.~F.}\ \bibnamefont {{Gonzalez-Zalba}}}, \bibinfo {author}
  {\bibfnamefont {J.}~\bibnamefont {Li}}, \bibinfo {author} {\bibfnamefont
  {Y.-M.}\ \bibnamefont {Niquet}}, \bibinfo {author} {\bibfnamefont
  {M.}~\bibnamefont {Vinet}},\ and\ \bibinfo {author} {\bibfnamefont
  {J.~J.~L.}\ \bibnamefont {Morton}},\ }\bibfield  {title} {\bibinfo {title}
  {Spin readout of a {{CMOS}} quantum dot by gate reflectometry and
  spin-dependent tunnelling},\ }\href {http://arxiv.org/abs/2005.07764}
  {\bibfield  {journal} {\bibinfo  {journal} {arXiv:2005.07764 [cond-mat,
  physics:quant-ph]}\ } (\bibinfo {year} {2020})}\BibitemShut {NoStop}%
\bibitem [{\citenamefont {Ansaloni}\ \emph {et~al.}(2020)\citenamefont
  {Ansaloni}, \citenamefont {Chatterjee}, \citenamefont {Bohuslavskyi},
  \citenamefont {Bertrand}, \citenamefont {Hutin}, \citenamefont {Vinet},\ and\
  \citenamefont {Kuemmeth}}]{ansaloni2020a}%
  \BibitemOpen
  \bibfield  {author} {\bibinfo {author} {\bibfnamefont {F.}~\bibnamefont
  {Ansaloni}}, \bibinfo {author} {\bibfnamefont {A.}~\bibnamefont
  {Chatterjee}}, \bibinfo {author} {\bibfnamefont {H.}~\bibnamefont
  {Bohuslavskyi}}, \bibinfo {author} {\bibfnamefont {B.}~\bibnamefont
  {Bertrand}}, \bibinfo {author} {\bibfnamefont {L.}~\bibnamefont {Hutin}},
  \bibinfo {author} {\bibfnamefont {M.}~\bibnamefont {Vinet}},\ and\ \bibinfo
  {author} {\bibfnamefont {F.}~\bibnamefont {Kuemmeth}},\ }\bibfield  {title}
  {\bibinfo {title} {Single-electron operations in a foundry-fabricated array
  of quantum dots},\ }\href {https://doi.org/10.1038/s41467-020-20280-3}
  {\bibfield  {journal} {\bibinfo  {journal} {Nature Communications}\ }\textbf
  {\bibinfo {volume} {11}},\ \bibinfo {pages} {6399} (\bibinfo {year}
  {2020})}\BibitemShut {NoStop}%
\bibitem [{\citenamefont {Zwerver}\ \emph {et~al.}(2021)\citenamefont
  {Zwerver}, \citenamefont {Kr{\"a}henmann}, \citenamefont {Watson},
  \citenamefont {Lampert}, \citenamefont {George}, \citenamefont
  {Pillarisetty}, \citenamefont {Bojarski}, \citenamefont {Amin}, \citenamefont
  {Amitonov}, \citenamefont {Boter}, \citenamefont {Caudillo}, \citenamefont
  {{Corras-Serrano}}, \citenamefont {Dehollain}, \citenamefont {Droulers},
  \citenamefont {Henry}, \citenamefont {Kotlyar}, \citenamefont {Lodari},
  \citenamefont {Luthi}, \citenamefont {Michalak}, \citenamefont {Mueller},
  \citenamefont {Neyens}, \citenamefont {Roberts}, \citenamefont {Samkharadze},
  \citenamefont {Zheng}, \citenamefont {Zietz}, \citenamefont {Scappucci},
  \citenamefont {Veldhorst}, \citenamefont {Vandersypen},\ and\ \citenamefont
  {Clarke}}]{zwerver2021}%
  \BibitemOpen
  \bibfield  {author} {\bibinfo {author} {\bibfnamefont {A.~M.~J.}\
  \bibnamefont {Zwerver}}, \bibinfo {author} {\bibfnamefont {T.}~\bibnamefont
  {Kr{\"a}henmann}}, \bibinfo {author} {\bibfnamefont {T.~F.}\ \bibnamefont
  {Watson}}, \bibinfo {author} {\bibfnamefont {L.}~\bibnamefont {Lampert}},
  \bibinfo {author} {\bibfnamefont {H.~C.}\ \bibnamefont {George}}, \bibinfo
  {author} {\bibfnamefont {R.}~\bibnamefont {Pillarisetty}}, \bibinfo {author}
  {\bibfnamefont {S.~A.}\ \bibnamefont {Bojarski}}, \bibinfo {author}
  {\bibfnamefont {P.}~\bibnamefont {Amin}}, \bibinfo {author} {\bibfnamefont
  {S.~V.}\ \bibnamefont {Amitonov}}, \bibinfo {author} {\bibfnamefont {J.~M.}\
  \bibnamefont {Boter}}, \bibinfo {author} {\bibfnamefont {R.}~\bibnamefont
  {Caudillo}}, \bibinfo {author} {\bibfnamefont {D.}~\bibnamefont
  {{Corras-Serrano}}}, \bibinfo {author} {\bibfnamefont {J.~P.}\ \bibnamefont
  {Dehollain}}, \bibinfo {author} {\bibfnamefont {G.}~\bibnamefont {Droulers}},
  \bibinfo {author} {\bibfnamefont {E.~M.}\ \bibnamefont {Henry}}, \bibinfo
  {author} {\bibfnamefont {R.}~\bibnamefont {Kotlyar}}, \bibinfo {author}
  {\bibfnamefont {M.}~\bibnamefont {Lodari}}, \bibinfo {author} {\bibfnamefont
  {F.}~\bibnamefont {Luthi}}, \bibinfo {author} {\bibfnamefont {D.~J.}\
  \bibnamefont {Michalak}}, \bibinfo {author} {\bibfnamefont {B.~K.}\
  \bibnamefont {Mueller}}, \bibinfo {author} {\bibfnamefont {S.}~\bibnamefont
  {Neyens}}, \bibinfo {author} {\bibfnamefont {J.}~\bibnamefont {Roberts}},
  \bibinfo {author} {\bibfnamefont {N.}~\bibnamefont {Samkharadze}}, \bibinfo
  {author} {\bibfnamefont {G.}~\bibnamefont {Zheng}}, \bibinfo {author}
  {\bibfnamefont {O.~K.}\ \bibnamefont {Zietz}}, \bibinfo {author}
  {\bibfnamefont {G.}~\bibnamefont {Scappucci}}, \bibinfo {author}
  {\bibfnamefont {M.}~\bibnamefont {Veldhorst}}, \bibinfo {author}
  {\bibfnamefont {L.~M.~K.}\ \bibnamefont {Vandersypen}},\ and\ \bibinfo
  {author} {\bibfnamefont {J.~S.}\ \bibnamefont {Clarke}},\ }\bibfield  {title}
  {\bibinfo {title} {Qubits made by advanced semiconductor manufacturing},\
  }\href {http://arxiv.org/abs/2101.12650} {\bibfield  {journal} {\bibinfo
  {journal} {arXiv:2101.12650 [cond-mat, physics:quant-ph]}\ } (\bibinfo {year}
  {2021})}\BibitemShut {NoStop}%
\bibitem [{\citenamefont {Hanson}\ \emph {et~al.}(2007)\citenamefont {Hanson},
  \citenamefont {Petta}, \citenamefont {Tarucha},\ and\ \citenamefont
  {Vandersypen}}]{hanson2007}%
  \BibitemOpen
  \bibfield  {author} {\bibinfo {author} {\bibfnamefont {R.}~\bibnamefont
  {Hanson}}, \bibinfo {author} {\bibfnamefont {J.~R.}\ \bibnamefont {Petta}},
  \bibinfo {author} {\bibfnamefont {S.}~\bibnamefont {Tarucha}},\ and\ \bibinfo
  {author} {\bibfnamefont {L.~M.~K.}\ \bibnamefont {Vandersypen}},\ }\bibfield
  {title} {\bibinfo {title} {Spins in few-electron quantum dots},\ }\href
  {https://doi.org/10.1103/RevModPhys.79.1217} {\bibfield  {journal} {\bibinfo
  {journal} {Reviews of Modern Physics}\ }\textbf {\bibinfo {volume} {79}},\
  \bibinfo {pages} {1217} (\bibinfo {year} {2007})}\BibitemShut {NoStop}%
\bibitem [{\citenamefont {Chatterjee}\ \emph {et~al.}(2020)\citenamefont
  {Chatterjee}, \citenamefont {Stevenson}, \citenamefont {De~Franceschi},
  \citenamefont {Morello}, \citenamefont {{de Leon}},\ and\ \citenamefont
  {Kuemmeth}}]{chatterjee2020}%
  \BibitemOpen
  \bibfield  {author} {\bibinfo {author} {\bibfnamefont {A.}~\bibnamefont
  {Chatterjee}}, \bibinfo {author} {\bibfnamefont {P.}~\bibnamefont
  {Stevenson}}, \bibinfo {author} {\bibfnamefont {S.}~\bibnamefont
  {De~Franceschi}}, \bibinfo {author} {\bibfnamefont {A.}~\bibnamefont
  {Morello}}, \bibinfo {author} {\bibfnamefont {N.}~\bibnamefont {{de Leon}}},\
  and\ \bibinfo {author} {\bibfnamefont {F.}~\bibnamefont {Kuemmeth}},\
  }\bibfield  {title} {\bibinfo {title} {Semiconductor {{Qubits In
  Practice}}},\ }\href {http://arxiv.org/abs/2005.06564} {\bibfield  {journal}
  {\bibinfo  {journal} {arXiv:2005.06564 [cond-mat, physics:quant-ph]}\ }
  (\bibinfo {year} {2020})}\BibitemShut {NoStop}%
\bibitem [{\citenamefont {Ladd}\ and\ \citenamefont
  {Carroll}(2018)}]{ladd2018}%
  \BibitemOpen
  \bibfield  {author} {\bibinfo {author} {\bibfnamefont {T.~D.}\ \bibnamefont
  {Ladd}}\ and\ \bibinfo {author} {\bibfnamefont {M.~S.}\ \bibnamefont
  {Carroll}},\ }\bibfield  {title} {\bibinfo {title} {Silicon {{Qubits}}},\
  }in\ \href {https://doi.org/10.1016/B978-0-12-803581-8.09736-8} {\emph
  {\bibinfo {booktitle} {Encyclopedia of {{Modern Optics}} ({{Second
  Edition}})}}},\ \bibinfo {editor} {edited by\ \bibinfo {editor}
  {\bibfnamefont {B.~D.}\ \bibnamefont {Guenther}}\ and\ \bibinfo {editor}
  {\bibfnamefont {D.~G.}\ \bibnamefont {Steel}}}\ (\bibinfo  {publisher}
  {{Elsevier}},\ \bibinfo {address} {{Oxford}},\ \bibinfo {year} {2018})\ pp.\
  \bibinfo {pages} {467--477}\BibitemShut {NoStop}%
\end{thebibliography}%


%

\end{document}